\pdfoutput=1
\documentclass [11pt]{article}

\usepackage{jheppub}

\usepackage[utf8]{inputenc}

\usepackage{amsfonts}

\usepackage{amsmath,amssymb}

\usepackage{mathtools}

\usepackage{graphicx,subfigure}
\usepackage[space]{grffile}
\usepackage{mathtools}

\usepackage{simplewick}

\usepackage{array}

\usepackage{float}

\usepackage{color}

\usepackage{mathrsfs}

\def\nn{\nonumber}

\newcommand{\n}{\mathbf{n}} 

\newcommand{\cO}{\mathcal{O}}

\newcommand{\<}{\langle}
\renewcommand{\>}{\rangle}

\newcommand{\myPrime}{{\; \prime}}
\newcommand{\myRho}{{\mathfrak{r}}}

\DeclareMathOperator*{\SumInt}{%
	\mathchoice%
	{\ooalign{$\displaystyle\sum$\cr\hidewidth$\displaystyle\int$\hidewidth\cr}}
	{\ooalign{\raisebox{.14\height}{\scalebox{.7}{$\textstyle\sum$}}\cr\hidewidth$\textstyle\int$\hidewidth\cr}}
	{\ooalign{\raisebox{.2\height}{\scalebox{.6}{$\scriptstyle\sum$}}\cr$\scriptstyle\int$\cr}}
	{\ooalign{\raisebox{.2\height}{\scalebox{.6}{$\scriptstyle\sum$}}\cr$\scriptstyle\int$\cr}}
}

\title{\boldmath Bootstrapping Massive Quantum Field Theories}

\author{Denis Karateev,}
\author{Simon Kuhn,}
\author{and Jo\~ao Penedones}

\affiliation{Fields and Strings Laboratory, Institute of Physics, EPFL, CH-1015 Lausanne, Switzerland}

\abstract{
We propose a new non-perturbative method for studying UV complete unitary quantum field theories (QFTs) with a mass gap in general number of spacetime dimensions.  
The method relies on unitarity formulated as  positive  semi-definiteness  of the matrix of inner products between asymptotic states (\emph{in} and \emph{out}) and states created by the action of local operators on the vacuum.
The corresponding matrix elements  involve scattering amplitudes,  form factors and  spectral densities of local operators. 
We test this method in two-dimensional QFTs by setting up  a linear optimization problem 
that gives a lower bound on the central charge of the UV CFT associated to a QFT with  a given mass spectrum of stable particles (and  couplings between them).
Some of our numerical bounds are saturated by known form factors in integrable theories like
the sine-Gordon, $E_8$ and $O(N)$ models.
	
}

\begin{document}
\maketitle

\section{Introduction}
One can define a quantum field theory (QFT) non-perturbatively as a renormalization group (RG) flow from the UV to the IR fixed point. The fixed points are assumed to have conformal invariance and are described by  conformal field theories (CFTs). In this work, we will focus on massive QFTs, \emph{i.e.} the IR CFT is trivial.
To specify a particular QFT  it is sufficient to provide the UV CFT 
and the relevant deformation triggering the RG flow. We would like to determine IR observables like the mass spectrum and scattering amplitudes from the UV data.
However, generically, the RG flow is strongly coupled and it is not possible to compute IR observables using perturbation theory around the UV CFT. 
In these cases, one has to  resort to  numerical methods (like lattice field theory, Hamiltonian truncation, tensor networks, etc) that require a UV cutoff and a costly extrapolation to the continuum limit.
This calls for  modern non-perturbative bootstrap methods that can constrain the space of QFTs directly in the continuum.
Unfortunately, the present  bootstrap methods study the UV and the IR separately.
Namely, one can use the conformal bootstrap \cite{Rattazzi:2008pe}\footnote{See also the review \cite{Poland:2018epd}.} to study the UV CFT data or the S-matrix bootstrap \cite{Paulos:2016fap,Paulos:2016but,Paulos:2017fhb,Homrich:2019cbt,Guerrieri:2019rwp,EliasMiro:2019kyf} to study scattering amplitudes of light particles.
Ideally, we would like to connect these two bootstrap approaches.
This work is a step in this direction.\footnote{The study of QFT in Anti-de Sitter (AdS) spacetime is another promising strategy to connect the conformal and the S-matrix bootstraps \cite{Paulos:2016fap}. However, this approach requires the introduction of an IR cutoff (the AdS radius).
The limit of large AdS radius leads to the usual conformal bootstrap for operators with large scaling dimension, which is very challenging with current methods.}

Our strategy is simple.  We consider a set of states that include asymptotic scattering states and states created by local operators acting on the vacuum. In the simplest setting, we consider the following three states\footnote{\label{note1}These formulas are schematic. The precise formulas are given in section \ref{sec:constraints_unitarity}.}
\begin{equation}
| \psi_1 \rangle =  |p_1, p_2 \rangle_{in}\,,\qquad
| \psi_2 \rangle = |p_1, p_2 \rangle_{out}\,,\qquad
| \psi_3 \rangle = \int dx e^{i(p_1+p_2)\cdot x} \mathcal{O}(x) |0 \rangle\,.
\end{equation}
Unitary  implies positive semi-definiteness  of the matrix\textsuperscript{\ref{note1}}
\begin{equation}
\langle \psi_a | \psi_b \rangle=
\begin{pmatrix}
1 & \mathcal{S}^*
& \mathcal{F}^*_2 \\
\mathcal{S} & 1
& \mathcal{F}_2 \\
 \mathcal{F}_2  
&  \mathcal{F}^*_2  &
\rho
\end{pmatrix} \succeq 0
\end{equation}
where $\mathcal{S}$ denotes the 2 to 2 scattering amplitude, $\mathcal{F}_2$ denotes the two-particle form factor of the operator $\mathcal{O}$ and $\rho$ its spectral density.

Notice that  the S-matrix bootstrap can be formulated in the same way if we only use scattering states ($| \psi_1 \rangle$ and $| \psi_2 \rangle$ in this case). 
The  presence of the spectral density $\rho \sim \langle \mathcal{O}\mathcal{O}\rangle$  in the setup makes it straightforward to establish a connection with the UV CFT: at large energies all the correlation functions should  coincide with the ones of the UV CFT. For example in the case of the two-point function of the stress-tensor,  conformal invariance in the UV fixes its form uniquely up to a constant known as the $C_T$ central charge \cite{Osborn:1993cr}. 
Notice that even if we knew the S-matrix for all energies it would not be easy
to extract  from it  information about the UV CFT.\footnote{In $d\ge 3$ using holography one can argue that the regime of hard scattering (high energy and fixed angle) should be directly related to the UV CFT \cite{Polchinski:2001tt}.
However, the present S-matrix bootstrap methods are not precise enough in this regime.}

In this work we test this strategy in 1+1 dimensional QFTs. In this case, we can write the central charge $c$ of the UV CFT as an integral over the spectral density of the trace of the stress tensor. This allows to address the following question: 
\emph{what is the minimal central charge of a UV CFT that can give rise to a massive QFT with a given set of masses and couplings\footnote{We define couplings from the physical S-matrix. For example, cubic couplings are given by residues of poles of the 2 to 2 scattering amplitudes.} of stable particles?}

In practice, we use analyticity of the amplitude and form factor, to  write a general ansatz for the amplitude, the form factor and the spectral density. 
Then we numerically optimize the parameters of the  ansatz such that  unitarity is obeyed at all  energies and the value of the central charge is as low as possible. 
In several cases we find that the optimal form factors are given by known integrable theories 
such as the sine-Gordon, the $E_8$ and the $O(N)$ models.

The structure of the paper is as follows. In section \ref{sec:review} we review all the basic ingredients in a consistent manner for generic number of dimensions and provide all the normalization conventions. 
In section \ref{sec:constraints_unitarity} we formulate unitarity as the semipositive definite condition on the three by three matrix $\langle \psi_a | \psi_b \rangle $ and discuss its implications. 
Then, we illustrate how this works with analytic examples from 2d integrable models like the sine-Gordon, the $E_8$ and $O(N)$ models, which we review in section  \ref{app:analytic_integrable_models}.
In section \ref{sec:numerics_2d} we define and set up the numerical linear optimization problem for 2d QFTs. We also present our results and compare them with the analytic formulas for integrable models. We conclude and briefly discuss applications to higher dimensions in section \ref{sec:conclusions}. We derive various auxiliary results in appendices \ref{app:auxiliary}, \ref{app:KL_euclidean} and \ref{sec:free_boson}.

\section{Review of basic ingredients}
\label{sec:review}
We work in $(1,d-1)$ Minkowski space with the mostly plus metric
\begin{equation}
\label{eq:metric}
\eta^{\mu\nu}=\{-,+,\ldots,+\}.
\end{equation}
The position and momentum in this $d$-dimensional space are denoted respectively by
\begin{equation}
x^\mu = \{x^0,\vec x\},\quad
p^\mu = \{p^0,\vec p\,\},
\end{equation}
where $\vec x$ and $\vec p$ are the position and momentum in the $(d-1)$ Euclidean subspace. We refer to $p^0$ as energy and $x^0$ as time.

We study unitary quantum field theories with restricted Poincar\'e symmetry group.\footnote{It is defined as the Poincar\'e group without parity and time-reversal discrete subgroups.} When working in $d=2$ in addition we will also assume parity. In what follows we summarize our conventions and review basic ingredients such as asymptotic states, scattering and partial amplitudes, spectral density and form factors. We will conclude with a  discussion of unitarity and its implications.

\subsection{States}
\label{eq:states}
The state of a system described by the unitary QFT is represented by a ``state'' vector in an infinite dimensional 
Hilbert space. 
In this space it is convenient to choose a basis of state vectors (or simply states) in such a way that they are eigenstates of the generators of translations $P^\mu$ with  eigenvalues $p^\mu$ and transform in the irreducible representation of the Little group $SO(d-1)$ which leaves invariant the d-vector $\{p^0,\vec 0\}$. We will always work with states which have a strictly positive energy $p^0>0$. We also restrict our attention to traceless symmetric representations of the  $SO(d-1)$ Little group. Then any state will have at least three labels
\begin{equation}
\label{eq:states_irreps}
|p,j,\mu\>.
\end{equation}
The label $j$ is a non-negative integer called spin $(j=0,1,2,\ldots)$. 
The label $\mu$ denotes the components of the spin $j$ irreducible representation of $SO(d-1)$. In the case of $d=4$, one can choose  $\mu=-j,\ldots,+j$ to be the helicity, \emph{i.e.} 
the projection of spin $j$ on the direction of $\vec p$. 
The normalization of the states \eqref{eq:states_irreps} is chosen as
\begin{align}
\label{eq:normalization_states}
\<p',j',\mu'|p,j,\mu\>
&=
(2\pi)^d\delta^{(d)}(p^\prime-p)
\delta_{j'j}\delta_{\mu'\mu}
\end{align}
except for the special case of one particle states that are normalized as in \eqref{eq:normalization_1PS}.  
We note that from this normalization condition it follows that the dimensionality of the state vector is
\begin{equation}
\label{eq:dimension}
\big[|p,j,\mu\>\big]=-\frac{d}{2}.
\end{equation}
For further discussion, see construction of irreducible unitary representations of the restricted Poincar\'e group \cite{Weinberg:1995mt,Tung:1985na}.

\subsubsection{Free particle states}

As we review in the next section, the asymptotic states of an interacting massive QFT are in one-to-one correspondence with the states of a non-interacting QFT.
For this reason, we first consider a   free QFT. 
There is a  special set of states, called the one particle states, which describe a single freely propagating particle. The tensor product of one particle states defines  multi particle states which describe a system of multiple non-interacting particles. The Hilbert space spanned by all the possible one and multi particle states is called the Fock space.

The states \eqref{eq:states_irreps} which obey the ``mass-shell'' condition
\begin{equation}
\label{eq:on-shell_condition}
p^2=-m^2\quad\Rightarrow\quad
p^0=\sqrt{m^2+\vec p\,^2},
\end{equation}
where $m$ is a discrete real non-negative number, called mass, are referred to as one particles states (1PS). We can denote them as
\begin{equation}
\label{eq:1PS}
|m,\vec p\,\>.
\end{equation}
We focus only on scalar particles in this work, thus we omit the spin labels $j=\mu=0$. The one particle states are normalized as
\begin{equation}
\label{eq:normalization_1PS}
\<m',\vec p\,'|m,\vec p\,\>=
2p^0\delta_{m'm}\times
(2\pi)^{d-1}\delta^{(d-1)}(\vec p\,^\prime-\vec p).
\end{equation}
From the above normalizations it is clear that the one particle states have the following mass dimensions
\begin{equation}
\label{eq:dimension_1PS}
\big[|m,\vec p\,\>\big]=-\frac{d-2}{2}.
\end{equation}

We define the $n$ particle state as
\begin{equation}
\label{eq:n-particle_state}
|\mathbf{n}\> \equiv |m_1,\vec p_1 \>
\otimes\ldots\otimes
|m_n,\vec p_n \>.
\end{equation}
The $n$ particle state has a well defined total $d$ momentum which reads as
\begin{equation}
\label{eq:total_p}
p^\mu = p_1^\mu+\ldots+p_n^\mu.
\end{equation}
Due to the very definition of the Fock space one can write the completeness relation in this space by summing over $n$ particle states and integrating over their phase space as
\begin{equation}
\label{eq:identity_operator}
\mathbb{I}= \displaystyle\SumInt_n|\mathbf{n}\>\<\mathbf{n}|,\quad
\displaystyle\SumInt_n \equiv \sum_{n=0}^{\infty}\int d\Phi_n,
\end{equation}
where the phase space $\Phi_n$ for $n$ identical particles is defined in \eqref{eq:phase_space}. 

Let us finish this section by focusing on  two particles states $|{\bf 2}\>$ of identical particles with mass $m$. Writing all the labels explicitly we denote it by
\begin{equation}
\label{eq:2PS}
|m,\vec p_1; m,\vec p_2\>\equiv
|m,\vec p_1\>\otimes
| m,\vec p_2\>.
\end{equation}
This state does not transform in the irreducible representation of the restricted Poincar\'e group (like any other $n$ particle state with $n\geq 2$) simply because it is not in the irreducible representation of the $SO(d-1)$ Little group. We can project it however to irreducible representations. For simplicity we focus on the two particle states in the center of mass frame defined as $\vec p_2 = - \vec p_1$ and the vector $\vec p_1$ has an angle $\theta_1$ with the $x^1$ axis and $\theta_2=\ldots =\theta_{d-2}=0$. See \eqref{eq:axis_1} for our conventions for spherical coordinates. The projection is done by integrating over the $(d-1)$ scalar spherical harmonics, which are the Gegenbauer polynomials, as\footnote{Strictly speaking \eqref{eq:irreps} holds only for $d\geq 4$ when the Little groups is non-Abelian. The $d=2$ and $d=3$ are special. In the former case the Little group is $Z_2$ and in the latter it is Abelian $SO(2)$.}
\begin{equation}
\label{eq:irreps}
|p,j\> = \Pi_j |m,\vec p_1; m,-\vec p_1\> \equiv \gamma_j\times\int d\Omega_{d-1}C_{j}^{(d-3)/2}(\cos\theta_1)\, |m,\vec p_1; m,-\vec p_1\>,
\end{equation}
where $\gamma_j$ is some coefficient fixed by the normalization, which we derive in \eqref{eq:coefficient_gamma}. In the left-hand side \eqref{eq:irreps} we dropped the label $\mu$ because we are considering states with zero spin projection along $\vec p_1$ and invariant under $SO(d-3)$ rotations  that leave the scattering plane ($\theta_2=\ldots =\theta_{d-2}=0$) invariant. 

The normalization of the two particle state \eqref{eq:2PS} is fixed by the normalization of one particle states \eqref{eq:normalization_1PS}. One has
\begin{multline}
\label{eq:normalization_2PS}
 \<m,\vec p_1^\myPrime;m, \vec p_2^\myPrime|m, \vec p_1;m, \vec p_2\>
= 4 p_1^0p_2^0\times
(2\pi)^{2\,(d-1)}\delta^{(d-1)}(\vec p_1^\myPrime - \vec p_1)\delta^{(d-1)}(\vec  p_2^\myPrime - \vec p_2) 
+(\vec p_1 \leftrightarrow \vec p_2)\\
= \mathcal{N}_d\times(2\pi)^{d}\delta^{(d)}(p_1'+p_2'-p_1-p_2)
\times(2\pi)^{d-2}\left(
\delta^{(d-2)}(\Omega'-\Omega)+\delta^{(d-2)}(\Omega'+\Omega)
\right).
\end{multline}
Notice that the normalization \eqref{eq:normalization_2PS} reflects explicitly that the system is symmetric under the permutation of particles 1 and 2.
In the second line of \eqref{eq:normalization_2PS} we have performed a change of variables, see \eqref{eq:change_of variables}. For identical particles the factor $\mathcal{N}_d$, derived in \eqref{eq:factor_N_full}, reads as
\begin{align}
\label{eq:factor_N}
\quad\mathcal{N}_d &\equiv 2^{d-1} \sqrt{s} \,\left(s-4m^2\right)^{(3-d)/2},\\
s &\equiv - (p_1+p_2)^2.
\label{eq:mandelstam}
\end{align}
In \eqref{eq:normalization_2PS} the spherical angles $\Omega$ and $\Omega'$ correspond to the $(d-1)$ vectors $\vec p_1$ and $\vec p_1^\myPrime$ respectively. The $\delta$-function in spherical coordinates is defined in \eqref{eq:delta_spherical}.\footnote{Given a spherical angle $\Omega$ of a $d-1$ vector $\vec p$, we schematically denote by $-\Omega$ the spherical angle of a $d-1$ vector $-\vec p$. If the former has the angles $(\theta_1,\ldots, \theta_{d-3},\theta_{d-2})$, the latter has all the angles shifted as $(\pi-\theta_1,\ldots,\pi-\theta_{d-3},\pi+\theta_{d-2})$. This is easy to see from \eqref{eq:axis_1}.}
The Mandelstam variable \eqref{eq:mandelstam} defines the square of the total energy for the two particle state in the center of mass frame.
We can now evaluate the value of the constant $\gamma_j$. Using \eqref{eq:normalization_states}, \eqref{eq:factor_N} and the orthogonality relation \eqref{eq:orthogonality_1}, we get\footnote{Without loss of generality the factor $\gamma_j$ is chosen to be purely real in the rest of the paper.}
\begin{equation}
\label{eq:coefficient_gamma}
|\gamma_j|^{-2} = (1+(-1)^j)\times\mathcal{N}_d(2\pi)^{d-2}\Omega_{d-2}\times\nu_j^{(d-3)/2},
\end{equation}
where the coefficient $\nu_j^{(d-3)/2}$ is defined in \eqref{eq:orthogonality_1}.

Finally, let us invert the projection \eqref{eq:irreps} by means of the orthogonality relation \eqref{eq:orthogonality_2}. One finds
\begin{equation}
\label{eq:decomposition}
|m, \vec p_1;m, -\vec p_1\> = \sum_{j=0}^\infty C_j(\cos\theta_1) |p,j\>,
\end{equation}
where the Clebsch-Gordan coefficient $C_j(\cos\theta_1)$ reads as
\begin{equation}
\label{eq:CG}
C_j(\cos\theta_1)= \left(\frac{\big(1+(-1)^j\big)\mathcal{N}_d(2\pi)^{d-2}}{\Omega_{d-2}\nu_j^{(d-3)/2}}\right)^{1/2}\times C_j^{(d-3)/2}(\cos\theta_1).
\end{equation}

\subsubsection{Asymptotic states}
This section is based on chapter 3.1 of \cite{Weinberg:1995mt}.

We work with states in the Heisenberg picture (states do not evolve in time) and describe the entire evolution of the system. They are defined however with an implicit choice of a reference frame $f$. Suppose we have another reference frame $f'$ with time $t'=t+\tau$.\footnote{If some event happens at $t=0$ in $f$, the very same event happens at $t'=\tau$ in $f'$.}  If a state $|\psi\>$ is seen by an observer in $f$, the same state will be seen by an observer in $f'$ as $|\psi'\>$. Due to time translation invariance these two states are related as
\begin{equation}
|\psi'\>=e^{-iH\tau}|\psi\>.
\end{equation}

In strongly interacting theories with a mass gap one can define a (complicated) state in the reference frame $f$, which for an observer in the reference frame $f'$ either in the far past ($\tau\rightarrow -\infty$) or in the far future ($\tau\rightarrow +\infty$) however will look like a set of non-interacting (free) particles. We call states with such a property the asymptotic {\it in} and {\it out} states. In what follows we will make this statement formal.

We assume that the strongly interacting Hamiltonian of our system can be written in the following way
\begin{equation}
\label{eq:hamiltonian}
H\equiv P^0,\quad H = H_0 + H_{int},
\end{equation}
where $H_0$ is a free Hamiltonian (with the mass spectrum including stable composite particles and bound states) and $H_{int}$ is the ``interaction part''. Note, that $H_{int}$ is not a small perturbation around $H_0$ and we do not know how to construct it explicitly.\footnote{Not all the systems can be written in such a way. Notable example are system with long-range interactions.} As a consequence the expression \eqref{eq:hamiltonian} is highly formal.
The eigenstates of the free Hamiltonian $H_0$ are nothing but the $n$ particle states defined in \eqref{eq:n-particle_state}, in other words
\begin{equation}
\label{eq:free_hamiltonian_eigenvalues}
H_0 |\mathbf{n}\> = p^0 |\mathbf{n}\>,
\end{equation}
where $p^\mu$ is the total $d$-momentum of the $n$-particle state \eqref{eq:total_p}.

We can now define the {\it in} state $|\mathbf{n}\>_{in}$ and the {\it out} state $|\mathbf{n}\>_{out}$ via the following conditions\footnote{The relation below should be understood in a sense of wave packets.}
\begin{equation}
\label{eq:condition_assymptotic_states}
\begin{aligned}
\lim_{\tau\rightarrow-\infty} e^{-iH\tau} |\mathbf{n}\>_{in\;\,} =
\lim_{\tau\rightarrow-\infty} e^{-iH_0\tau} |\mathbf{n}\>,\\
\lim_{\tau\rightarrow+\infty} e^{-iH\tau} |\mathbf{n}\>_{out} =
\lim_{\tau\rightarrow+\infty} e^{-iH_0\tau} |\mathbf{n}\>.
\end{aligned}
\end{equation}
Here the asymptotic and the free $n$ particle states are defined in the reference frame $f$ and are required to match in the reference frame $f'$.
Given the condition \eqref{eq:condition_assymptotic_states} one can express the {\it in} and {\it out }  states in terms of the free $n$ particle states as
\begin{equation}
\label{eq:moller_operator}
|\mathbf{n}\>_{in} =  \Omega(-\infty)|\mathbf{n}\>,\qquad
|\mathbf{n}\>_{out} = \Omega(+\infty)|\mathbf{n}\>,
\end{equation}
where we have defined the operator 
\begin{equation}
\Omega(\tau)\equiv e^{+iH\tau}e^{-iH_0\tau},
\end{equation}
known as the M{\o}ller operator. For details see \cite{taylor2012scattering}. Clearly, the M{\o}ller operator is unitary
\begin{equation}
\label{eq:unitarity_moller_operators}
\Omega^\dagger(\tau)\Omega(\tau) = \Omega(\tau)\Omega^\dagger(\tau) = 1.
\end{equation}
From this it follows that the normalization of the asymptotic states is the same as the  one of the $n$ particle states
\begin{equation}
\label{eq:normalization_as}
{}_{in}\<\mathbf{m}|\mathbf{n}\>_{in} =
{}_{out}\<\mathbf{m}|\mathbf{n}\>_{out} =
\<\mathbf{m}|\mathbf{n}\>.
\end{equation}

Let us from now assume that all the asymptotic states in the theory span a complete basis of states. Then the completeness relation \eqref{eq:identity_operator} can also be written for the asymptotic states. Multiplying \eqref{eq:identity_operator} by $\Omega(\mp\infty)\Omega^\dagger(\mp\infty)$ and using \eqref{eq:unitarity_moller_operators} and \eqref{eq:moller_operator} we simply get
\begin{equation}
\label{eq:identity_operator_asymptotic}
\mathbb{I}= \displaystyle\SumInt_n
|\mathbf{n}\>_{in}\;{}_{in}\<\mathbf{n}|=
\displaystyle\SumInt_n
|\mathbf{n}\>_{out}\;{}_{out}\<\mathbf{n}|.
\end{equation}

\subsection{Scattering and partial amplitudes}
\label{sec:amplitudes}
The scattering process of $n$ free particles in the far past and $m$ free particles in the far future 
is described by the $n\rightarrow m$ scattering amplitude defined as follows.
\begin{equation}
\label{eq:amplitude}
\mathcal{S}
(p_1,\ldots,p_n;p'_1,\ldots,p'_m)\times (2\pi)^d\delta^{(d)}(p'-p)
\equiv{}_{out}\<\mathbf{m}|\n\>_{in} = \<\mathbf{m}|S|\n\>.
\end{equation}
Here $p_i$ and $p'_i$ describe the $d$-momenta of the one particle states constituing $|\n\>$ and $|\mathbf{m}\>$, $p$ and $p'$ denote the total incoming and outgoing momenta. In \eqref{eq:amplitude} we have explicitly extracted the overall $\delta$-function.
The scattering operator $S$ due to \eqref{eq:moller_operator} reads as
\begin{equation}
\label{eq:s_matrix_definition}
S\equiv \Omega^\dag(+\infty) \Omega(-\infty).
\end{equation}
It can be split into the trivial and the interacting part as
\begin{equation}
\label{eq:S_split}
S = 1 + iT.
\end{equation}

From now on let us focus on the $2\rightarrow 2$ processes of identical scalar particles. The expression \eqref{eq:amplitude} then reads as
\begin{equation}
\label{eq:amplitude_2to2}
\mathcal{S}(s,t,u)\times (2\pi)^d\delta^{(d)}(p_1'+p_2'-p_1-p_2)= \<m,\vec p_1^\myPrime;m,\vec p_2^\myPrime|S|m,\vec p_1;m,\vec p_2\>,
\end{equation}
where we have defined the three Mandelstam variables as
\begin{equation}
\label{eq:mandelstam_variable}
s\equiv -(p_1+p_2)^2,\quad
t\equiv -(p_1-p_1')^2,\quad
u\equiv -(p_1-p_2')^2,
\end{equation}
which obey the standard constraint
\begin{equation}
\label{eq:relationsSTU}
s+t+u=4 m^2.
\end{equation}
Notice, that the $s$ variable has already appeared in \eqref{eq:mandelstam}.

The partial amplitude of the $2\rightarrow 2$ process is defined as a matrix element of the $S$ operator between the states \eqref{eq:irreps}, namely
\begin{equation}
\label{eq:partial_amplitude}
\mathcal{S}_j(s) \times \delta_{j'j}(2\pi)^d \delta^{(d)}(p-p') \equiv
\<p',j'|S|p,j\>.
\end{equation}
We would now like to write the relation between the partial amplitude $\mathcal{S}_j(s)$ defined in \eqref{eq:partial_amplitude} and the scattering amplitude \eqref{eq:amplitude_2to2}. In principle this can be done by simply plugging
\eqref{eq:irreps} into \eqref{eq:partial_amplitude}, however it is easier to derive this relation in the following way.
Take the two particle states in the center of mass frame $\<m,\vec p\,';m,-\vec p\,'|$ and $|m,\vec p;m,-\vec p\>$. We align the incoming particle $\vec p$ with the $x^1$ axis. The outgoing particle $\vec p^\myPrime$ will have an angle $\theta_1$ with the $x^1$ axis. All the other angles for incoming and outgoing particles are chosen to be zero.  In this frame it is very convenient, instead of using $(s,t,u)$ obeying the constraint \eqref{eq:relationsSTU}, to use the variables $(s,\,\cos\theta_1)$. The variables $t$ and $u$ can then be written as
\begin{equation}
\label{eq:mandelstam_variables_angle}
t = - \frac{s-4m^2}{2}\,(1-\cos\theta_1),\quad
u = - \frac{s-4m^2}{2}\,(1+\cos\theta_1).
\end{equation}
We consider the scattering amplitude \eqref{eq:amplitude_2to2} and apply the decomposition of states \eqref{eq:decomposition}. Using the definition \eqref{eq:partial_amplitude} we can write
\begin{equation}
\mathcal{S}(s,\,\cos\theta_1) = 
\sum_{j=0}^\infty C_{j}(1)C_{j}(\cos\theta_1) \mathcal{S}_j(s).
\end{equation}
This relation can be inverted by means of \eqref{eq:orthogonality_1} and leads to
\begin{align}
\label{eq:partial_scattering}
\mathcal{S}_j(s) &= \kappa_j\times
\int_{-1}^{+1}dx \,(1-x^2)^{\frac{d-4}{2}} C_j^{(d-3)/2}(x)\, \mathcal{S}(s,x),\\
x &\equiv \cos\theta_1,
\end{align}
where the coefficient $\kappa_j$ reads as
\begin{equation}
\label{eq:kappa}
\kappa_j\equiv
\frac{\Omega_{d-2}}{2\mathcal{N}_d(2\pi)^{d-2}\,C_j^{(d-3)/2}(+1)}=  \frac{j!\,\Gamma\left(\frac{d-3}{2}\right)}{4(4\pi)^{(d-1)/2}\Gamma(d-3+j)}
\times\frac{\left(s-4\,m^2\right)^{(d-3)/2}}{\sqrt{s}}.
\end{equation}
Notice, that for identical particles only the partial amplitudes with even spin $j$ exist.  The partial amplitudes \eqref{eq:partial_scattering} with odd spin $j$ vanish due to the $x \leftrightarrow -x$ symmetry of the scattering amplitude and the antisymmetry of the Gegenbauer polynomial.

To conclude let us address the consequences of \eqref{eq:S_split}. The scattering amplitude \eqref{eq:amplitude_2to2} can be split into the connected and the disconnected parts according to \eqref{eq:S_split}. Using  \eqref{eq:normalization_2PS} we can write
\begin{equation}
\label{eq:amplitude_2to2_expanded}
\mathcal{S}(s,t,u)= \mathcal{N}_d
\times(2\pi)^{d-2}\left(
\delta^{(d-2)}(\Omega'-\Omega)+\delta^{(d-2)}(\Omega'+\Omega)
\right)
+ i \mathcal{T}(s,t,u),
\end{equation}
where we have defined
\begin{equation}
\label{eq:connected_part_scat_amp}
\mathcal{T}(s,t,u)\times (2\pi)^d\delta^{(d)}(p_1'+p_2'-p_1-p_2)\equiv
\<m,\vec p_1^\myPrime;m,\vec p_2^\myPrime|T|m,\vec p_1;m,\vec p_2\>.
\end{equation} 
The connected (interacting) part of the amplitude $\mathcal{T}(s,t,u)$ should not be confused with time-reversal operator $\mathcal{T}$ which we unfortunately denote in the same way.
Combining \eqref{eq:partial_scattering} and \eqref{eq:amplitude_2to2_expanded} we can also write a similar expression for the partial amplitude
\begin{align}
\label{eq:partial_amplitude_final}
\mathcal{S}_j(s) &= 1 + i\,\kappa_j\mathcal{T}_j(s),\\
\mathcal{T}_j(s) &\equiv
\int_{-1}^{+1}dx \,(1-x^2)^{\frac{d-4}{2}} C_j^{(d-3)/2}(x)\, \mathcal{T}(s,x).
\end{align}
This matches precisely the expression given in equation (10) of \cite{Paulos:2017fhb}.

\subsection{Spectral density}
In this section we will discuss two important instances of two-point correlation functions, namely the two-point Wightman and time-ordered correlation functions. We will define the notion of spectral density and show how both types of two-point functions can be rewritten in terms of the spectral density.

\paragraph{Wightman correlation functions}
Let us consider a local operator $\cO(x)$ and study the Wightman two-point correlation function
\begin{equation}
\label{eq:wightman_function}
\<0| \cO^\dagger(x_1^0-i\epsilon_1,\,\vec x_1) \cO(x_2^0-i\epsilon_2,\,\vec x_2)|0\>,
\quad \epsilon_1>\epsilon_2,
\end{equation}
where $\epsilon_i$ are infinitesimal positive numbers. See \eqref{eq:wightman} for slightly more details. In what follows we will not display the $i\epsilon$'s in order not to complicate the notation. They are however always present and must be taken into account when we deal with  Wightman functions.

By using translation operators we can write\footnote{Translations by $a^\mu$ are given by the operator $U(a)=e^{-iP\cdot a}$. We follow the conventions of chapter 2 \cite{Weinberg:1995mt}.}
\begin{equation}
\label{eq:operator_transformation}
\cO(x) = e^{-iP\cdot x} \cO(0) e^{+iP\cdot x}.
\end{equation}
Assuming that the basis of asymptotic states is complete, we can inject the completeness relation \eqref{eq:identity_operator_asymptotic} into \eqref{eq:wightman_function} and using \eqref{eq:operator_transformation} we find
\begin{equation}
\label{eq:spectral_representation}
\<0| \cO^\dagger(x_1) \cO(x_2)|0\> =
\int \frac{d^dp}{(2\pi)^d}\; e^{ip\cdot x_{12}} \; (2\pi)\theta(p^0)\rho(-p^2),
\quad x_{ij}^\mu\equiv x_i^\mu-x_j^\mu,
\end{equation}
where $\theta$ is the step function and $\rho$ is the spectral density defined via\footnote{Notice that $\<0|\cO(0)|{\bf n}\>_{in}={}_{in}\<{\bf n}|\cO^\dagger(0)|0\>^*$ and $\<0|\cO(0)|{\bf n}\>_{out}={}_{out}\<{\bf n}|\cO^\dagger(0)|0\>^*$.}

\begin{align}
\label{eq:spectral_density}
(2\pi)\theta(p^0)\rho(-p^2)
&= \displaystyle\SumInt_n\,  (2\pi)^d\delta^{(d)}(p-p_n)
|\<0|\cO^\dagger(0)|\n\>_{in}|^2,\\
\label{eq:spectral_density_out}
&= \displaystyle\SumInt_n\,  (2\pi)^d\delta^{(d)}(p-p_n)
|\<0|\cO^\dagger(0)|\n\>_{out}|^2.
\end{align}

The Fourier transform of the Wightman function \eqref{eq:wightman_function} is related to the spectral density in the following simple way
\begin{equation}
\label{eq:fourier_spectral_representation}
\xi (k^2) \equiv \int d^d x\;e^{-ik\cdot x}\<0| \cO^\dagger(x) \cO(0)|0\>= (2\pi)\theta(k^0)\rho (-k^2).
\end{equation}
When taking the Fourier transform we integrate over coincident points which is potentially dangerous. The presence of $i\epsilon$'s in the time components ensures that the integral always converges since it gives a dumping prefactor $e^{-H\epsilon}$ with $\epsilon>0$.

The spectral representation \eqref{eq:spectral_representation} can be further rewritten by injecting an additional $\delta$-function and integrating over it. One then has
\begin{equation}
\label{eq:spectral_representation_2}
\<0| \cO^\dagger(x_1) \cO(x_2)|0\> = \int_0^\infty d\mu^2 \rho(\mu^2) \Delta_{+}(x_{12};\mu^2),
\end{equation}
where we have defined the Wightman propagator as
\begin{equation}
\label{eq:wightman_propagator}
\Delta_{+}(y;\mu^2) \equiv \int \frac{d^dp}{(2\pi)^d}\;
e^{ip\cdot y}\; (2\pi)\theta(p^0)\delta(p^2+\mu^2),
\end{equation}
which satisfies the Klein-Gordon equation
\begin{equation}
(\partial^2_{y}-\mu^2) \Delta_{+}(y;\mu^2) = 0.
\end{equation}

\paragraph{Time-ordered correlation functions}
Now let us consider the time-ordered correlation function
\begin{equation}
\<0| \cO^\dagger(x_1) \cO(x_2)|0\>_T \equiv
\theta(x_1^0-x_2^0)\<0| \cO^\dagger(x_1) \cO(x_2)|0\>+
\theta(x_2^0-x_1^0)\<0| \cO(x_2) \cO^\dagger(x_1)|0\>.
\end{equation}
Plugging here the expression \eqref{eq:spectral_representation_2} we obtain the K\"{a}ll\'en-Lehmann spectral representation of the time-ordered two-point correlation function
\begin{equation}
\label{eq:spectral_representation_3}
\<0| \cO^\dagger(x_1) \cO(x_2)|0\>_T = -i\int_0^\infty d\mu^2 \rho(\mu^2) \Delta_F(x_{12};\mu^2),
\end{equation}
where the Feynman propagator is defined as
\begin{equation}
\label{eq:f_prop_inter}
-i\Delta_F(x_{12};\mu^2) \equiv
\theta(x_1^0-x_2^0) \Delta_{+}(x_{12};\mu^2)+
\theta(x_2^0-x_1^0) \Delta_{+}(x_{21};\mu^2).
\end{equation}
Equivalently, we can write 
the Feynman propagator \eqref{eq:f_prop_inter} in its standard form\footnote{To see the equivalence between \eqref{eq:feynman_propagator} and \eqref{eq:f_prop_inter} just integrate over $q^0$ in  \eqref{eq:feynman_propagator} using the residue theorem.}
\begin{equation}
\label{eq:feynman_propagator}
\Delta_F(y;\mu^2) =
\lim_{\epsilon\rightarrow 0^+}
\int \frac{d^d q}{(2\pi)^d}\;e^{iq\cdot y}\; \frac{1}{q^2+\mu^2-i\epsilon}.
\end{equation}
From the above expression it is clear that the Feynman propagator satisfies the Klein-Gordon equation with a source
\begin{equation}
(\partial^2_{y}-\mu^2) \Delta_{F}(y;\mu^2) = -\delta^{(d)}(y).
\end{equation}
Finally, the Fourier transform of the time-ordered two-point function reads as
\begin{equation}
\label{eq:fourier_kallen-lehman}
\xi_T (k^2) \equiv \int d^d x_{12}\;e^{-ik\cdot x_{12}}\<0| \cO^\dagger(x_1) \cO(x_2)|0\>_T=
\int_0^\infty d\mu^2 \rho(\mu^2)\; \frac{-i}{k^2+\mu^2-i\epsilon}.
\end{equation}

\paragraph{High energy behavior}

In the UV, due to the presence of conformal symmetry, the Wigthman function \eqref{eq:wightman_function} is completely fixed and reads as\footnote{For the  special cases of scaling dimensions $\Delta=d/2+n$, where $n$ is a non-negative integer we may also have   contact terms like $\partial^{2n}\delta^{(d)}(x_{12})$. We disregard these cases. }
\begin{equation}
\label{eq:two_point_function_CFT}
\< 0|\cO^\dagger(x_1)\cO(x_2) |0\>_{CFT} = 
\frac{c_{\<\cO^\dagger\cO\>}}{\left(
-(x_1^0-x_2^0-i\epsilon)^2+(\vec x_1-\vec x_2)^2\,
\right)^{\Delta}},
\end{equation}
where $\Delta$ is the scaling dimension of $\cO$ and $c_{\<\cO^\dagger\cO\>}$ is a normalization constant. One can straightforwardly establish the relation between the two-point function \eqref{eq:wightman_function} at generic energies and the two-point function \eqref{eq:two_point_function_CFT} in the UV (at extremely large energies) via their Fourier transforms \eqref{eq:fourier_spectral_representation} as
\begin{align}
\lim_{s\rightarrow +\infty}\rho(s) &= \rho_{\text{CFT}}(s),\\
\label{eq:spectral_density_CFT}
\rho_{\text{CFT}}(s) &= \text{const}\times s^{\delta},\\
\delta&\equiv\Delta-d/2.
\label{eq:delta}
\end{align}
where $\rho_{\text{CFT}}(s)$ simply follows from \eqref{eq:two_point_function_CFT}.
The precise value of the constant factor is irrelevant for this work. It is found straightforwardly by performing the Fourier transform carefully. For its value see (2.4) in \cite{Gillioz:2018mto} and section 2 in \cite{Bautista:2019qxj}.

\subsection{Form factors}
\label{sec:form_factors}
Consider the following matrix elements called the form factors
\begin{equation}
\label{eq:fourier_transformed_form_factors}
\mathcal{F}_n(p_1,\ldots,p_n) \equiv {}_{out}\<\n|\cO(0)|0\>,\qquad
\mathcal{G}_n(p_1,\ldots,p_n) \equiv \<0|\cO^\dagger(0)|\n\>_{in}.
\end{equation}
Using \eqref{eq:operator_transformation} and the definition \eqref{eq:fourier_transformed_form_factors} we have
\begin{equation}
\label{eq:matrix_el_ff}
\begin{aligned}
{}_{out}\<\n|\,\cO(x)\,|0\>& =
e^{-ip\cdot x}\mathcal{F}_n(p_1,\ldots,p_n),\\
\<0|\cO^\dagger(x)|\n\>_{in}&=
e^{+ip\cdot x}\mathcal{G}_n(p_1,\ldots,p_n),
\end{aligned}
\end{equation}
where $p$ is the total $d$-momentum of the {\it in} and {\it out} asymptotic states. The Fourier transform of the matrix elements \eqref{eq:matrix_el_ff} reads as
\begin{align}
\int d^d x e^{-ik\cdot x}{}_{out}\<\n|\,\cO(x)\,|0\>=
(2\pi)^d\delta^{(d)}(k+p)\mathcal{F}_n(p_1,\ldots,p_n),\\
\int d^d x e^{-ik\cdot x}\<0|\cO^\dagger(x)|\n\>_{in}=
(2\pi)^d\delta^{(d)}(k-p)\mathcal{G}_n(p_1,\ldots,p_n).
\end{align}

Let us now discuss the structure of $\mathcal{F}_n(p_1,\ldots,p_n)$ for $n=0,1,2$. When $n=0$ or $n=1$ we simply get
\begin{equation}
\label{eq:ff_01}
\mathcal{F}_0,\,
\mathcal{F}_1=\text{const}.
\end{equation}
since it is impossible in these cases to construct a scalar function out of zero or one $d$-momenta.
When $n=2$ we can form only one scalar object out of two $d$-momenta $p_1$ and $p_2$ which is simply the $s$ Mandelstam variable \eqref{eq:mandelstam_variable}. We can write then
\begin{equation}
\mathcal{F}_2(p_1,p_2)=\mathcal{F}_2(s).
\end{equation}
Analogous statements hold for ${\mathcal{G}}_n$.

\paragraph{CPT invariance}
As a consequence of the CPT theorem \cite{Streater:1989vi} there is always an anti-unitary operator $\Sigma$ in the theory which acts on scalar local operators and asymptotic states (of scalar neutral particles) as
\begin{equation}
\Sigma \mathcal{O}(x) \Sigma^\dagger = \mathcal{O}^\dagger(-x),\quad
\Sigma |\n\>_{in} = {}_{out}\<\n|. 
\end{equation}
We can use this fact to write the following equality
\begin{equation}
\<0|\cO^\dagger(0)|\n\>_{in} = 
\<0|\Sigma^\dagger\Sigma\cO^\dagger(0)\Sigma^\dagger\Sigma|\n\>_{in}=
{}_{out}\<\n|\cO(0)|0\>,
\end{equation}
which equates the two form factors \eqref{eq:fourier_transformed_form_factors}, in other words
\begin{equation}
\label{eq:ff_equality}
\mathcal{F}_n(p_1,\ldots,p_n) = \mathcal{G}_n(p_1,\ldots,p_n).
\end{equation}

\paragraph{Relation to spectral density}
Using the definitions of the Fourier transformed form factors \eqref{eq:fourier_transformed_form_factors} one can rewrite the spectral density \eqref{eq:spectral_density} in the following way
\begin{equation}
\label{eq:spectral_density_form_factors}
(2\pi)\theta(p^0)\rho(-p^2)
= \displaystyle\SumInt_n  (2\pi)^d\delta^{(d)}(p-p_n)
|\mathcal{F}_n|^2.
\end{equation}

From the definition of the spectral density \eqref{eq:spectral_density} we see that the one particle state according to \eqref{eq:phase_space} gives the following contribution
\begin{equation}
\nn
\int \frac{d^dp_1}{(2\pi)^d}
(2\pi)\theta(p_1^0)\delta(p_1^2+m^2)\times
(2\pi)^d\delta^{(d)}(p-p_1)
|{}_{out}\<{\bf 1}|\cO(0)|0\>|^2=
|\mathcal{F}_1|^2\times (2\pi)\delta(s-m^2).
\end{equation}
Analogously the two particle states contribute to the spectral density as
\begin{multline}
\nn
\frac{1}{2}\int \frac{d^{d-1}p_1}{(2\pi)^{d-1}}\frac{1}{2p^0_1}
\int \frac{d^{d-1}p_2}{(2\pi)^{d-1}}\frac{1}{2p^0_2}\;
(2\pi)^d\delta^{(d)}(p-p_1-p_2)
|{}_{out}\<{\bf 2}|\cO^\dagger(0)|0\>|^2 = \\
|\mathcal{F}_2(s)|^2\times
\frac{\Omega_{d-1}}{2\,\mathcal{N}_d(2\pi)^{d-2}}\times\theta(s-4m^2),
\end{multline}
where we have performed the change of variables according to \eqref{eq:change_of variables}. Combining the above we get the following expression for spectral density
\begin{equation}
\label{eq:spectral_density_ff}
\rho(s) = \rho_1(s)+\rho_2(s)+\ldots,
\end{equation}
where the one and two particle contributions read as
\begin{equation}
\label{eq:spectral_2part}
\rho_1(s)\equiv|\mathcal{F}_1|^2\times\delta(s-m^2),\quad
\rho_2(s) =
|\mathcal{F}_2(s)|^2\times\frac{\Omega_{d-1}}{2\,\mathcal{N}_d(2\pi)^{d-1}}
\times\theta(s-4m^2)
\end{equation}
and the dots denote the contribution of $n\geq 3$ particle form factors.

\paragraph{Crossing symmetry}
Let us now consider the following matrix element ${}_{out}\<p_1|\cO(x)|p_2\>_{in}$,
where both $p_1$ and $p_2$ satisfy the ``mass-shell'' condition \eqref{eq:on-shell_condition}. We demand that this matrix element satisfies crossing\footnote{The crossing equations for the form factors in 2d are discussed for example in \cite{Karowski:1978vz} and \cite{doi:10.1142/1115}. In general dimensions they can be derived in the QFT framework using the LSZ procedure.
For the derivation of crossing equations in the case of scalar form factors in 4d see chapter 7.2 in \cite{barton1965introduction}.
For the derivation of crossing equations in the case of scattering amplitudes in 4d see section 5.3.2 in \cite{Itzykson:1980rh}. } which can be written as
\begin{equation}
\label{eq:crossing_FF}
{}_{out}\<p_1|\cO(x)|p_2\>_{in}=
{}_{out}\<p_1,-p_2|\cO(x)|0\> = \mathcal{F}_2(p_1,-p_2).
\end{equation}
Notice, that the right-hand side of \eqref{eq:crossing_FF} is not the usual form factor defined in \eqref{eq:fourier_transformed_form_factors}, because it has a negative energy $-p_2^0$. Thus, the expression in the right-hand side of \eqref{eq:crossing_FF} is related to the usual two particle form factor by an analytic continuation.

\paragraph{Constraint from the UV} Due to the relation of the form factors with the spectral density \eqref{eq:spectral_2part} and the UV behavior of the spectral density \eqref{eq:spectral_density_CFT} one obtains the following bound on the large $s$ behaviour of the two particle form factor
\begin{equation}
\label{eq:bound_ff}
\lim_{s\rightarrow +\infty} \mathcal{F}_2(s) \lesssim s^{1+\frac{\Delta-d}{2}}.
\end{equation}

\subsection{Unitarity constraints}
\label{sec:analytic_structure}

We are now ready to discuss the implications of unitarity for scattering amplitudes, partial amplitudes and form factors. In order to do this we will exploit the unitarity of the $S$ operator which follows from \eqref{eq:s_matrix_definition} and \eqref{eq:unitarity_moller_operators}. Taking into account \eqref{eq:S_split} it reads as
\begin{equation}
\label{eq:S_unitarity}
SS^\dagger = 1
\quad\Leftrightarrow\quad T-T^\dagger=i T T^\dagger.
\end{equation}

\subsubsection{Appearance of poles}
\label{sec:poles}
The main goal of this section is to argue that the interacting part of the scattering amplitude and the two particle form factor contain simple poles and show how they are related.\footnote{Strictly speaking the presence of poles cannot be deduced from the pure $S$-matrix approach and should be accepted as an additional assumption. One can however trade this assumption for another one, namely the existence of the relation \eqref{eq:unitarity_relation} for complex values of external momenta, which in turn allows for all the derivations in this section. In order to discuss rigorously the presence of poles one needs to appeal to a higher level framework, e.g. quantum field theory, see section 10.2 in \cite{Weinberg:1995mt}.}

\paragraph{Scattering amplitude}
We focus here on the interacting part of the two to two scattering amplitude. Using \eqref{eq:S_unitarity} and the completeness relation \eqref{eq:identity_operator} we can write
\begin{multline}
\label{eq:unitarity_relation}
\<m,\vec p_1^\myPrime;m,\vec p_2^\myPrime|T|m,\vec p_1;m,\vec p_2\> -
\<m,\vec p_1^\myPrime;m,\vec p_2^\myPrime|T^\dagger|m,\vec p_1;m,\vec p_2\> =\\
i\displaystyle\SumInt_n\,\<m,\vec p_1^\myPrime;m,\vec p_2^\myPrime|T|\n\> \<\n|T^\dagger|m,\vec p_1;m,\vec p_2\>.
\end{multline}
Let us focus on the left-hand side  of \eqref{eq:unitarity_relation} and evaluate it in the center of mass configuration where $\vec p_2=-\vec p_1$ and $\vec p_2^\myPrime=-\vec p_1^\myPrime$. Furthermore we use the following spherical angles $(0,0,\ldots,0)$ for the vector $\vec p_1$ and $(\theta_1,0,\ldots,0)$ for the vector $\vec p_1^\myPrime$. Labeling the states for transparency by the square of the total energy $s$ and the angle $\theta_1$ we have
\begin{equation}
\nn
\<s,\theta_1|T|s,0\> - \<s,\theta_1|T^\dagger|s,0\>=
\<s,\theta_1|T|s,0\> - \<s,0|T|s,\theta_1\>^* = 
\<s,\theta_1|T|s,0\> - \<s,-\theta_1|T|s,0\>^*.
\end{equation}
In the last equality we have used rotational invariance. Using the fact that the matrix element depends on the angle via $\cos\theta_1$ we conclude that the left-hand side of \eqref{eq:unitarity_relation} reads as
\begin{equation}
\label{eq:lhs_sa}
2i \text{Im}  \mathcal{T}(s,t) \times (2\pi)^d \delta^{(d)}(0).
\end{equation}

Let us now discuss the right-hand side of \eqref{eq:unitarity_relation} in a generic frame. We focus on the special case of $n=1$ where
\begin{equation}
\label{eq:rhs_sa}
\int \frac{d^dp}{(2\pi)^d} (2\pi)\theta(p^0)\delta(p^2+m^2)\times
\<m,\vec p_1^\myPrime;m,\vec p_2^\myPrime|T|p\>
\<p|T^\dagger|m,\vec p_1;m,\vec p_2\>.
\end{equation}
Due to translation invariance we can extract an overall delta function of the matrix elements entering \eqref{eq:rhs_sa} as
\begin{equation}
\label{eq:3pe}
g(s)\times (2\pi)^d\delta^{(d)}(p_1+p_2-p) \equiv
\<p|T|m,\vec p_1;m,\vec p_2\>,
\end{equation}
where $s=-p^2=-(p_1+p_2)^2$ is the total energy. Notice that \eqref{eq:3pe} does not exist for physical $(d-1)$-momenta and is defined via an analytic continuation. Plugging \eqref{eq:3pe} into \eqref{eq:rhs_sa} we perform the integral and get the final form of \eqref{eq:rhs_sa} which reads as
\begin{equation}
\label{eq:rhs_la}
|g|^2\times(2\pi)\delta(s-m^2)\times(2\pi)^d \delta^{(d)}(p_1+p_2-p_1'-p_2'),\quad
g\equiv g(m^2).
\end{equation}

We can now evaluate \eqref{eq:rhs_la} in the center of mass frame and plug it into \eqref{eq:unitarity_relation} together with \eqref{eq:lhs_sa}. Dropping the overall delta function we get the following expression
\begin{equation}
\label{eq:disc}
2i \text{Im} \mathcal{T}(s,t) = 2\pi i|g|^2\times\delta(s-m^2) +  \ldots,
\end{equation}
where $\ldots$ denote the continuous part due to $n\geq 2$ particle states. 
This corresponds to a pole in the $s$ complex plane, 
\begin{equation}
\label{eq:pole_amplitude}
\mathcal{T}(s,t)  =
-\frac{|g|^2}{s-m^2} + \ldots.
\end{equation}
Due to presence of $n\geq2$ particle states the imaginary part of the amplitude \eqref{eq:disc} is non-zero for $s\geq 4m^2$. This implies that the amplitude itself develops a discontinuity or equivalently has a branch cut\footnote{More precisely a set of branch cuts.} along the real axis for $s\geq 4m^2$.

\paragraph{Form factors}
Given the definitions of the form factors, we can use the completeness relation \eqref{eq:identity_operator_asymptotic} to write the following equality
\begin{equation}
\label{eq:main_relation_ff}
{}_{out}\<\n|\cO(0)|0\> = \displaystyle\SumInt_m\,
{}_{out}\<\n|{\bf m}\>_{in}\,
{}_{in}\<{\bf m}|\cO(0)|0\>.
\end{equation}
Using the definition of the scattering amplitude \eqref{eq:amplitude} and its splitting into the trivial and interacting part \eqref{eq:S_split} we can re write the above relation as
\begin{equation}
\label{eq:ff_relation}
{}_{out}\<\n|\cO(0)|0\> =
{}_{in}\<\n|\cO  (0)|0\>  
+ i\,\displaystyle\SumInt_m\,
\<\n|T|{\bf m}\>\,
\<0|\cO^\dagger(0)|{\bf m}\>_{in}^*.
\end{equation}
In the first term of the right-hand side \eqref{eq:ff_relation} we have used the normalization of multi particle states which removes the sum over $m$ and integration over the phase space. (This normalization follows from \eqref{eq:normalization_1PS}. See the first line of \eqref{eq:normalization_2PS} for the example of two identical particles). 
Let us focus on the $n=2$ case, using \eqref{eq:ff_equality} this allows to write \eqref{eq:ff_relation} as
\begin{equation}
\mathcal{F}_2(s)-\mathcal{F}^*_2(s) = i\,
\int \frac{d^dp}{(2\pi)^d} (2\pi)\theta(p^0)\delta(p^2+m^2)\times
\<m,\vec p_1^\myPrime;m,\vec p_2^\myPrime|T|p\>\,
\<0|\cO^\dagger(0)|p\>_{in}^*+\ldots.
\end{equation}
Plugging here \eqref{eq:3pe}, analogously to \eqref{eq:rhs_la}, we get
\begin{equation}
\label{eq:im_ff}
\mathcal{F}_2(s)-\mathcal{F}^*_2(s) =
 2\pi ig^*\mathcal{F}^*_1\times\delta(s-m^2) +  \ldots,
\end{equation}
where $\mathcal{F}^*_1$ is a constant as discussed below \eqref{eq:ff_01}. We thus get an analogous expression to \eqref{eq:disc}. Assuming analyticity in some region of the complex plane we obtain the pole structure of the form factor
\begin{equation}
\label{eq:analytic_structure_FF}
\mathcal{F}_2 (s) = -\frac{g^* \mathcal{F}_1^*}{s-m^2} + \ldots.
\end{equation}
Analogously to the discussion below \eqref{eq:pole_amplitude} the form factor develops a branch cut along the real axis for $s\geq 4m^2$ due to the contribution of two particles states (and higher) in \eqref{eq:im_ff}.

\subsubsection{Watson's equation}
Let us consider the two particle contribution to the completeness relation \eqref{eq:identity_operator}. Using the change of variables to spherical coordinates \eqref{eq:change_of variables} and \eqref{eq:decomposition} we can write
\begin{align}
\nn
\frac{1}{2}\,\int
\frac{d^{d-1}p_1}{(2\pi)^{d-1}}\frac{1}{2p^0_1}
&\frac{d^{d-1}p_2}{(2\pi)^{d-1}}\frac{1}{2p^0_2}\,
|m,\vec p_1;m,\vec p_2\>\<m,\vec p_1;m,\vec p_2|\\
\nn
&=\sum_{j,j'=0}^\infty
\int
\frac{1}{2\mathcal{N}_d}\,\theta(p^0)\,
\frac{d^d p}{(2\pi)^d}\,
\frac{d\Omega_{d-1}}{(2\pi)^{d-2}}
C_j(\cos\theta_1)C_{j'}(\cos\theta_1)
|p,j\>\<p,j'|\\
&=\sum_{j=0}^\infty
\int
\frac{d^d p}{(2\pi)^d}\,\theta(p^0)\,
|p,j\>\<p,j|.
\label{eq:2particle_contribution}
\end{align}
In the last equality we have used the explicit expression of the Clebsch-Gordan coefficient \eqref{eq:CG} and the orthogonality of the Gegenbauer polynomial \eqref{eq:orthogonality_1}.

We can now consider matrix elements of the first unitarity condition \eqref{eq:S_unitarity} with two particle states of definite spin \eqref{eq:irreps}. Injecting the completeness relation \eqref{eq:identity_operator} and focusing on the two particle contribution \eqref{eq:2particle_contribution} we get the standard unitarity condition on the partial amplitudes
\begin{equation}
\label{eq:unitarity_partial_amp}
\mathcal{S}_j(s)\mathcal{S}_j^*(s)+\ldots = 1,
\end{equation}
where the dots denote the $n\geq 3$ particle contribution which starts at $s\geq (3m)^2$. 

Let us now consider the two particle form factor and use the decomposition of the two particle state into irreducible representations \eqref{eq:decomposition} to write the following equality
\begin{equation}
\label{eq:rel}
{}_{out}\<m,\vec p_1;m,-\vec p_1|\cO(0)|0\> = \sum_{j=0}^\infty C_j(\cos\theta_1)\;
{}_{out}\<p,j|\cO(0)|0\> = C_0\;{}_{out} \<p,0|\cO(0)|0\>.
\end{equation}
In the last equality we have used the rotation invariance and the fact that the local operator $\mathcal{O}(0)$ is a scalar which selects $j=0$ representations only. 
In \eqref{eq:rel} $C_0$ is simply a real constant which follows from \eqref{eq:CG} since the Gegenbauer polynomial is one for $j=0$. 

Consider now the relation \eqref{eq:main_relation_ff} where instead of the two particle asymptotic state we use the two particle state projected into $j=0$ irreducible representation. We also rewrite the two particle contribution in the completeness relation according to \eqref{eq:2particle_contribution} in order to get
\begin{equation}
{}_{out} \<p,0|\cO(0)|0\> = \mathcal{S}_0(-p^2) \<0|\cO(0)|p,0\>_{in}^*+\ldots
\end{equation}
where $\mathcal{S}_0(s)$ is the zero spin partial amplitude and the dots denote the $n\geq 3$ contribution which starts at $s\geq(3m)^2$. This can be simply rewritten by using \eqref{eq:rel} to get the final expression of interest
\begin{equation}
\mathcal{F}_2(s)=\mathcal{S}_0(s)\mathcal{G}_2^*(s)+\ldots
\end{equation}
Using the equality \eqref{eq:ff_equality} we obtain the Watson's equation \cite{Watson:1954uc}
\begin{equation}
\label{eq:watson's_equations}
\mathcal{S}_0(s) = \frac{\mathcal{F}_2(s)}{\mathcal{F}_2^*(s)},\quad
s\in[4m^2,\,9m^2].
\end{equation}
In integrable models, there is no particle production and the Watson's equation is valid to all  energies.  One can solve the Watson's equation \eqref{eq:watson's_equations} and obtain the form factor in terms of the partial amplitude up to an analytic function which is real on the real axis $s$, see  for example \cite{Monin:2018lee}.\footnote{See also \cite{Monin:2018lee} for a nice application of pion form factors in phenomenology.}${}^{,}$\footnote{In the $S$-matrix literature it is common to use the term real analytic function which means an analytic function which takes real values on some interval of the real axis. In mathematics instead the term real analytic function means a function on (an interval of) the real axis which allows analytic extension to its neighborhood. The function itself may be real or complex on the real axis, see for example \cite{krantz2002primer}.}

\subsection{Stress-energy tensor}
\label{sec:stress_tensor}

We discuss here the  stress-tensor in general dimensions. We show that there is an integral of the two-point function of the stress-tensor which gives the central charge $C_T$ of the UV CFT. We will work in Euclidean signature within this section. The final results however are independent of the signature.

Consider the stress-energy tensor $T^{\mu\nu}(x)$ operator, which satisfies the following constraints
\begin{equation}
\label{eq:constraints}
T^{\mu\nu}(x) = T^{\nu\mu}(x),\qquad
\partial_\mu T^{\mu\nu}(x)=0.
\end{equation}
The most general form of the two-point function of the stress-tensor \cite{Cardy:1988cwa} with appropriate mass dimensions which respects Lorentz invariance is\footnote{In $d=2$ and $d=3$ there are additional parity odd tensor structures which we omit here.}
\begin{equation}
\label{eq:stress-tensor}
\<0|T^{\mu\nu}(x)T^{\lambda\sigma}(0)|0\> =
\sum_{i=1}^5\frac{1}{x^{2d}}\;  h_i(r) \mathbb{T}_i^{(\mu\nu),(\lambda\sigma)},
\quad r\equiv |x|,
\end{equation}
where $h_i(r)$ are scalar dimensionless functions and $\mathbb{T}_i$ are dimensionless linearly independent tensor structures which read as
\begin{align}
\nn
\mathbb{T}_1^{(\mu\nu),(\lambda\sigma)} &=\frac{x^\mu x^\nu x^\lambda x^\sigma}{r^4},\\
\nn
\mathbb{T}_2^{(\mu\nu),(\lambda\sigma)} &=
\frac{x^\mu x^\nu \delta^{\lambda\sigma}+x^\lambda x^\sigma \delta^{\mu\nu}}{r^2},\\
\mathbb{T}_3^{(\mu\nu),(\lambda\sigma)} &=
\frac{
	x^\mu x^\lambda \delta^{\nu\sigma}+
	x^\nu x^\lambda \delta^{\mu\sigma}+
	x^\mu x^\sigma \delta^{\nu\lambda}+
	x^\nu x^\sigma \delta^{\mu\lambda}}{r^2},\\
\nn
\mathbb{T}_4^{(\mu\nu),(\lambda\sigma)} &=\delta^{\mu\nu}\delta^{\lambda\sigma},\\
\nn
\mathbb{T}_5^{(\mu\nu),(\lambda\sigma)} &=\delta^{\mu\lambda}\delta^{\nu\sigma}+\delta^{\nu\lambda}\delta^{\mu\sigma}.
\end{align}
Notice, that away from fixed points one is required to have dimensionful parameters to construct dimensionless functions $h_i(x)$. At fixed points there are no dimensionful parameters and thus all the functions $h_i(x)$ are simply constants.
Let us also define the following three contracted two-point functions
\begin{equation}
\label{eq:contractions}
\begin{aligned}
\frac{A(r)}{r^{2d}} &\equiv\langle 0| T^\mu_\mu(x) T^\nu_\nu(0)|0\rangle ,\\
\frac{I(r)}{r^{2d}} &\equiv\langle 0| T^\mu_\nu(x) T_\mu^\nu(0)|0\rangle,\\
\frac{J(r)}{r^{2d}} &\equiv\frac{x^\lambda x^\sigma}{r^2}\times\langle 0| T^\mu_\lambda(x) T_{\mu\sigma}(0)|0\rangle.
\end{aligned}
\end{equation}
Comparing these expressions with \eqref{eq:stress-tensor} we can write
\begin{align}
\nn
A(r) &= h_1(r) +2d\,h_2(r)+4h_3(r)+d^2h_4(r)+2d\,h_5(r),\\
I(r) &= h_1(r) +2h_2(r)+2(d+1)h_3(r)+d\,h_4(r)+d(d+1)h_5(r),\\
J(r) &= h_1(r) +2h_2(r)+(d+3)h_3(r)+h_4(r)+(d+1)h_5(r).
\nn
\end{align}

Let us apply the conservation equation \eqref{eq:constraints} to the two-point \eqref{eq:stress-tensor} using  \eqref{eq:der}. 
Setting to zero coefficients of three independent tensor structures we get the following three conditions
\begin{align}
\nn
rh'_1(r)+rh'_2(r)+2rh'_3(r) &= (d+1) h_1(r)+2(d+1)h_2(r)+4(d+1)h_3(r),\\
\label{eq:conservation_equations}
rh'_2(r)+rh'_4(r) &= (d+1) h_2(r)-2h_3(r)+2d\,h_4(r),\\
\nn
rh'_3(r)+rh'_5(r) &= -h_2(r)+d\,h_3(r)+2d\,h_5(r).
\end{align}
Here we have used the fact that
\begin{equation}
\label{eq:der}
\partial_\mu h_i(r) = rh'(r)\times \frac{x_\mu}{x^2}.
\end{equation}
There are five functions $h_i(r)$ with three differential constraints on them. There are thus only two independent functions which define the two-point function of the stress-tensor. We can take various linear combinations of three equations \eqref{eq:conservation_equations} to form a single differential equation. Following \cite{Cardy:1988cwa} we can write for example 
\begin{align}
C(r)   &\equiv h_1(r)+\frac{d^2+d+2}{2}\,h_2(r)+(d+3)h_3(r)+\frac{d(d+1)}{2}\,h_4(r)+(d+1)h_5(r),\\
rC'(r) &= (d + 1)\left(A(r)+\frac{(d - 1)(d-2)}{2}h_2(r)\right).
\label{eq:differential_2}
\end{align}
Another expression, more convenient for $d\geq 3$ is as follows
\begin{align}
H(r)   &\equiv h_1(r)+2h_2(r)+(d+3)h_3(r)+h_4(r)+(d+1)h_5(r),\\
rH'(r) &= I(r)+d\,J(r).
\label{eq:differential_d}
\end{align}

\paragraph{Integral expressions for the central charges}
In the presence of conformal symmetry the form of the two-point function \eqref{eq:stress-tensor} is severely restricted. According to \cite{Osborn:1993cr} it reads as\footnote{\label{foot:TT_parity_odd}In $d=2$ the two-point function of the stress-tensor has an extra parity odd tensor structure with a new independent coefficient.}
\begin{align}
\label{eq:stress-tensor_CFT}
\<0|T_{\mu\nu}(x)T_{\lambda\sigma}(0)|0\>_{CFT}
&=\frac{C_T}{x^{2d}}\times\left( \frac{1}{2}\left(\mathcal{I}_{\mu\lambda}(x)\mathcal{I}_{\nu\sigma}(x)+\mathcal{I}_{\mu\sigma}(x)\mathcal{I}_{\nu\lambda}(x)\right)-\frac{1}{d}\,\delta_{\mu\nu}\delta_{\lambda\sigma}\right),\\
\mathcal{I}_{\mu\nu}(x) &\equiv \delta_{\mu\nu}-\frac{2x_\mu x_\nu}{x^2}.
\end{align}
Here $C_T$ is one of the central charges of the UV CFT. 
Comparing this form with \eqref{eq:stress-tensor} we deduce that in CFT
\begin{equation}
\label{eq:matching_CFT}
h_1(r)=4 C_T,\quad
h_2(r)=0,\quad
h_3(r)=- C_T,\quad
h_4(r)=- C_T/d,\quad
h_5(r)=C_T/2.
\end{equation}
Provided that our QFT is defined as a flow between the UV and IR fixed points (which are reached at $r=0$ and $r=\infty$) we can write the differential conditions \eqref{eq:differential_2} and \eqref{eq:differential_d} in an integral form using \eqref{eq:matching_CFT}. We get two equivalent expressions
\begin{align}
\label{eq:c_2d}
C_T^{UV}-C_T^{IR} &=(d+1)\,\int_0^\infty \frac{dr}{r}
\left(\frac{1}{d-1}\,A(r)+\frac{d-2}{2}h_2(r)\right),\\
\label{eq:c_d}
C_T^{UV}-C_T^{IR} &=\frac{2d}{(d-1)(d-2)}\,\int_0^\infty \frac{dr}{r}
\Big(I(r)+d\,J(r)\Big).
\end{align}
Notice, that the latter holds only for $d\geq 3$.
In a massive QFT, the theory in the IR is empty and thus we have
\begin{equation}
C_T^{IR} = 0.
\end{equation}

\paragraph{Stress-tensor form factor}
Let us consider the two particle form factor of the stress-tensor. It has the following most generic form
\begin{align}
\nn
{}_{out}\<m,\vec p_1;m,\vec p_2|T^{\mu\nu}(0)|0\> =
-\mathcal{F}_2^{(0)}(s)&\times
\left(\delta^{\mu\nu}-\frac{(p_1+p_2)^\mu(p_1+p_2)^\nu}{(p_1+p_2)^2}\right)\\
+\,\mathcal{F}_2^{(2)}(s)&\times \frac{(p_1-p_2)^\mu(p_1-p_2)^\nu}{(p_1-p_2)^2},
\label{eq:stress-tensor_form_factor}
\end{align}
which is symmetric in both indices and satisfies the conservation condition \eqref{eq:constraints} written as
\begin{equation}
(p_1+p_2)_\mu\;{}_{out}\<m,\vec p_1;m,\vec p_2|T^{\mu\nu}(0)|0\> =0.
\end{equation}
Taking trace of \eqref{eq:stress-tensor_form_factor} we obtain the form factor of the trace of the stress-tensor $\Theta\equiv T_\mu^\mu$
 \begin{equation}
\mathcal{F}_2^{\Theta}(s)=
(1-d) \mathcal{F}_2^{(0)}(s)+
\mathcal{F}_2^{(2)}(s).
\label{formfactorTheta}
\end{equation}
In $d=2$ the two structures in \eqref{eq:stress-tensor_form_factor} are linearly dependent, we take it into account  by setting $\mathcal{F}_2^{(0)}(s)=0$.

\paragraph{Normalization of the stress-tensor}
We can form the following conserved charges
\begin{equation}
\label{eq:poincare_generators}
P^\mu \equiv \int d^{d-1} x T^{0\mu}(x),\quad
M^{\mu\nu} \equiv \int d^{d-1} x\,
\Big(x^\mu T^{0 \nu}(x) - x^\nu T^{0 \mu}(x)\Big)
\end{equation}
which are the generators of translations and Lorentz transformations respectively. 
In particular the Hamiltonian is $H =P^0$ as in \eqref{eq:hamiltonian}.
Let us now evaluate the matrix elements of $P^\mu$ with one particle states. Since they are the eigenstates of $P^\mu$ we get the following expression
\begin{equation}
\label{eq:rel2}
\<m,\vec p_1| P^\mu |m,\vec p_2\> = 2p_1^0p_1^\mu \times(2\pi)^{d-1}\delta^{(d-1)}(\vec p_1-\vec p_2),
\end{equation}
where $p_1^0$ satisfies the ``mass-shell'' condition \eqref{eq:on-shell_condition}.
On the other hand we have
\begin{align}
\nn
\<m,\vec p_1| P^\mu |m,\vec p_2\>
&= \<m,\vec p_1|T^{0\mu}(0)|m,\vec p_2\> \times \int_{-\infty}^{+\infty} d^{d-1}x\, e^{i(p_2-p_1)\cdot x} \\
&= \<m,\vec p_1|T^{0\mu}(0)|m,\vec p_2\>\times(2\pi)^{d-1}\delta^{(d-1)}(\vec p_1-\vec p_2),
\label{eq:rel3}
\end{align}
where we have used \eqref{eq:operator_transformation}. 
 Combining together \eqref{eq:rel2} and \eqref{eq:rel3} we get
\begin{equation}
\label{eq:normalization_1}
\Big(\<m,\vec p_1|T^{0\mu}(0)|m,\vec p_2\>-2p_1^0p_1^\mu\Big)
\times(2\pi)^{d-1}\delta^{(d-1)}(\vec p_1-\vec p_2) = 0.
\end{equation}

We can now compare the matrix elements in \eqref{eq:normalization_1} and \eqref{eq:stress-tensor_form_factor} taking into account the crossing relation \eqref{eq:crossing_FF} which effectively makes a replacement $p_2^\mu\rightarrow -p_2^\mu$. This leads to
\begin{equation}
\label{eq:normalization_2FF_2d}
\mathcal{F}_2^{(2)}(s=0) = -2m^2,\qquad
\mathcal{F}_2^\Theta(s=0) = -2m^2
\end{equation}
and $\mathcal{F}_2^{(0)}(s)\sim s$ around $s=0$. For more details see appendix G in \cite{Karateev:2020axc}.
The Lorentz generators in \eqref{eq:poincare_generators} do not provide any further conditions.

\paragraph{Special case of 2d}
Let us focus now on the specific case of 2d \cite{Belavin:1984vu}. It is conventional to use complex coordinates defined as 
\begin{equation}
z\equiv x^1+i x^2,\quad
\bar z\equiv x_1-i x_2.
\end{equation}
In these coordinates we can write the components of the stress-tensor as
\begin{align}
\nn
T(z,\bar z) &\equiv (2\pi)\times T_{zz}(z,\bar z)=(2\pi)\times \frac{1}{4}\,\left(T_{11}(x)-T_{22}(x)-2i\, T_{12}(x)\right),\\
\label{eq:rel_2d}
\Theta(z,\bar z) &\equiv 4 T_{z\bar z}(z,\bar z)=T_{11}(x)+T_{22}(x),\\
\nn
\overline T(z,\bar z) &\equiv (2\pi)\times T_{\bar z \bar z}(z,\bar z)=(2\pi)\times \frac{1}{4}\,\left(T_{11}(x)-T_{22}(x)+2i\, T_{12}(x)\right).
\end{align}
Notice the presence of $2\pi$ factors in the definitions \eqref{eq:rel_2d}.
Conservation implies
\begin{equation}
\partial_{\bar z} T(z,\bar z)+\frac{\pi}{2}\,\partial_{z}\Theta(z,\bar z)= \partial_z \overline T(z,\bar z) +\frac{\pi}{2}\,\partial_{\bar z}\Theta(z,\bar z) = 0.
\end{equation}
At the critical point we have
\begin{equation}
\Theta(z,\bar z)=0,\quad
\langle 0 | T(z) T(0) |0\rangle = \frac{c/2}{z^4},\quad
\langle 0 | \overline T(\bar z) \overline T(0) |0\rangle = \frac{\bar c/2}{\bar z^4},
\end{equation}
where $c=\bar c$ is the standard central charge in parity preserving 2d CFTs. Using \eqref{eq:rel_2d} we can compare this form with \eqref{eq:stress-tensor_CFT}. We conclude that
\begin{equation}
\label{eq:relation_c_conventions}
c=(2\pi)^2\times C_T/2.
\end{equation}
In this convention the central charge of a free boson is $c=1$, see \eqref{eq:c_free_boson}.

We can rewrite the integral expression \eqref{eq:c_2d} using \eqref{eq:contractions}, \eqref{eq:rel_2d} and \eqref{eq:relation_c_conventions} in the following form
\begin{equation}
\label{eq:c_theorem}
c_{UV} - c_{IR} = (2\pi)^2\times\frac{3}{4\pi}
\int d^2x_E \, x_E^2\, \<0|\Theta(x_E )\Theta(0)|0\>_T,\quad c_{IR}=0.
\end{equation}
Due to reflection positivity of the two-point function of the stress-tensor, we can conclude that $c_{UV} > c_{IR}$. This is   Zamolodchikov's $c$-theorem \cite{Zamolodchikov:1986gt,Cardy:1988tj}. No such statement can be made about $C_T$ in higher dimensions.

The Euclidean two-point function in \eqref{eq:c_theorem} is time-ordered. We can then use the Euclidean K\"{a}ll\'en-Lehmann spectral representation \eqref{eq:spectral_representation_3_euclidean} to relate the central charge $c_{UV}$ with the spectral density $\rho$ of the trace of the stress-tensor. We have
\begin{align}
c_{UV} &= (2\pi)^2\times\frac{3}{4\pi}
\int_0^\infty d\mu^2 \rho_\Theta(\mu^2)
\int \frac{d^2q_E}{(2\pi)^2} \frac{1}{q^2_E+\mu^2}\,\left(-\partial_{q_E}^2\,(2\pi)^2\delta^{(2)}(q_E) \right).
\end{align}
Using the integration by parts we arrive at the final expression
\begin{equation}
\label{eq:c_final}
c_{UV} = (2\pi)^2\times\frac{3}{\pi}
\int_0^\infty ds\, \frac{\rho_\Theta(s)}{s^2}=
(2\pi)^2\times\frac{3}{\pi}
\left(
m^{-4}\,|\mathcal{F}_1^\Theta|^2 + 
\int_{4m^2}^\infty ds\, \frac{ \rho_\Theta(s)}{s^2}
\right),
\end{equation}
where in the second equality we have used \eqref{eq:spectral_density_ff} and \eqref{eq:spectral_2part}.


\section{Unitarity as positive semidefiniteness}
\label{sec:constraints_unitarity}

We are now ready to present the main idea of this paper. We will construct a hermitian matrix which must be semipositive definite in a unitary theory. This requirement intertwines the partial amplitudes, the form factors and the spectral density and puts constraints on them.

\subsection{General spacetime dimension}
\label{sec:unitarity_general_d}

We will work with the simplest case of identical particles with mass $m$. Let us define the following three states
\begin{align}
\label{eq:state_1}
|\psi_1\> &\equiv \Pi_j|{\bf 2}\>_{in}\;=\Omega(-\infty)\Pi_j|m,\vec p_1;m,\vec p_2\>,\\
\label{eq:state_2}
|\psi_2\> &\equiv \Pi_j|{\bf 2}\>_{out}=\Omega(+\infty)\Pi_j|m,\vec p_1;m,\vec p_2\>,\\
\label{eq:state_3}
|\psi_3\> &\equiv m^{-\delta}\times\int d^d x e^{+ip\cdot x}\cO(x)|0\>,
\end{align}
where $p^\mu$ is the total $d$ momentum
\begin{equation}
p^\mu = p_1^\mu+p_2^\mu,\quad p_i^0 = \sqrt{m^2+\vec p_i}.
\end{equation}
The first two states $|\psi_1\>$ and $|\psi_2\>$ are the {\it in} and {\it out} two particle states projected to the irreducible spin representation according to \eqref{eq:irreps} with the total $d$-momentum $p^\mu$. The third state is the Fourier transform of the state generated by the local operator $\cO(x)$ acting on the vacuum. It also has $p^\mu$ total $d$-momentum. The extra factor $m^{-\delta}$ is injected in order to make all three states to be of the same mass dimension
\begin{equation}
\big[|\psi_a\>\big]=-\frac{d}{2}
\quad\Rightarrow\quad
\delta= \Delta_\cO-d/2.
\end{equation}
This follows from \eqref{eq:dimension} and \eqref{eq:dimension_1PS}. Notice, that the parameter $\delta$ has already appeared in \eqref{eq:spectral_density_CFT}. Let us now construct a 3 by 3 matrix out of all possible inner products of the states \eqref{eq:state_1} - \eqref{eq:state_3}, we have
\begin{equation}
\label{eq:matrix}
B^{ab}_{j}\times(2\pi)^d\delta^{(d)}(p-p')\equiv \<\psi_a|\psi_b\>,
\end{equation}
where $a,b=1,2,3$ and the total $d$-momentum of the states $|\psi_a\>$ and $\<\psi_b|$ are $p^\mu$ and $p^{\prime\mu}$ respectively.

\paragraph{Entries of the $B$-matrix}
Let us now inspect the entries of the matrix \eqref{eq:matrix}. The entries 11 and 22 on the diagonal are simply fixed by the normalization condition \eqref{eq:normalization_states} since the M{\o}ller operators are unitary, see \eqref{eq:normalization_as}, and thus read as
\begin{align}
\<\psi_1|\psi_1\> &= \<\psi_2|\psi_2\> = (2\pi)^d\delta^{(d)}(p^\prime-p).
\end{align}
Using \eqref{eq:fourier_spectral_representation} we can write the entry 33 as
\begin{align}
\nn
\<\psi_3|\psi_3\>
&= m^{-2\delta}\int d^d x d^d y\, e^{+ip\cdot x} e^{-ip'\cdot y}
\<0|\cO^\dagger(y)\cO(x)|0\>\\
&= m^{-2\delta}\times (2\pi)^d\delta^{(d)}(p-p')\times 2\pi\theta(p^0)\rho(s).
\end{align}
Let us address now the off-diagonal elements. Since the matrix \eqref{eq:matrix} is hermitian we will only need to discuss the elements $12$, $13$ and $23$. The element 12 reads as
\begin{align}
\<\psi_1|\psi_2\> &= 
\left(\<m,\vec p_1';m,\vec p_2'|\Pi_j\right)
\;\Omega^\dagger(-\infty)\Omega(+\infty)\;
\left(\Pi_j|m,\vec p_1;m,\vec p_2\>\right)\\
&= (2\pi)^d\delta^{(d)}(p-p') \times \mathcal{S}_j^*(s),
\end{align}
where we have used the definition of the $S$ operator \eqref{eq:s_matrix_definition} and partial amplitude \eqref{eq:partial_amplitude}. The element 13 reads as
\begin{align}
\<\psi_1|\psi_3\>
&= m^{-\delta}\times\int d^d x e^{+ip\cdot x}
\left(\<m,\vec p_1';m,\vec p_2'|\Pi_j\right)
\;\Omega^\dagger(-\infty)\cO(x)\;
|0\>\\
&=(2\pi)^d\delta^{(d)}(p-p') \times m^{-\delta}\,\omega\,\delta_{j0}\,\mathcal{G}^*_2(s),
\end{align}
where in the second line we have used \eqref{eq:irreps}, \eqref{eq:operator_transformation} and the results of section \ref{sec:form_factors}. The coefficient $\omega$ is defined as
\begin{align}
\omega\,\delta_{j0} \equiv \gamma_j\, \Omega_{d-2}\times \int_{-1}^{+1} dx\, (1-x^2)^{(d-4)/2}C_j^{(d-3)/2}(x)
=\gamma_0\, \Omega_{d-2}\times\frac{\sqrt{\pi}\,\Gamma\left(\frac{d-2}{2}\right)}{\Gamma\left(\frac{d-1}{2}\right)}\,\delta_{j0}.
\end{align}
Simplifying we get the following compact result
\begin{equation}
\label{eq:omega}
\omega^2=\frac{\Omega_{d-1}}{2\,\mathcal{N}_d(2\pi)^{d-2}}.
\end{equation}
Analogously, the element $23$ can be written as
\begin{flalign}
\qquad\qquad\qquad
\<\psi_2|\psi_3\>
&= m^{-\delta}\times\int d^d x e^{+ip\cdot x}
\left(\<m,\vec p_1';m,\vec p_2'|\Pi_j\right)
\;\Omega^\dagger(+\infty)\cO(x)\;
|0\>&&\\
&=(2\pi)^d\delta^{(d)}(p-p') \times m^{-\delta}\,\omega\,\delta_{j0}\,\mathcal{F}_2(s).
\end{flalign}

\paragraph{Positivity constraint}
Plugging all these expression into \eqref{eq:matrix} and using \eqref{eq:ff_equality} we recover the final form of the $B$-matrix
\begin{equation}
\label{eq:matrix_B}
B_j(s)\equiv
\begin{pmatrix}
1 & \mathcal{S}^*_j(s)
& m^{-\delta} \omega\, \mathcal{F}^*_2(s)\delta_{j0}\\
\mathcal{S}_j(s) & 1
& m^{-\delta} \omega\, \mathcal{F}_2(s)\delta_{j0}\\
m^{-\delta} \omega\, \mathcal{F}_2(s)\delta_{j0} 
& m^{-\delta} \omega\, \mathcal{F}^*_2(s)\delta_{j0} &
m^{-2\delta}\,2\pi\rho(s)
\end{pmatrix}.
\end{equation}
The matrix $B$ is hermitian by construction and must be positive semidefinite in unitary theories. This can be easily seen as follows. The matrix $B$ is positive semi-definite if and only if its eigenvalues are non-negative. One can show that the latter is the case by taking a linear combination of states \eqref{eq:state_1} - \eqref{eq:state_3} for which the $B$-matrix is diagonal. The elements of this matrix are simply the norms of the news states. Unitarity of the theory requires these norms to be non-negative. Thus, 
\begin{equation}
\label{eq:condition_semipositiveness}
B_j(s)\succeq 0,\quad \forall s \geq 4m^2
\quad{\text{\bf and}}\quad \forall j.
\end{equation}
The necessary and sufficient condition for the matrix to be positive semidefinite is the Sylvester's criterion. It states that $B\succeq0$ if and only if all its principal minors are non-negative  (including the determinant of the $B$ matrix itself).

\paragraph{Consequences of the positivity constraint}
Let us start with the minor associated to removing the third row and column. The Sylvester's criterion leads to
\begin{equation}
\label{eq:bound_1}
|\mathcal{S}_j(s)|^2\leq 1.
\end{equation}
This is the standard unitarity constraint for the partial amplitude already obtained in \eqref{eq:unitarity_partial_amp}.
Now consider instead the minor associated to removing the first row and column. The Sylvester's criterion leads then to
\begin{equation}
\label{eq:bound_2}
2\pi\rho(s)\geq \omega^2\,|\mathcal{F}_2(s)|^2.
\end{equation}
This inequality also follows straightforwardly from \eqref{eq:spectral_density_ff} for $s\geq 4m^2$. The minor associated to removing the first two rows and columns leads to the following requirement
\begin{equation}
\label{eq:rho_bound}
\rho(s)\geq 0,
\end{equation}
which was already obvious from the definition \eqref{eq:spectral_density}. Finally, the determinant of the $B$ matrix must be non-negative,
\begin{equation}
\label{eq:bound_3}
2\pi\rho(s) \left(1-|\mathcal{S}_0(s)|^2\right)
-2w^2|\mathcal{F}_2(s)|^2 +
w^2\mathcal{F}_2^{*2}(s)\mathcal{S}_0(s) +
w^2\mathcal{F}_2^2(s)\mathcal{S}_0^*(s)\geq 0.
\end{equation}
This provides a non-trivial positivity condition which mixes together the amplitudes, the form factors and the spectral density.

\paragraph{Degenerate situation}
Let us now investigate a very particular situation when only one state out of the three \eqref{eq:state_1} - \eqref{eq:state_3} is linearly independent. This for instance happens in the energy range
\begin{equation}
\label{eq:range_of_energies}
4m^2\leq s \leq 9m^2.
\end{equation}
We refer to this situation as the absence of ``particle production'' in the range of energies \eqref{eq:range_of_energies}. 
One can imagine even a more extreme case when there is no ``particle production'' for the whole range of energies $s\in[4m^2,+\infty)$. In $d\geq 3$ according to the Aks theorem \cite{Aks:1965} this situation leads to a trivial theory. Theories in $d=2$ escape this constraint however and we enter the realm of integrable models. 

In what follows we investigate the consequence of having only a single linearly independent state among \eqref{eq:state_1} - \eqref{eq:state_3} or equivalently the situation when
\begin{equation}
\label{eq:rank}
\text{rank} B_j(s) = 1.
\end{equation}
The characteristic polynomial in $\lambda$ is then required to have the following form
\begin{equation}
\label{eq:character_pol_required_form}
\det(B_j(s)-\lambda\, I_{3\text{x}3})= - \lambda^2(\lambda-\lambda_0),
\end{equation}
where $\lambda_0$ is the only non-zero eigenvalue of the matrix $B$. Let us now compute the characteristic polynomial for the $B$ matrix \eqref{eq:matrix_B}, it gets the required form \eqref{eq:character_pol_required_form} with
\begin{equation}
\lambda_0 = 2 + m^{-2\delta}\,2\pi \rho(s),
\end{equation}
if the following conditions are fulfilled
\begin{align}
\label{eq:condition_1}
|\mathcal{S}_j(s)|^2 &= 1, \\
\label{eq:condition_2}
|\mathcal{F}_2(s)|^2 &= 2\pi\omega^{-2}\rho(s), \\
\nn
2|\mathcal{F}_2(s)|^2 &=
\mathcal{S}_0^*(s) \mathcal{F}_2^2(s) + 
\mathcal{S}_0(s) \mathcal{F}_2^{*2}(s). 
\end{align}
The latter equation is solved by
\begin{equation}
\label{eq:condition_3}
\mathcal{F}_2(s) = \mathcal{S}_0(s)\mathcal{F}_2^*(s),
\end{equation}
which is the already familiar Watson's equation \eqref{eq:watson's_equations}.
We see that these conditions simply saturate the bounds \eqref{eq:bound_1}, \eqref{eq:bound_2} and \eqref{eq:bound_3}.

\subsection{Special case of 2d}
\label{sec:unitarity_constraint_2d}
Let us summarize here the unitarity constraints for the special case of 2d. We will then generalize them to include the $O(N)$ global symmetry.

Let us start with partial amplitudes. Since the Little group is the discrete $Z_2$ group, effectively we have a single partial wave with spin $j=0$. From now on we denote it as
\begin{equation}
\hat{\mathcal{S}}(s)\equiv \mathcal{S}_0(s).
\end{equation}
Moreover, in 2d there is not much difference between the partial and the scattering amplitude. In our conventions they simply differ by a normalization
\begin{equation}
\label{eq:inteeracting_part_2d}
\hat{\mathcal{S}}(s) = \mathcal{N}_2^{\,-1}\mathcal{S}(s) =1+i\,\mathcal{N}_2^{-1} \mathcal{T}(s),
\end{equation}
where $\mathcal{T}(s)$ as before is the interacting part of the scattering amplitude and the normalization factor is given by \eqref{eq:factor_N_full} and reads in 2d as
\begin{equation}
\mathcal{N}_2 = 2\sqrt{s}\sqrt{s-4m^2}.
\end{equation}

From now on we will measure every dimensional quantity in  units of mass $m$. This is equivalent to setting
\begin{equation}
m = 1.
\end{equation} 
The unitarity constraint \eqref{eq:matrix_B} and \eqref{eq:condition_semipositiveness} read in 2d as
\begin{equation}
\label{eq:matrix_B_2d}
B(s)\equiv
\begin{pmatrix}
1 & \hat{\mathcal{S}}^*(s)
& \omega\, \mathcal{F}^*_2(s)\\
\hat{\mathcal{S}}(s) & 1
&  \omega\, \mathcal{F}_2(s)\\
\omega\, \mathcal{F}_2(s)
&  \omega\, \mathcal{F}^*_2(s) &
2\pi\rho(s)
\end{pmatrix}\succeq 0,\quad
\omega = \mathcal{N}_2^{-1/2}.
\end{equation}
For the future purposes it is also convenient to rewrite this expression in the following way
\begin{align}
\nn
B(s) &=
\begin{pmatrix}
1 & \;1\; & \;0 \\
1 & \;1\; & \;0 \\
0 & \;0\; & \;0
\end{pmatrix}
+
\mathcal{N}_2^{-1} \times
\begin{pmatrix}
0 & -i\,\mathcal{T}^*(s) & \;\;\;0\\
+i\, \mathcal{T}(s) & 0 &  \;\;\;0\\
0 &  0 & \;\;\;0
\end{pmatrix}\\
&+\mathcal{N}_2^{-1/2}\times
\begin{pmatrix}
0 & 0 & \mathcal{F}^*_2(s)\\
0 & 0 & \mathcal{F}_2(s)\\
\mathcal{F}_2(s)
&  \mathcal{F}^*_2(s) &
0
\end{pmatrix}+
2\pi\rho(s)\times
\begin{pmatrix}
0 & \;0\; & \;0 \\
0 & \;0\; & \;0 \\
0 & \;0\; & \;1
\end{pmatrix}\succeq 0.
\label{eq:matrix_B_2d_final}
\end{align}

\paragraph{$O(N)$ global symmetry}
\label{sec:unitarity_ON}
Let us consider the case when the system has a global $O(N)$ symmetry.  We will require our asymptotic states to transform in the vector representation of  $O(N)$. They will thus carry an extra label $a=1\ldots N$. The one particle states are normalized as before with an addition of a  Kronecker delta due to presence of $O(N)$ vector indicies
\begin{align}
\label{eq:normalization_global}
{}_b\<m,\vec p_2|m,\vec p_1\>_a
=2p^0 \delta_{ab}\times 2\pi\delta(\vec p_2-\vec p_1).
\end{align}
The full scattering amplitude can be decomposed into three independent scattering amplitudes $\sigma_i(s)$, $i=1,2,3$. In the notation of \cite{Zamolodchikov:1978xm} we have
\begin{align}
\nn
{}_{cd}\< m,\vec p_3;m,\vec p_4|S|m,\vec p_1;m,\vec p_2\>_{ab}=&
(2\pi)^{2}\delta^{(2)}(p_1+p_2-p_3-p_4)\times\\
&\big(\sigma_1(s)\delta_{ab}\delta_{cd}+
\sigma_2(s)\delta_{ac}\delta_{bd}+
\sigma_3(s)\delta_{ad}\delta_{bc}
\big).
\label{eq:scattering_amplitudes_ZZ}
\end{align}
The st-crossing symmetry (under exchanging particles 1 and 3) relates the amplitudes $\sigma_i$ as
\begin{equation}
\label{eq:crossing_sigma}
\sigma_1(s)=\sigma_3(4m^2-s),\qquad
\sigma_2(s)=\sigma_2(4m^2-s).
\end{equation}

The two-particle state is in the reducible $O(N)$ representation and can be further decomposed into three irreducible representations as
\begin{equation}
\label{eq:global_decomposition}
|m,\vec p_1;m,\vec p_2\>_{ab}=
\frac{\delta_{ab}}{\sqrt{N}}|m,\vec p_1;m,\vec p_2\>^{\bullet}+
|m,\vec p_1;m,\vec p_2\>^{\textbf{S}}_{(ab)}+
|m,\vec p_1;m,\vec p_2\>^{\textbf{A}}_{[ab]},
\end{equation}
where we have defined
\begin{align}
\label{eq:bullet}
|m,\vec p_1;m,\vec p_2\>^{\bullet} &\equiv \frac{1}{\sqrt{N}}\,\sum_{a=1}^N|m,\vec p_1;m,\vec p_2\>_{aa},\\
|m,\vec p_1;m,\vec p_2\>^{\textbf{S}}_{(ab)} &\equiv \frac{1}{2}\,\Big(
|m,\vec p_1;m,\vec p_2\>_{ab}+|m,\vec p_1;m,\vec p_2\>_{ba}\Big)-\frac{\delta_{ab}}{\sqrt{N}}|m,\vec p_1;m,\vec p_2\>^{\bullet},\\
|m,\vec p_1;m,\vec p_2\>^{\textbf{A}}_{[ab]} &\equiv \frac{1}{2}\,\Big(
|m,\vec p_1;m,\vec p_2\>_{ab}-|m,\vec p_1;m,\vec p_2\>_{ba}\Big).
\end{align}
The labels $\bullet$, $\textbf{S}$ and $\textbf{A}$ stand for trivial, symmetric traceless and antisymmetric representations. Taking into account \eqref{eq:global_decomposition} alternatively to \eqref{eq:scattering_amplitudes_ZZ} we can rewrite the full scattering amplitude in terms of independent scattering amplitudes $S_{\bullet}(s)$, $S_{\textbf{S}}(s)$ and $S_{\textbf{A}}(s)$, as
\begin{align}
\nn
{}_{cd}\< m,\vec p_3;m,\vec p_4|S|m,\vec p_1;m,\vec p_2\>_{ab} =&
(2\pi)^{2}\delta^{(2)}(p_1+p_2-p_3-p_4)\times\\
&\big(S_{\bullet}(s)T^{ab,cd}_{\bullet}+S_{\textbf{S}}(s)T^{ab,cd}_{\textbf{S}}+S_{\textbf{A}}(s)T^{ab,cd}_{\textbf{A}}\big),
\label{eq:scattering_amplitudes}
\end{align}
where the tensor structures associated to the three irreducible representations are defined as
\begin{equation}
T^{ab,cd}_{\bullet}\equiv \frac{1}{N}\delta_{ab}\delta_{cd},\quad
T^{ab,cd}_{\textbf{S}}\equiv \frac{\delta_{ac}\delta_{bd}+\delta_{ad}\delta_{bc}}{2}-\frac{1}{N}\delta_{ab}\delta_{cd},\quad
T^{ab,cd}_{\textbf{A}}\equiv \frac{\delta_{ac}\delta_{bd}-\delta_{ad}\delta_{bc}}{2}.
\end{equation}
The relation between two sets of amplitudes $\sigma_1$, $\sigma_2$, $\sigma_3$ and $S_{\bullet}$, $S_{\textbf{S}}$, $S_{\textbf{A}}$ simply reads as
\begin{align}
\nn
S_{\bullet}(\theta) &=\sigma_2(\theta)+\sigma_3(\theta)+N \sigma_1(\theta),\\
\label{eq:definitions}
S_{\textbf{S}}(\theta) &=\sigma_2(\theta)+\sigma_3(\theta),\\
\nn
S_{\textbf{A}}(\theta) &=\sigma_2(\theta)-\sigma_3(\theta),
\end{align}
The normalization of two particle states in the irreducible representation of the $O(N)$ group follows from \eqref{eq:normalization_global}. We have
\begin{align}
\label{eq:normalization_trivial}
{}^{\bullet}\langle m,\vec p_3;m,\vec p_4|m,\vec p_1;m,\vec p_2\>^{\bullet}
&=\mathcal{N}_2\times(2\pi)^{2}\delta^{(2)}(p_1+p_2-p_3-p_4),\\
{}_{(cd)}^{\textbf{S}}\langle m,\vec p_3;m,\vec p_4|m,\vec p_1;m,\vec p_2\>^{\textbf{S}}_{(ab)}
&=\mathcal{N}_2\, T^{ab,cd}_{\textbf{S}}\times(2\pi)^{2}\delta^{(2)}(p_1+p_2-p_3-p_4),\\
{}_{[cd]}^{\textbf{A}}\langle m,\vec p_3;m,\vec p_4|m,\vec p_1;m,\vec p_2\>^{\textbf{A}}_{[ab]}
&=\mathcal{N}_2\, T^{ab,cd}_{\textbf{A}}\times(2\pi)^{2}\delta^{(2)}(p_1+p_2-p_3-p_4).
\end{align}
The crossing equations \eqref{eq:crossing_sigma} in terms of the amplitudes \eqref{eq:definitions} read as
\begin{equation}
\label{eq:crossing_ON}
\begin{pmatrix}
{\mathcal{S}}_{\bullet}(s) \\
{\mathcal{S}}_{\textbf{S}}(s)\\
{\mathcal{S}}_{\textbf{A}}(s)
\end{pmatrix}=
\begin{pmatrix}
\frac{1}{N}  && \frac{1}{2}-\frac{1}{N}+\frac{N}{2} && \frac{1}{2}-\frac{N}{2} \\
\frac{1}{N}  && \frac{1}{2}-\frac{1}{N} &&  \frac{1}{2}\\
-\frac{1}{N}  && \frac{1}{2}+\frac{1}{N} && \frac{1}{2}
\end{pmatrix}
\begin{pmatrix}
{\mathcal{S}}_{\bullet}(4m^2-s) \\
{\mathcal{S}}_{\textbf{S}}(4m^2-s)\\
{\mathcal{S}}_{\textbf{A}}(4m^2-s)
\end{pmatrix}.
\end{equation}

Let us now consider the unitarity constraints. We have three states transforming in irreducible representations of $O(N)$. They cannot mix with each other, in other words non-zero inner products can be formed only between the states in the same representation. Let us start with the trivial representation
\begin{equation}
\Omega(-\infty)|m,\vec p_1;m,\vec p_1\rangle^{\bullet},\quad
\Omega(+\infty)|m,\vec p_1,i;m,\vec p_1,j\rangle^{\bullet},\quad
\int d^2 x e^{ip\cdot x} \mathcal{O}(x)|0\rangle,
\end{equation}
where the local operator $\mathcal{O}(x)$ does not transform under the $O(N)$ group. 
Analogously to the discussion of section \ref{sec:unitarity_general_d} we conclude
\begin{equation}
\label{eq:unitarity_t}
\begin{pmatrix}
1 & \hat{\mathcal{S}}_{\bullet}^*(s)
& \omega\, \mathcal{F}^*_{\bullet 2}(s)\\
\hat{\mathcal{S}}_{\bullet}(s) & 1
&  \omega\, \mathcal{F}_{\bullet 2}(s)\\
\omega\, \mathcal{F}_{\bullet 2}(s)
&  \omega\, \mathcal{F}^*_{\bullet 2}(s) &
2\pi\rho(s)
\end{pmatrix}\succeq 0,
\end{equation}
where the hatted amplitudes are defined according to \eqref{eq:inteeracting_part_2d} and the form factor is defined as
\begin{equation}
\label{eq:form_factor_trivial}
\mathcal{F}_{\bullet 2}(s)\equiv 
\<0|\mathcal{O}(0)|m,\vec p_1;m,\vec p_2\>^{\bullet}.
\end{equation}
For the symmetric and antisymmetric representations we consider only the {\it in} and {\it out} states because they do not overlap with the state created by the $O(N)$ invariant local operator.\footnote{Another natural local operator to consider is the conserved current of the $O(N)$ global group which we can denote by $J^\mu_{[ab]}(x)$. It transform in the adjoint representation of the $O(N)$ or equivalently in the antisymmetric representation. Its form factor and the spectral density can thus mix with $\mathcal{S}_{\textbf{A}}(s)$ partial amplitudes.} The unitarity conditions then simply read as
\begin{equation}
\label{eq:unitarity_sa}
\begin{pmatrix}
1 & \hat{\mathcal{S}}_{\textbf{S}}^*(s)\\
\hat{\mathcal{S}}_{\textbf{S}}(s) & 1
\end{pmatrix}\succeq 0,\qquad
\begin{pmatrix}
1 & \hat{\mathcal{S}}_{\textbf{A}}^*(s)\\
\hat{\mathcal{S}}_{\textbf{A}}(s) & 1
\end{pmatrix}\succeq 0.
\end{equation}

\section{Analytic examples in 2d}
\label{app:analytic_integrable_models}
In this section we provide a uniform summary of the exact analytic expressions of the partial amplitudes and form factors in several 2d integrable models, namely the sine-Gordon, the $E_8$ model (also known as the 2d Ising model with magnetic deformation) and
the $O(N)$ $\sigma$-model with $N\geq 3$.

\paragraph{$\theta$ variable}
In 2d instead of the Mandelstam variable $s$ it is convenient to use the rapidity  variable $\theta$. Given a particle with the 2-momentum $p_i^\mu$ and the mass $m_i$ we can define
\begin{equation}
p^0_i \equiv m_i \cosh\theta_i,\quad
p^1_i \equiv m_i \sinh\theta_i.
\end{equation}
For scattering of two particles with masses $m_i$ and $m_j$ the Mandelstam $s$ variable reads as
\begin{equation}
s = m_i^2 + m_j^2 +2m_im_j\cosh\theta,\quad
\theta \equiv \theta_i - \theta_j. 
\end{equation}
In case of identical particles $m_1=m_2=m$ the above relation reduces to
\begin{equation}
\label{eq:map}
s = 4m^2 \cosh^2(\theta/2).
\end{equation}
When $s$ and $\theta$ are complex variables, the map \eqref{eq:map} can be depicted as on figure 6 in \cite{Paulos:2016but}.

\paragraph{Partial amplitudes}
The 2d integrable models possess   an infinite number of conserved charges which allow for factorization of any scattering amplitude into a product of $2\rightarrow2$ scattering amplitudes $\mathcal{S}(s)$. The consistency of this factorization leads to the Yang–Baxter factorization equations on $2\rightarrow2$ scattering amplitudes $\mathcal{S}(s)$. Instead of $\mathcal{S}(s)$ it is convenient to work with partial amplitudes $\hat{\mathcal{S}}(s)$ which differ by a simple normalization, see section \ref{sec:unitarity_constraint_2d}. The unitarity and crossing conditions then read
\begin{equation}
\label{eq:2d_crossing_unitarity}
\hat{\mathcal{S}}(\theta)\hat{\mathcal{S}}(-\theta)=1,\qquad
\hat{\mathcal{S}}(\theta) = \hat{\mathcal{S}}(i\pi-\theta).
\end{equation}

The Yang–Baxter equations together with unitarity and crossing \eqref{eq:2d_crossing_unitarity} allow to obtain exact analytic expressions for partial amplitudes up to a CDD ambiguity \cite{Castillejo:1955ed}. The latter states that given the solution to all the above constraints, one can obtain another solution by multiplying it with any number of CDD factors (and their inverses)\footnote{An inverse of the CDD factor \eqref{eq:CDD} introduces   zeros in the amplitude.} defined as
\begin{equation}
\label{eq:CDD}
t_\alpha(\theta)\equiv \frac
{\tanh\frac{\theta+i \pi \alpha}{2}}
{\tanh\frac{\theta-i \pi \alpha}{2}} =
\frac
{\sinh\theta+i\sin(\pi\alpha)}
{\sinh\theta-i\sin(\pi\alpha)}.
\end{equation}
Here $\alpha\in(0,1)$ is a real parameter.\footnote{One can consider the CDD factors with negative or even complex values of the parameter $\alpha$. For a discussion see page 12 of \cite{Paulos:2016but}.} Notice the factors of $\pi$ in \eqref{eq:CDD} compared to the standard definition. The CDD factor \eqref{eq:CDD} satisfies automatically both constraints \eqref{eq:2d_crossing_unitarity}. It contains a pole at $\theta=i \pi \alpha$ and thus encodes the contribution of a given asymptotic state to the amplitude. The correct choice of the CDD factors is usually postulated and then gets checked in perturbation theory for  some range of parameters in the model when it is applicable.

\paragraph{Form factors} In 2d the form factors satisfy the following equations
\begin{equation}
\label{eq:ff_equations_2d}
\mathcal{F}_2(\theta) = \mathcal{F}_2(-\theta) \hat{\mathcal{S}}(\theta),\qquad
\mathcal{F}_2(i\pi-\theta) = \mathcal{F}_2(i\pi+\theta).
\end{equation}
The former is the familiar Watson's equation and the latter encodes crossing symmetry. Given the analytic expression for the partial amplitude, the equations \eqref{eq:ff_equations_2d} can be solved analytically \cite{Karowski:1978eg}. The solution reads as
\begin{equation}
\label{eq:ff_general_form}
F(\theta)=R(\theta)F_{min}(\theta),
\end{equation}
where $R(\theta)$ is an arbitrary rational function of $\cosh(\theta)$ since $\cosh(\theta)$ automatically satisfies the second condition in \eqref{eq:ff_equations_2d}. More precisely it can be written as
\begin{equation}
\label{eq:R}
R(\theta)= \frac{K_{\alpha_1}(\theta)K_{\alpha_2}(\theta)\ldots}{K_{\beta_1}(\theta)K_{\beta_2}(\theta)\ldots}=(A+B \cosh\theta+C\cosh^2\theta\ldots)K_{\alpha_1}(\theta)K_{\alpha_2}(\theta)\ldots,
\end{equation}
where we have defined
\begin{equation}
\label{eq:pole_ff}
K_\alpha(\theta) \equiv -\frac{\cos^2(\pi\alpha/2)}{\sinh\frac{\theta-i \pi\alpha}{2}\sinh\frac{\theta+i \pi\alpha}{2}}=
\frac{2\cos^2(\pi\alpha/2)}{\cos(\pi\alpha)-\cosh\theta}.
\end{equation}
In \eqref{eq:R} the parameters $\alpha_i$ define the positions of poles and parameters $\beta_i$ (or equivalently $A$, $B$, $C$, etc) define the positions of zeros. The constant factors in the numerator of \eqref{eq:pole_ff} are introduced for convenience, they allow for the following normalization $K_\alpha(i\pi)=1$.
The ``minimal'' form factor $F_{min}(\theta)$ in \eqref{eq:ff_general_form}, also known as the Omn\`es solution in higher dimensions \cite{Omnes:1958hv}, is defined as a function without poles or zeros. According to \cite{Karowski:1978eg} in 2d due to \eqref{eq:ff_equations_2d} it can be expressed in terms of the partial amplitude as follows
\begin{equation}
\ln F_{min}(\theta) =\frac{1}{4\pi i}
\int_{-\infty}^{+\infty} dz \left(\coth\left(\frac{z-\theta}{2}\right)-
\coth\left(\frac{z}{2}\right)\right)
\ln \hat{\mathcal{S}}.
\end{equation}

The functions \eqref{eq:pole_ff} can be thought of as analogues of the CDD factors \eqref{eq:CDD} for the form factors. The choice   \eqref{eq:R} is usually postulated first and then gets checked with perturbation theory when applicable.
Finally the form factor \eqref{eq:ff_general_form} for a given operator must obey the bound \eqref{eq:bound_ff} which reads in 2d as\footnote{In 2d this bound was first derived in \cite{Delfino:2003yr}, see formulas (3.33) and (3.34).}
\begin{equation}
\label{eq:bound_2d}
\mathcal{F}_{2}(\theta) \lesssim \left(\exp\theta\right)^{\Delta/2}.
\end{equation}

\subsection{sine-Gordon model}
\label{sec:SG}
The quantum sine-Gordon model is defined as the renormalization group (RG) flow triggered by the deformation of the free scalar UV CFT by the vertex operator
\begin{equation}
\label{eq:vertex_operator}
V_\beta(x)\equiv\; :e^{i\beta \phi}:\,,\quad
\Delta_{V_\beta}=\frac{\beta^2}{4\pi},
\end{equation}
in the following way
\begin{align}
\mathcal{L}_{SG} &= - \frac{1}{2} (\partial_\mu \phi)^2 + \frac{m_0^2}{2\,\beta^2}\,
\big(V_\beta(x)+V_\beta^*(x)\big).
\end{align}
Here $\phi(x)$ is the real scalar field, $\Delta_{V_\beta}$ is the UV scaling dimension of the vertex operator\footnote{The scaling dimension of the vertex operator can be straightforwardly deduced from the Euclidean two point function computed in the free massless theory (which posses conformal invariance for $m=0$)
	\begin{equation}
	\nn
	\<0|V_\beta(x_E)V^*_\beta(0)|0\>\sim \exp\left(\beta^2\,\<0|\phi(x_E)\phi(0)|0\>\right)\sim
	\left(x_E^{2}\right)^{-\frac{\beta^2}{4\pi}}.
	\end{equation}	
	In the second step we have used the propagator $\<0|\phi(x_E)\phi(0)|0\>=-\frac{1}{4\pi}\log x_E^2$.
}, $m_0$ is a mass-like parameter and $\beta$ is a real coupling constant.
The sine-Gordon model possesses  several remarkable properties. First, the model is dual to the Thirring model \cite{Coleman:1974bu}.\footnote{See also chapter 6 of \cite{Coleman:1985rnk}.} 
Second, it possesses the $O(2)$ topological symmetry  \cite{Zamolodchikov:1978xm} and thus can also be regarded as the $O(2)$ $\sigma$-model.

The mass spectrum of the sine-Gordon model was first found with semi-classical methods \cite{Korepin:1975vd,Dashen:1975hd} and later argued to be exact \cite{Schroer:1976if,Nussinov:1975kp}. It consists of a soliton and an antisoliton with  mass $m$ and a number of breathers (soliton - antisoliton pairs) denoted by $b_n$ with the masses
\begin{equation}
m_n = 2 m \sin\frac{n\gamma}{16},\quad
\gamma\equiv\frac{\beta^2}{1-\frac{\beta^2}{8\pi}},\quad
n=1,2,\ldots,
\left\lfloor \frac{8\pi}{\gamma}\right\rfloor.
\end{equation}
Here $\left\lfloor x \right\rfloor$ denotes the greatest integer not larger than $x$.
The breathers exist only in the following range of parameters
\begin{equation}
\gamma\in[0,8\pi]
\quad\Leftrightarrow\quad
\beta^2\in[0,\,4\pi].
\end{equation}
It is interesting to study the sine-Gordon model in the regime when at least two breathers exist. Then one can define a ratio of masses for the first two lightest breathers
\begin{equation}
\label{eq:ratio}
R\equiv\frac{m_2}{m_1} =2 \cos \frac{\gamma}{16}.
\end{equation}
The range of parameters which allow for this is
\begin{equation}
\label{eq:parameters_2breathers}
\gamma\in[0,4\pi]
\quad\Leftrightarrow\quad
\beta^2\in[0,8\pi/3]
\quad\Leftrightarrow\quad
R\in[2,\sqrt{2}].
\end{equation}

Let us now discuss partial amplitudes for the scattering of asymptotic states in the sine-Gordon model. The soliton - (anti)soliton scattering was computed in \cite{,Zamolodchikov:1978xm}. The (anti)soliton - breather and breather - breather scattering was computed in \cite{Karowski:1977fu}. The uniformed treatment for all these cases was done in \cite{Zamolodchikov:1977py}. In this work we are concerned only with the   lightest breather - breather scattering  in the parameter range \eqref{eq:parameters_2breathers} which reads as
\begin{equation}
\label{eq:b1b1-b1b1}
\hat{\mathcal{S}}_{b_1b_1\rightarrow b_1b_1} (\theta) =\frac
{\sinh\theta+i\sin\gamma}
{\sinh\theta-i\sin\gamma}=
t_{\frac{\gamma}{8\pi}}(\theta).
\end{equation}
It is given by the single CDD factor \eqref{eq:CDD} and thus contains a single pole at $\theta=i \gamma/8$ or equivalently at $\sqrt{s}=2m_1\cos(\gamma/16)$. From \eqref{eq:ratio} we see that this pole is simply at the mass of the second breather $\sqrt{s}=m_2$.

Let us now address the form factors of a scalar operator. The scalar soliton - (anti)soliton form factors were computed in \cite{Weisz:1977ii}. The scalar breather - breather form factors were found in \cite{Karowski:1978vz}.\footnote{For some more recent work see \cite{Babujian:1998uw}.} The latter form factor corresponds to the partial amplitude \eqref{eq:b1b1-b1b1} and reads as
\begin{equation}
\label{eq:theta-b1b1}
\mathcal{F}_{b_1b_1}(\theta) = A\,
K_{\frac{\gamma}{8\pi}}(\theta)\, F_{min}^{SG}(\theta),\quad
F_{min}^{SG}(\theta)\equiv
\cosh\left(\frac{i\pi-\theta}{2}\right)T_{\frac{\gamma}{8\pi}}(\theta),
\end{equation}
where $A$ is a normalization constant, $K$ is given by \eqref{eq:pole_ff} and $F_{min}^{SG}$ is the minimal form factor of the sine-Gordon model where we have defined
\begin{equation}
\label{eq:object_T}
T_\alpha(\theta) \equiv \exp\left(2\int_0^\infty \frac{dx}{x}
\frac{\cosh\left((\alpha-1/2)x\right)}{\cosh(x/2)\sinh(x)} \,\sin^2\left(\frac{(i\pi-\theta)x}{2\pi}\right)
\right).
\end{equation}
At large energies the object \eqref{eq:object_T} behaves as\footnote{In order to show this, one can make a variable redefinition $x\rightarrow x'\equiv x \theta$. Keeping $x'$ fixed, we can then consider only the leading behavior of the integrand at large $\theta$ and perform the integration.}
\begin{equation}
\label{eq:T_asymptotics}
\lim _{\theta\rightarrow +\infty}T_\alpha(\theta) \sim \exp(\theta/2).
\end{equation}
The form factor \eqref{eq:theta-b1b1} for the vertex operator \eqref{eq:vertex_operator} is the most general solution (with a single pole due to $b_2$) which satisfies the bound \eqref{eq:bound_2d} for the whole range of parameters \eqref{eq:parameters_2breathers} since
\begin{equation}
\lim _{\theta\rightarrow +\infty}\mathcal{F}_{b_1b_1}(\theta) \sim \text{const}.
\end{equation}
and $\Delta_V\in[0,\,2/3]$. 

The form factor of the trace of the stress-tensor is proportional to the UV deforming operator (vertex operator in our case). It is thus also given by \eqref{eq:theta-b1b1}. The value of the constant $A$ follows from the normalization convention \eqref{eq:normalization_2FF_2d} and reads as
\begin{equation}
A = -2m_1^2.
\end{equation}

\begin{figure}[tb]
	\begin{center}
		\includegraphics[scale=0.65]{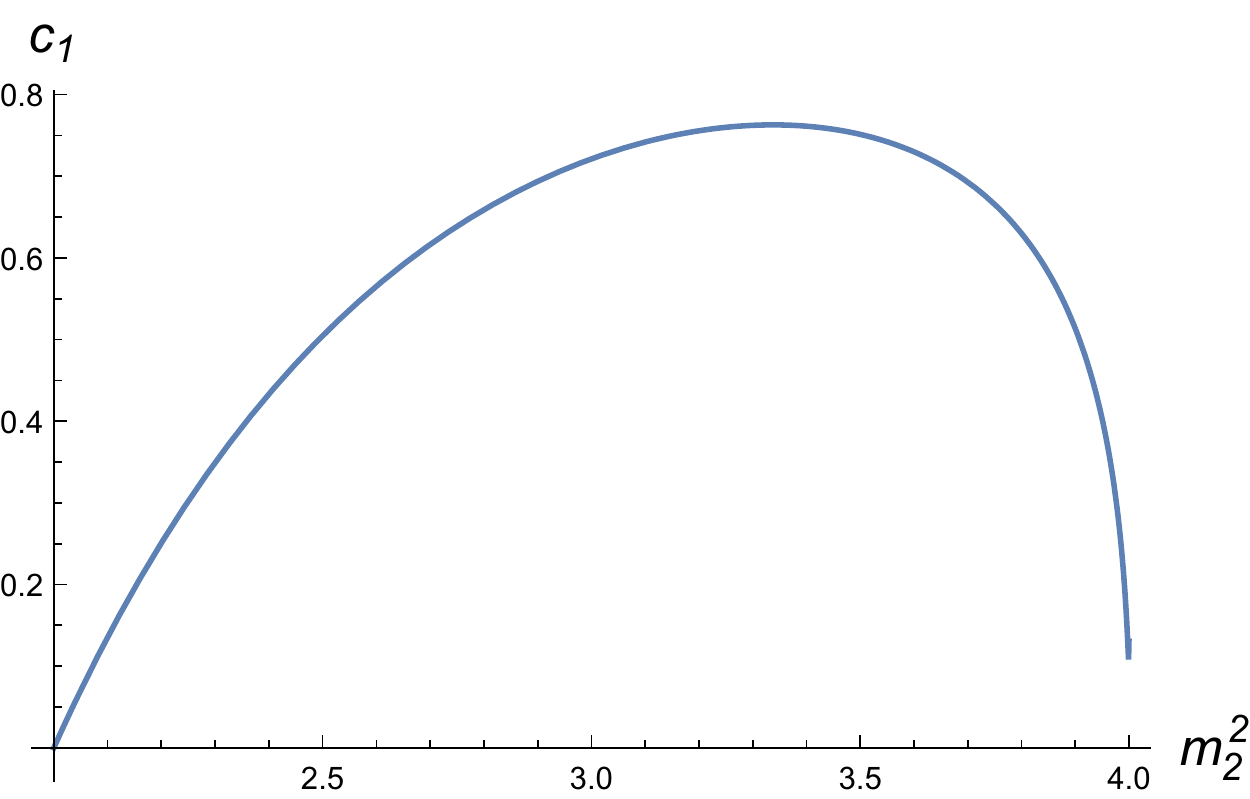}
		\caption{Contribution of the one particle states of the second breather into the UV central charge as a function of $m_2^2$. On the horizontal axis the mass $m_2$ is given in the units of $m_1$. The range of parameters which allow for the existence of the second breather is provided in \eqref{eq:parameters_2breathers}. For $m_2^2=3$ we have $c_1\approx 0.72126$.}
		\label{fig:c1SG}
	\end{center}
\end{figure}

Let us discuss now the interacting part of the scattering amplitude in $s$ variable which according to \eqref{eq:inteeracting_part_2d} can be written as
\begin{equation}
\label{eq:interacting_part_2d}
\mathcal{T}(s) = -i\mathcal{N}_2\left(\hat{\mathcal{S}}(s)-1\right).
\end{equation}
Given the exact expression of the partial amplitude \eqref{eq:b1b1-b1b1} we have
\begin{equation}
\label{eq:coupling_SG}
\mathcal{T}(s) =
-\frac{g^2}{s-m_2^2}+\ldots,\quad
g^2 = \frac{4m_2^3}{m_2^2-2}\times\left(4-m_2^2\right)^{3/2}.
\end{equation}
Here we wrote explicitly only the pole and denoted by $\ldots$ the finite part at $s=m_2^2$. Similarly for the form factor we have 
\begin{equation}
\label{eq:ff_pole_SG}
\mathcal{F}_{b_1b_1}(s) =
-\frac{g \mathcal{F}_{b_2}}{s-m_2^2}+\ldots,\quad
\mathcal{F}_{b_2} = -\frac{2m_2^2}{g}\,F_{min}^{SG}(s=m_2^2).
\end{equation} 
In case of the trace of the stress-tensor we can use \eqref{eq:spectral_2part}, \eqref{eq:c_final} and the explicit expressions for the form factor \eqref{eq:theta-b1b1} to estimate the contribution to the total central charge of the one particle state of the second breather $c_{b_2}$ and of the two particle state of the first breather $c_{b_1b_1}$. The total central charge then reads as
\begin{equation}
\label{eq:c_SG}
c= c_{b_2}+c_{b_1b_1}+\ldots.
\end{equation}
For concreteness, on figure \ref{fig:c1SG} we provide the numerical value of $c_{b_2}$ as a function of $m_2^2$. All the contributions in \eqref{eq:c_SG} should sum up to $c=1$ which is the central charge of a free boson, see \eqref{eq:c_free_boson}. For more detailed investigation of the sine-Gorden model see \cite{Delfino:2003ia}, in particular figures 6 and 7.

\subsection{$E_8$ model} The 2d Ising model is a 2d conformal field theory with a $Z_2$ symmetry. It contains only two relevant operators $\sigma$ and $\epsilon$ with the scaling dimensions $\Delta_{\sigma}=1/8$ and $\Delta_{\epsilon}=1$ respectively. The former is $Z_2$ odd and the latter is $Z_2$ even. In the lattice formulation of the 2d Ising model the operator $\sigma$ couples to the magnetic field. We consider here a QFT obtained by deforming the 2d Ising model with the operator $\sigma$. This QFT defines an integrable model \cite{Zamolodchikov:1989fp} which we refer to as the $E_8$ model. It contains eight asymptotic states $m_1,\ldots,m_8$. Given the value of $m_1$, the spectrum in the $E_8$ model reads as
\begin{equation}
m_2 = 2m_1\cos(\pi/5),\quad
m_3 = 2m_1\cos(\pi/30),
\end{equation}
where we have ignored particles with masses $m_4,\ldots,m_8$ since their masses lie above the two particle threshold $2m_1$ and are thus invisible to the techniques of section \ref{sec:numerics_2d}.

The partial amplitude for the scattering of the lightest asymptotic state reads as
\begin{equation}
\label{eq:E8_11-11}
\hat{\mathcal{S}}_{11\rightarrow 11} (\theta) =
t_{2/3}(\theta)t_{2/5}(\theta)t_{1/15}(\theta),
\end{equation}
where $t$ are the CDD factors \eqref{eq:CDD}. The form factor for a scalar relevant operator with the lightest asymptotic states was computed in \cite{Delfino:2003yr}, it reads as
\begin{equation}
\mathcal{F}_{11}(\theta) = (A+B\cosh\theta)
K_{2/3}(\theta)K_{2/5}(\theta)K_{1/15}(\theta)\,F_{min}^{E_8}(\theta),
\label{eq:ff_E8}
\end{equation}
where $A$ and $B$ are independent parameters and the minimal form factor for the $E_8$ model reads as
\begin{equation}
F_{min}^{E_8}(\theta)\equiv
\cosh\left(\frac{i\pi-\theta}{2}\right)
T_{2/3}(\theta)T_{2/5}(\theta)T_{1/15}(\theta).
\end{equation}
At large energies due to \eqref{eq:T_asymptotics} the  form factor \eqref{eq:ff_E8} behaves as
\begin{equation}
\lim _{\theta\rightarrow +\infty}\mathcal{F}_{11}(\theta) \sim \text{const}.
\end{equation}
It has thus the  most general form which obeys the bound \eqref{eq:bound_2d} for both $\sigma$ and $\epsilon$ operators. 
According to \cite{Delfino:2003yr} the expression \eqref{eq:ff_E8} provides the two particle form factor for the $\sigma$ and $\epsilon$ operators, given the following ratios of parameters
\begin{equation}
\label{eq:sigma_epsilon_coefficients}
A_\sigma/B_\sigma = 4.86984066\ldots,\quad
A_\epsilon/B_\epsilon = 1.25558515\ldots.
\end{equation}
The overall normalization of the form factor depends as usually on the chosen normalization of the operators $\sigma$ and $\epsilon$ and is not important in our work. We set for convenience
\begin{equation}
\label{eq:norm_sigma_epsilon}
B_\sigma = B_\epsilon = 1.
\end{equation}

Consider now the form factor of the trace of the the stress-tensor $\Theta$. It is proportional to the form factor of the deforming operator which is $\sigma$ in our case. Thus, the coefficients $A_\Theta$ and $B_\Theta$ for the trace of the stress-tensor $\Theta$ are completely fixed by  the following conditions
\begin{equation}
\label{eq:values_AB}
A_\Theta/B_\Theta =A_\sigma/B_\sigma,\quad
A_\Theta-B_\Theta = -2m_1^2,
\end{equation}
where the second equation follows from the normalization condition \eqref{eq:normalization_2FF_2d}.

Let us discuss now the interacting part of the scattering amplitude in $s$ variable which is related to the partial amplitude via \eqref{eq:interacting_part_2d}.
Given \eqref{eq:E8_11-11}, we can write its pole structure   as
\begin{equation}
\mathcal{T}(s) =
-\frac{g_1^2}{s-m_1^2}-\frac{g_2^2}{s-m_2^2}-\frac{g_3^2}{s-m_3^2}+\ldots.
\end{equation}
The values of the trilinear couplings read as
\begin{equation}
\label{eq:couplings_E8}
g_1 \approx 26.922055,\quad
g_2 \approx 38.527928,\quad
g_3 \approx 0.611666.
\end{equation}
Similarly for the form factor of the trace of the stress-tensor we can write
\begin{equation}
\mathcal{F}^\Theta_{11}(s) =
-\frac{g_1 \mathcal{F}^\Theta_1}{s-m_1^2}
-\frac{g_2 \mathcal{F}^\Theta_2}{s-m_2^2}
-\frac{g_3 \mathcal{F}^\Theta_3}{s-m_3^2}+\ldots,
\end{equation} 
where the one particle form factors read as
\begin{equation}
\mathcal{F}_1^\Theta \approx -0.111898 ,\quad
\mathcal{F}_2^\Theta \approx 0.059131,\quad
\mathcal{F}_3^\Theta \approx -0.032590.
\end{equation}
In case of the trace of the stress-tensor, using \eqref{eq:spectral_2part}, \eqref{eq:c_final} and the explicit expressions for the form factor \eqref{eq:ff_E8}, \eqref{eq:values_AB} we can estimate the central charge contribution of the one particle states of the first three lightest asymptotic states $c_i$ and the two particle contribution of the very first asymptotic state $c_{11}$. The total central charge reads
\begin{equation}
\label{eq:c_Ising_approx}
c= c_1+c_2+c_3+c_{11}\ldots,
\end{equation}
where we provide for completeness the numerical values
\begin{equation}
\label{eq:c_E8}
c_1 \approx 0.472038,\quad
c_2 \approx 0.0192313,\quad
c_3 \approx 0.0025581.
\end{equation}
All the contributions in \eqref{eq:c_Ising_approx} should sum up to $c=1/2$ which is the central charge of the 2d Ising model.

\subsection{Non-linear sigma model}
\label{sec:NLSM}
The $O(N)$ non-linear sigma model (NLSM) with $N\geq 3$ is defined in the UV via the Lagrangian density
\begin{equation}
\label{eq:ON_def}
\mathcal{L}_{NLSM} = \frac{1}{2g_0} \sum_{i=1}^N (\partial_\mu n_i)^2,\quad
\sum_{i=1}^N n_i^2 =1,
\end{equation}
where $n_i(x)$ is a $O(N)$ vector of real scalar fields and $g_0$ is a dimensionless coupling. 
This model can be seen as a marginally relevant deformation of a theory of $N-1$ free massless  scalar fields.\footnote{One way to see this is to solve the constraint on the scalar fields in \eqref{eq:ON_def}, and write
\begin{equation}
\label{eq:ON_def_2}
\mathcal{L}_{O(N)} = \frac{1}{2g_0} \left(\sum_{i=1}^{N-1} (\partial_\mu n_i)^2+
\mathcal{O}(x)\right),\quad
\mathcal{O}(x)\equiv\frac
{\sum_{i,j=1}^{N-1}n_in_j(\partial_\mu n_i)(\partial^\mu n_j)}
{1-\sum_{k=1}^{N-1}n_k^2}.
\end{equation}
The operator $\mathcal{O}(x)$ is marginal since its UV scaling dimension is $\Delta_{\mathcal{O}}=2$. }
The NLSM is asymptotically free in the UV and is gapped in the IR. Away from the UV fixed point, its spectrum consists of a single asymptotic state of mass $m$ transforming in the vector representation of the $O(N)$ group. 

The scattering of asymptotic states is described according to \eqref{eq:scattering_amplitudes_ZZ} by three amplitudes $\sigma_1$, $\sigma_2$ and $\sigma_3$ or equivalently by  $\hat S_{\bullet}$, $\hat S_{\textbf{S}}$ and $\hat S_{\textbf{A}}$ according to \eqref{eq:scattering_amplitudes}. The relation between two sets of amplitudes is given in \eqref{eq:definitions}.
The analytic expressions for $\sigma_1$, $\sigma_2$ and $\sigma_3$ in the NLSM were found in \cite{Zamolodchikov:1978xm}, they read as
\begin{equation}
\label{eq:contraints}
\sigma_1(\theta) = - \frac{i \lambda}{i\pi-\theta}\sigma_2(\theta),\quad
\sigma_3(\theta) = - \frac{i \lambda}{\theta}\sigma_2(\theta),\quad
\lambda\equiv \frac{2\pi}{N-2},
\end{equation}
where $\sigma_2(\theta)$ is given by the ``plus'' part of (3.17) and (3.18) in \cite{Zamolodchikov:1978xm}.
The results of \cite{Zamolodchikov:1978xm} can be rewritten in a compact integral form \cite{Karowski:1978vz} as
\begin{equation}
\label{eq:anti_integral}
\hat S_{\textbf{A}}(\theta) =
\exp\left(2\int_0^{+\infty}\frac{dx}{x}\frac{\exp(-x\lambda/\pi)-1}{1+\exp(x)}
\sinh\frac{x \theta}{i\pi}\right)
\end{equation}
together with
\begin{equation}
\hat S_{\bullet}(\theta) =-\frac{\pi-i\theta}{\pi+i\theta}\times\hat S_{\textbf{A}}(\theta),\quad
\hat S_{\textbf{S}}(\theta) = \frac{\theta-i\lambda}{\theta+i\lambda}\times\hat S_{\textbf{A}}(\theta).
\end{equation}

To characterize the strength of the interaction in the NLSM one can evaluate the partial amplitudes at the crossing symmetric point $s=2$ (which corresponds to $\theta=i\pi/2$). We have the following values of the partial amplitudes then
\begin{equation}
\label{eq:pa_cros_sym_1}
\hat S_{\bullet}(i\pi/2) = -3\,\hat S_{\textbf{A}}(i\pi/2),\quad
\hat S_{\textbf{S}}(i\pi/2) = \frac{N-6}{N+2}\,\hat S_{\textbf{A}}(i\pi/2),
\end{equation}
together with
\begin{equation}
\label{eq:pa_cros_sym_2}
\hat S_{\textbf{A}}(i\pi/2) = \frac{N+2}{N-2}\times\left(
\frac{\Gamma\left(\frac{3}{4}\right)\Gamma\left(\frac{1}{4}+\frac{1}{N-2}\right)}
{\Gamma\left(\frac{1}{4}\right)\Gamma\left(\frac{3}{4}+\frac{1}{N-2}\right)}
\right)^2.
\end{equation}
We notice that crossing equations have the simplest form for $\sigma_i(\theta)$ partial amplitudes. At the crossing symmetric point they lead to the equality $\sigma_1(i\pi/2)=\sigma_3(i\pi/2)$.

The form factor associated to the antisymmetric partial amplitude was computed in \cite{Karowski:1978vz}. It reads as
\begin{equation}
\label{eq:ff_antisymetri}
\mathcal{F}_{\textbf{A}}(\theta) =
\exp\left(2\int_0^{+\infty}\frac{dx}{x \sinh x}\frac{\exp(-x\lambda/\pi)-1}{1+\exp(x)}
\sin^2\frac{x\,(i\pi-\theta)}{2\pi}\right).
\end{equation}
The form factor associated to the scalar operator was reported in \cite{Babujian:2013roa} and reads as\footnote{This paper is extremely hard to read. The formula of interest contains multiple typos. We checked however that the result we present here satisfies the system of equations \eqref{eq:ff_equations_2d} and is thus correct.}
\begin{equation}
\label{eq:ff_trivial}
\mathcal{F}_{\bullet}(\theta) = A\times\frac{\sinh \theta}{i\pi-\theta}\times
\mathcal{F}_{\textbf{A}}(\theta),
\end{equation}
where $A$ is the normalization constant. One can estimate the asymptotic behavior of expressions \eqref{eq:ff_antisymetri} and \eqref{eq:ff_trivial} at large energies.\footnote{At large $\theta$ the integrand in \eqref{eq:ff_antisymetri} has a highly oscillating piece. In order to study the asymptotics of such integrals one needs to rewrite them as a (generalized) Fourier integral. In our case we have
	\begin{equation*}
	\frac{d}{d\theta} \ln\mathcal{F}_{\textbf{A}}(\theta) =
	-\frac{1}{\pi} \text{Im}\left( \int_0^\infty dx\, g(x) \exp\left(-\frac{ix\theta}{\pi}\right) \right),\quad
	g(x)\equiv \frac{\exp(-x)}{\sinh(x)}\frac{\exp(-\lambda x/\pi)-1}{1+\exp(x)}.
	\end{equation*}
	Riemann-Lebesgue lemma states then that such integral vanishes at large $\theta$. Its leading behavior can be estimated by using integration by parts, where the leading behavior comes from the boundary term. One has then $\frac{d}{d\theta}\ln \mathcal{F}_{\textbf{A}}(\theta) \sim -\frac{\lambda}{2\pi} \frac{1}{\theta}$. } One has
\begin{equation}
\label{eq:asymptotics_ON}
\lim_{\theta\rightarrow +\infty} \mathcal{F}_{\textbf{A}}(\theta) \sim \theta^{-\frac{\lambda}{2\pi}},\quad
\lim_{\theta\rightarrow +\infty} \mathcal{F}_{\bullet}(\theta) \sim \exp(\theta)\,\theta^{-\frac{\lambda}{2\pi}-1}.
\end{equation}
From these asymptotics one sees for instance that the form factor \eqref{eq:ff_trivial} is the most general expression which satisfies the bound \eqref{eq:bound_2d} for a relevant scalar operator
since $-(\frac{\lambda}{2\pi}+1)<0$ for $N\geq 3$.

\begin{figure}[tb]
	\begin{center}
		\includegraphics[scale=0.7]{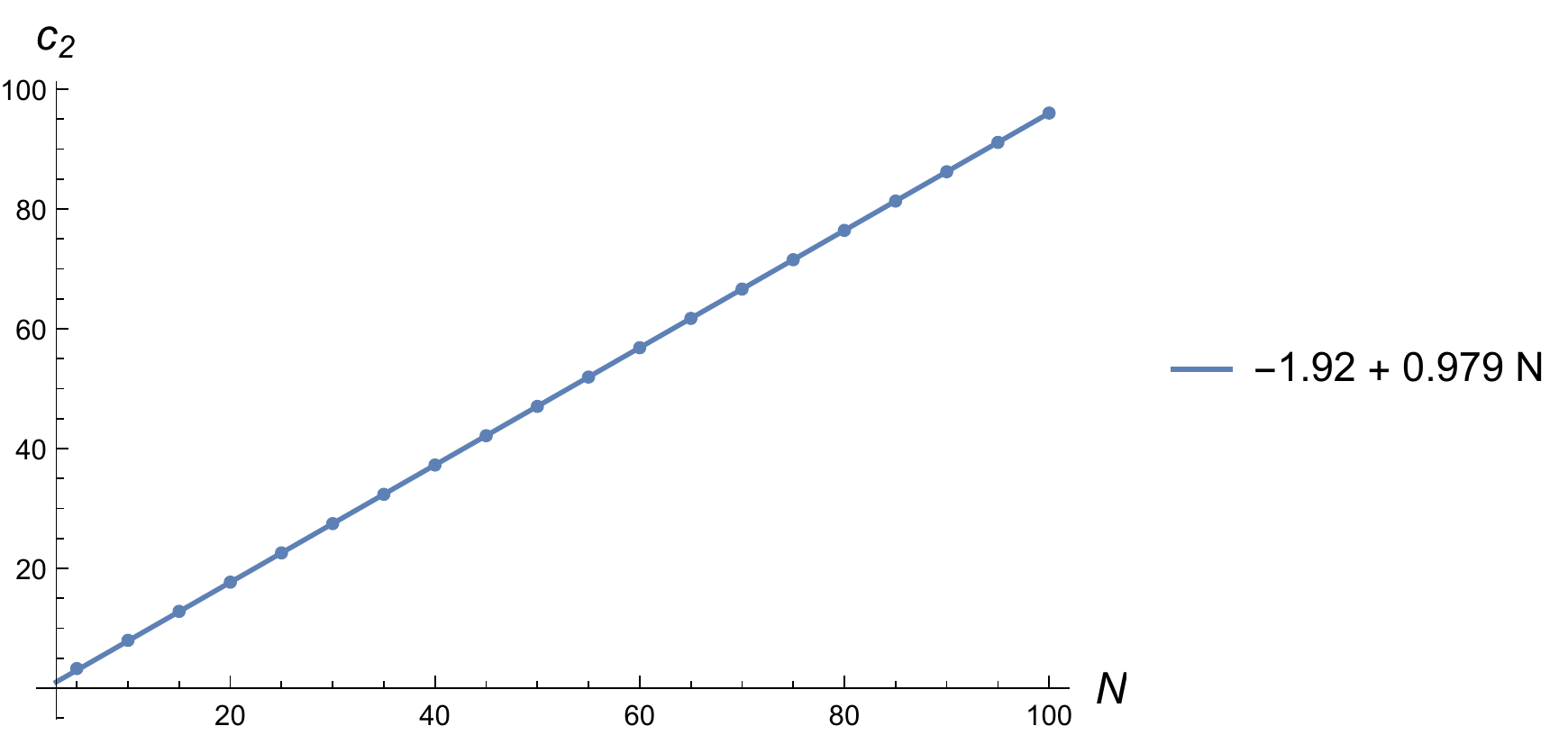}
		\caption{Contribution of the two particle states to the total central charge as a function of $N$. Dots represent the numerical values and the solid line represents the best linear fit applied for points with $N\geq 20$ only. The values of $c_2$ for small values of $N$ differ notably from the linear asymptotics. For examples for $N=3, 4, 5$ we have $c_2\approx 1.6,\; 2.39,\; 3.26$.}
		\label{fig:c2ON}
	\end{center}
\end{figure}

The form factor for the trace of the stress-tensor takes the same form (one can say that it is proportional to  the operator $\mathcal{O}$ in \eqref{eq:ON_def_2}). It is thus given by the expression \eqref{eq:ff_trivial} with the following value of normalization constant
\begin{equation}
A = -2m^2\sqrt{N},
\end{equation}
which follows from the normalization condition \eqref{eq:normalization_2FF_2d} and the definition of $\bullet$ states  given in \eqref{eq:bullet}. Notice that there are no poles in the scattering amplitudes \eqref{eq:definitions} or in the form  form factor \eqref{eq:ff_trivial}. Also the one particle form factor is zero, $\mathcal{F}_1^\Theta=0$. These follow from the $O(N)$ symmetry (some matrix elements simply cannot be constructed). One can use the form factor \eqref{eq:ff_trivial} to compute the two particle contribution to the central charge $c_2$ via \eqref{eq:spectral_2part} and \eqref{eq:c_final}. The numerical values of $c_2$ are presented on figure \ref{fig:c2ON}. For large values of $N$ we get the following approximate expression
\begin{equation}
\label{eq:c_ON}
c = c_2 + \ldots,\quad
c_2\approx 0.98 N - 1.92,
\end{equation}
where the dots represent four and higher particle contributions (notice that the odd number of particles in the majority of cases does not contribute due to $O(N)$ symmetry). In order to obtain \eqref{eq:c_ON} we have evaluated \eqref{eq:ff_trivial} numerically for multiple values of $\theta$, we have interpolated the results to obtain a continuous function and integrated it numerically to get the value of the central charge for different values of $N$.\footnote{It is important to change the integration variable to $\theta$ in \eqref{eq:c_final} in order to perform the numerical integration. The reason for that is the very slow convergence of the integral for large values of $s$.} The contribution coming from two particles and all the multi particle states should some up to the central charge of $N-1$ free bosons which is $c=N-1$.

\section{Numerical bootstrap in 2d}
\label{sec:numerics_2d}
We are now in position to formulate the numerical bootstrap problem which allows to obtain the partial amplitude, the form factor and the spectral density of a UV complete massive unitary QFT. We will focus on two dimensions in this section for two reasons: to avoid technical complications due to  spin in higher dimensions and to be able to compare our results with analytic results for 2d integrable models discussed in section \ref{app:analytic_integrable_models}.

Given a QFT which has at least one asymptotic state with a non-zero mass, we can consider the following three functions
\begin{equation}
\label{eq:functions}
\hat S(s),\quad
\mathcal{F}_{2}^\Theta(s),\quad
\rho_\Theta(s),
\end{equation}
which are the partial amplitude for 2 to 2 scattering of the lightest particle, the  form factor and the spectral density of the trace of the stress-tensor respectively. The form factor of the trace of the stress-tensor is normalized according to \eqref{eq:normalization_2FF_2d}. The spectral density of the trace of the stress-tensor is related to the UV central charge according to \eqref{eq:c_final}. 

We can write the most general ansatz for the functions \eqref{eq:functions} with real coefficients entering linearly. In case of the $\hat S(s)$ function, these coefficients are further restricted to satisfy crossing. We can then solve the following problem: determine the parameters of the ansatze leading to the minimal possible UV central charge such that the functions \eqref{eq:functions} obey the  unitarity condition \eqref{eq:matrix_B_2d}.

In section \ref{sec:optimization_problem} we will provide the details of the numerical setup. In section \ref{sec:numerical_results} we will present the numerical results. We will consider three different cases: partial amplitude with a single pole, partial amplitude with three poles and a partial amplitude with no poles but with a global $O(N)$ symmetry. We will see that these cases will reproduce numerically the known results in the sine-Gordon, $E_8$ and $O(N)$ integrable models respectively.

\subsection{Setting up the optimization problem}
\label{sec:optimization_problem}
It is convenient to introduce the $\myRho$ variable defined as 
\begin{equation}
\myRho(s;s_0) \equiv \frac{\sqrt{4-s_0}-\sqrt{4-s}}{\sqrt{4-s_0}+\sqrt{4-s}}.
\end{equation}
It maps an $s$ complex plane with one branch cut $s\in[4,\infty)$ into the unit disc (with the cut mapped to the boundary). The point $s_0<4$ is a free parameter which is mapped to the center of the disc.\footnote{The physical domain is defined via $
s+i\epsilon$ with $\epsilon>0$.	We can thus rotate the cuts using the identity
\begin{equation}
\nn
\sqrt{4-s} = -i\,\sqrt{s-4}.
\end{equation}}
Another useful variable is $\phi(s_0)$ defined as
\begin{equation}
e^{i\phi(s_0)} \equiv \myRho(s;s_0)
\quad\Rightarrow\quad
s = s_0+\frac{8-2s_0}{1+\cos \phi(s_0)}.
\end{equation}
In what follows we will often use the $\phi$ variable defined as
\begin{equation}
\phi\equiv\phi(0)\quad\Rightarrow\quad
s = \frac{8}{1+\cos \phi},
\end{equation}
which maps a ray into an interval, more precisely
\begin{equation}
s\in [4,\infty)
\quad\Leftrightarrow\quad
\phi\in[0,\pi].
\end{equation}

Let us discuss now the situation when our QFT has $k$ asymptotic states below the two-particle threshold
\begin{equation}
\label{eq:masses_asymptotic_states}
m_1=1,\;
m_2,\;\ldots,\; m_k.
\end{equation}
According to the discussion of section \ref{sec:poles} these asymptotic states will appear as simple poles in the interacting part of the amplitude and the form factor. We can then write the following ansatze \cite{Paulos:2017fhb},
\begin{align}
\label{eq:ansatz_1}
\mathcal{T}(s) &= -\sum_{i=1}^k\frac{g_i^2}{s-m_i^2} + \sum_{n=0}^{N_{max}} a_n\times \myRho(s;2)^n+(s\leftrightarrow 4-s),\\
\label{eq:ansatz_2}
\mathcal{F}_{2}^\Theta(s) &= -\sum_{i=1}^k\frac{\lambda_i}{s-m_i^2} + \sum_{n=0}^{N_{max}} b_n \times\myRho(s;0)^n,\\
\label{eq:ansatz_3}
\rho_\Theta(s) &= 2\sum_{n=0}^1c_n\times\cos(n\phi)-2\sum_{n=1}^{N_{max}}
d_n\times\sin(n\phi),
\end{align}
where we have defined for convenience
\begin{equation}
\lambda_i\equiv g_i \mathcal{F}_{1,i}.
\end{equation}
These ansatze depend on the set of real parameters $a_n$, $b_n$, $c_n$ and $d_n$ which enter linearly. The form factor of the trace of the stress-tensor obeys the normalization \eqref{eq:normalization_2FF_2d}. This leads to the linear constraint for the unknown coefficient
\begin{equation}
\label{eq:FF_condition}
\sum_{i=1}^k \lambda_i\,m_i^{-2} +b_0= -2.
\end{equation}
Taking it into account we can write the final ansatz for the form factor as
\begin{equation}
\label{eq:ansatz_2_corrected}
\mathcal{F}_2^\Theta(s) = -2
-\sum_{i=1}^k\lambda_i\times\left(\frac{1}{m_i^2}+\frac{1}{s-m_i^2}\right)
+\sum_{n=1}^{N_{max}} b_n \times\myRho(s;0)^n.
\end{equation}

\paragraph{Unitarity constraints}
The unitarity constraint is given by \eqref{eq:matrix_B_2d_final}. It should be obeyed for any value of $s\in [4m_1^2,+\infty)$. To implement this requirement in practice we discretize $s$ and choose a large set of sample values. All the plots are made with 200 sample points distributed on the Chebyshev grid $\phi\in[0,\pi]$. The entries of  the $3\times3$ matrix \eqref{eq:matrix_B_2d_final} are complex. we can rewrite however the semipositive definite condition \eqref{eq:matrix_B_2d_final} in terms of $6\times6$ matrices with purely real coefficients by defining
\begin{equation}
\label{eq:real_imaginary}
R(s)\equiv \text{Re} B(s),\quad
I(s)\equiv \text{Im} B(s),\quad
R^T=R,\quad I^T=-I.
\end{equation}
The semipositive definite constraint reads as
\begin{equation}
z^\dagger B(s) z \geq 0,
\end{equation}
where $z$ are some complex 3 dimensional vectors. Due to~\eqref{eq:real_imaginary} this is equivalent to
\begin{equation}
\begin{pmatrix}
R(s) & -I(s) \\
I(s) &  R(s)
\end{pmatrix}
\succeq 0.
\end{equation}

\paragraph{Central charge bound}
Let us consider now the expression of the UV central charge in terms of the spectral density \eqref{eq:c_final}. Generalizing it to the case of multiple asymptotic states we can write
\begin{equation}
\label{eq:c}
c_{UV}=12\pi\left(
\sum_{i=1}^km_i^{-4}\,|\mathcal{F}_{1,i}^\Theta|^2 + 
\int_{4m_1^2}^\infty ds\, s^{-2} \rho_\Theta(s)
\right).
\end{equation}
There is in principle no bound on how big the spectral density $\rho_\Theta(s)$ can be. However there is certainly at least a trivial lower bound $\rho_\Theta(s)\geq 0$ due to \eqref{eq:rho_bound} which implies $c_{UV}\ge 0$.

The expression \eqref{eq:c} has the unknown constants $\mathcal{F}_{1,i}$ entering in a quadratic way. Thus, we cannot directly apply methods of linear programming to minimize \eqref{eq:c}. We can however use a simple trick to rewrite \eqref{eq:c} in a linear way. Consider the following inequality
\begin{equation}
\label{eq:cT_bound}
c_{UV} \leq c_{UV}^{\text{bound}},\quad
c_{UV}^{\text{bound}} \equiv
12\pi\,\left(
\sum_{i=1}^km_i^{-4}\,u_i + \int_{4m_1^2}^\infty ds\;s^{-2}\rho_\Theta(s)\right),
\end{equation}
where we have introduced new non-negative real parameters $u_i$ which obey the following constraints
\begin{equation}
\forall i:\quad 
0\leq|\mathcal{F}_{1,i}|^2 \leq u_i
\quad\Rightarrow\quad
0\leq|\lambda_i|^2 \leq g_i^2 u_i.
\end{equation}
The latter inequality is equivalent to
\begin{equation}
\label{eq:additional_constraint}
\forall i:\quad 
\begin{pmatrix}
g_i^2 & \lambda_i^* \\
\lambda_i & u_i
\end{pmatrix}\succeq 0.
\end{equation}
Now instead of minimizing $c_{UV}$ we can minimize $c_{UV}^{\text{bound}}$ defined in \eqref{eq:cT_bound} given that the condition \eqref{eq:additional_constraint} is satisfied. At the minimum we will simply get $u_i=|\mathcal{F}_{1,i}|^2$ and $c_{UV}=c_{UV}^{\text{bound}}$.

\paragraph{Central charge minimization}
We are finally ready to formulate the numerical bootstrap problem: given a set of asymptotic states and their masses \eqref{eq:masses_asymptotic_states}, determine the linear coefficients
\begin{equation}
\nn
g_i^2,\;
\lambda_i,\;
u_i,\;
a_n,\;
b_n,\;
c_n,\;
d_n,\;
\end{equation}
in the ansazte \eqref{eq:ansatz_1} - \eqref{eq:ansatz_3} and \eqref{eq:cT_bound} such that the semipositive conditions \eqref{eq:matrix_B_2d_final}, \eqref{eq:additional_constraint} are satisfied and the central charge $c_{UV}^{\text{bound}}$ in \eqref{eq:cT_bound} has the minimal possible value. Sometimes we will also be fixing the values of $g_i^2$ in order to single out known integrable models. To perform the numerics we use the semipositive program solver SDPB \cite{Simmons-Duffin:2015qma,Landry:2019qug}.

\subsection{Numerical results}
\label{sec:numerical_results}

We now solve the optimization problem of the central charge minimization defined in section \ref{sec:optimization_problem} in three different cases. First, in the presence of a single pole. We find a special point on the central charge bound which corresponds to the sine-Gordon model. We will recover its partial amplitude and the two particle form factor. We will then investigate the dependence of the central charge on the parameter of the sine-Gordon model. Second, in the presence of three poles. Injecting the values of the masses and the residue of the lightest asymptotic state in the $E_8$ we recover numerically the partial amplitude of the $E_8$ model. We also obtain the form factor consistent with the analytic results. Third, we address the case of no poles in the presence of $O(N)$ global symmetry. We will minimize the central charge by scanning over different values of the partial amplitudes at the crossing symmetric point.

\subsubsection{One pole}
We assume that the system is described at least by two asymptotic states with masses
\begin{equation}
\label{eqspectrum_SG}
m_1=1,\quad m_2\in[\sqrt{2},2].
\end{equation}
This parameter range is chosen to mimic the sine-Gordon behavior, see \eqref{eq:ratio} and \eqref{eq:parameters_2breathers}.
We consider the scattering of the $m_1$ asymptotic state and assume that there is only one simple pole in the scattering amplitude due to the second asymptotic state.\footnote{In other words there is no self coupling of $m_1$ state. This can be justified by requiring for example a $Z_2$ symmetry.}

\begin{figure}[t]
	\begin{center}
		\includegraphics[scale=0.5]{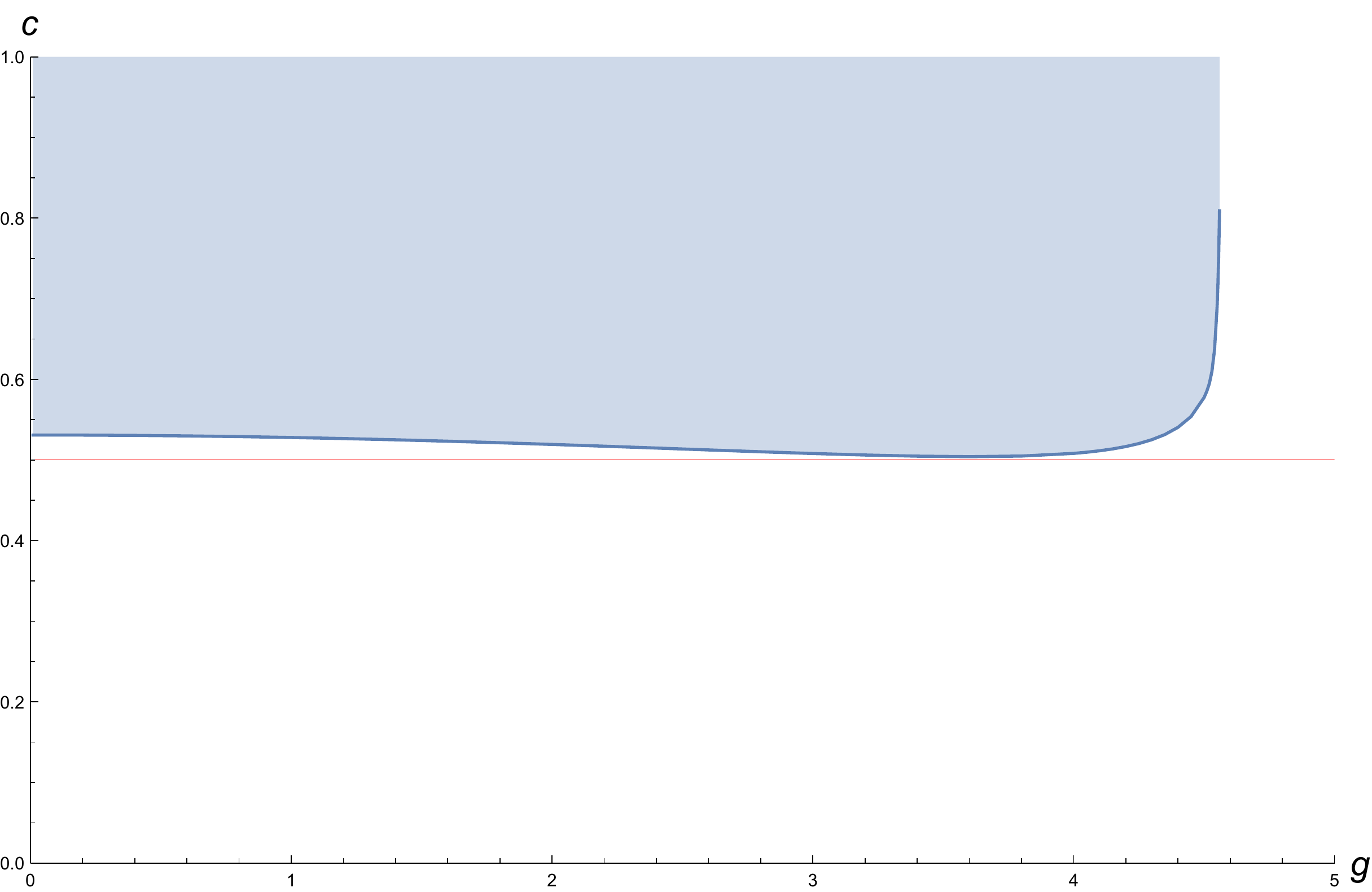}
		\caption{Lower bound on the UV central charge as a function of the cubic coupling $g$ between particles of mass $m_1=1$, $m_1=1$ and $m_2=\sqrt{3}$.
		The allowed region is depicted in blue. The bound extends up to $g=4.55901$ which is a critical value for which the optimization problem is feasible.
		The bound was obtained with $N_{max}=50$. 
		  The red horizontal line at $c=1/2$ is added for convenience.}
		\label{fig:bound}
	\end{center}
\end{figure}

We can now look for a minimum of the UV central charge fixing the value of the trilinear coupling $g=g_2$. We also set the mass of the second asymptotic state to be $m_2=\sqrt{3}$. The numerical results are presented on figure \ref{fig:bound}. On the plot there appears a special value of the trilinear coupling for which the optimization problem becomes unfeasible. This critical value is
\begin{equation}
\label{eq:critical_g}
g
\approx 4.55901.
\end{equation}
The value \eqref{eq:critical_g} is in a perfect agreement with the results of \cite{Paulos:2016but} where it was found that there is an upper bound on the trilinear coupling, see figure 4. It was also found that this value corresponds to the $b_1b_1b_2$ trilinear coupling of the sine-Gordon model \eqref{eq:coupling_SG}, where $b_1$ and $b_2$ stand for the first and the second breathers.

\begin{figure}[tb]
	\begin{center}
		\includegraphics[scale=0.6]{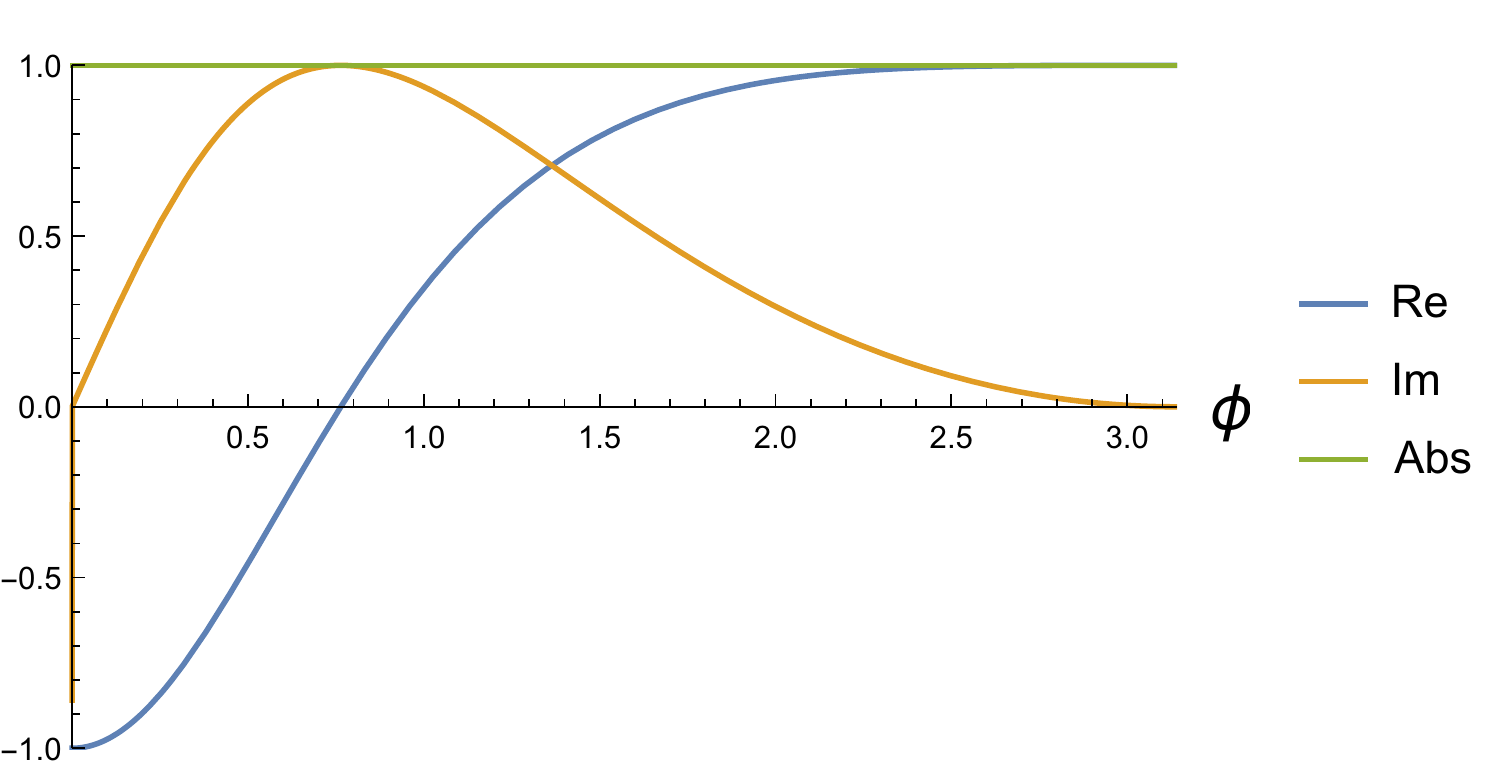}
		\caption{The real part, the imaginary part and the absolute value of the partial amplitude for the scattering of the lightest asymptotic state with $m_1=1$ given the mass of the second asymptotic state $m_2=\sqrt{3}$ and the value of the trilinear coupling \eqref{eq:critical_g}. The plot is constructed with $N_{max}=50$.}
		\label{fig:amplitude}
	\end{center}
\end{figure}

\begin{figure}[tb]
	\begin{center}
		\includegraphics[scale=0.6]{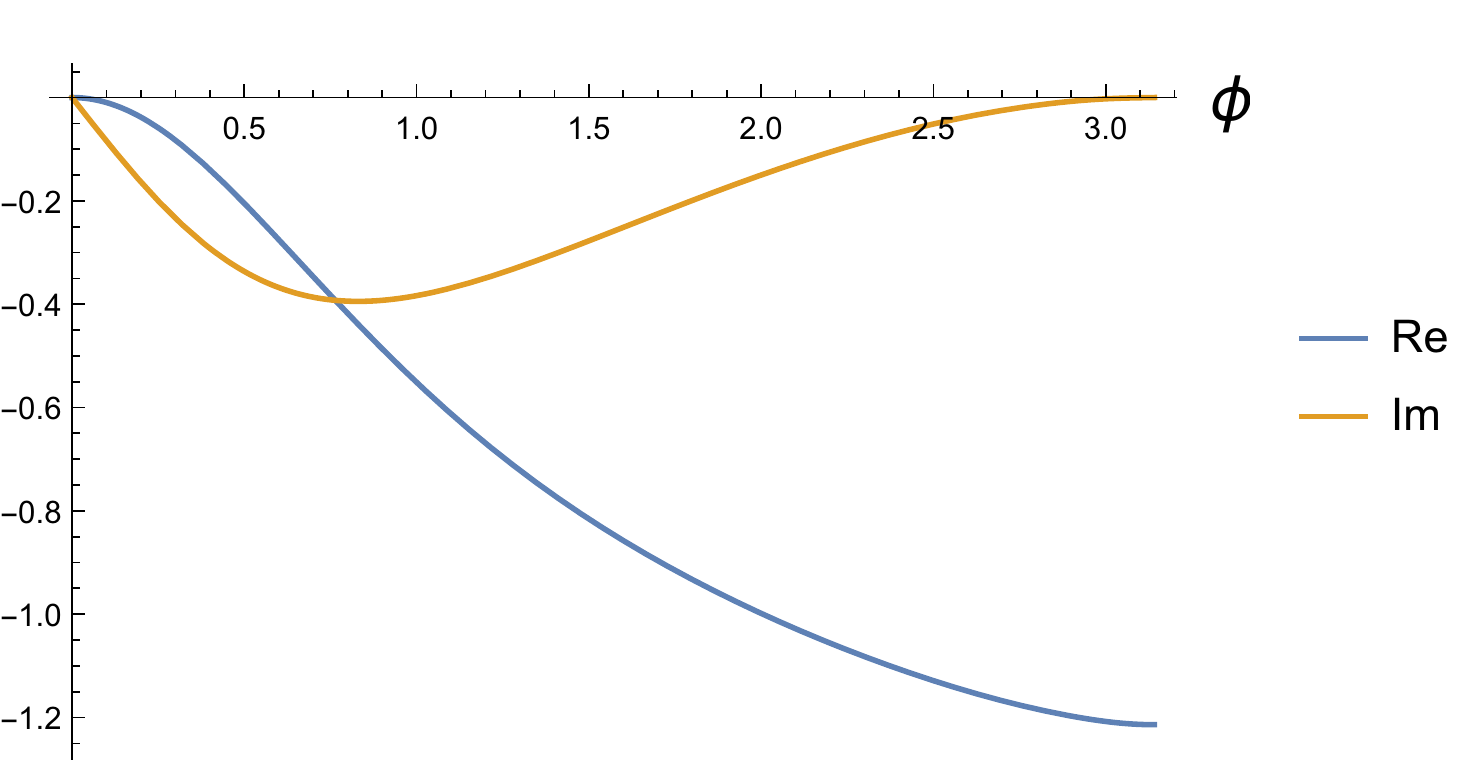}
		\caption{The real and imaginary parts of the two particle form factor for the masses $m_1=1$ and $m_2=\sqrt{3}$ and the value of the trilinear coupling \eqref{eq:critical_g}. The plot is constructed with $N_{max}=50$.}
		\label{fig:ff}
	\end{center}
\end{figure}

\begin{figure}[tb]
	\begin{center}
		\includegraphics[scale=0.5]{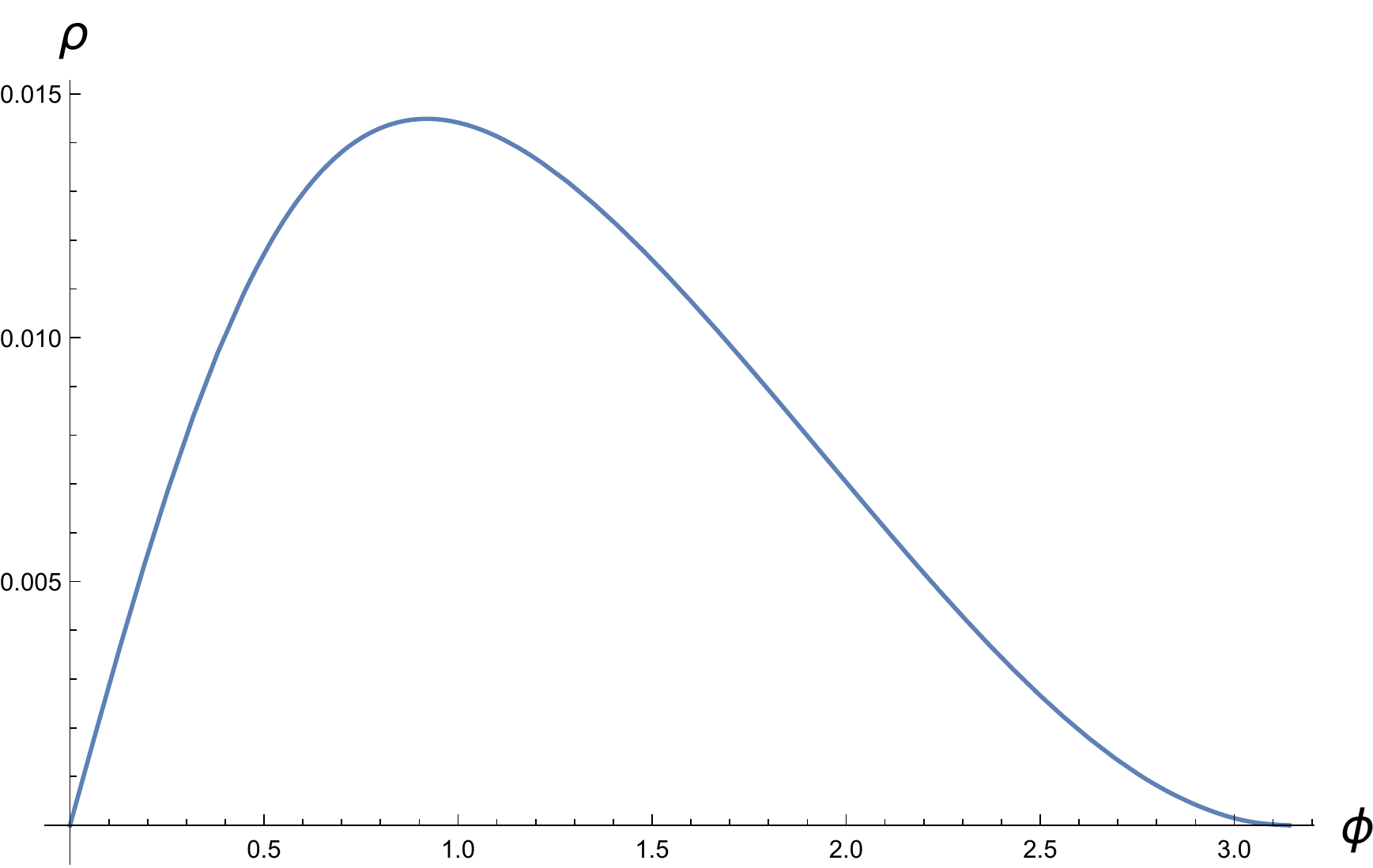}
		\caption{The two particle contribution to the spectral density for the masses $m_1=1$ and $m_2=\sqrt{3}$ and the value of the trilinear coupling \eqref{eq:critical_g}. The plot is constructed with $N_{max}=50$.}
		\label{fig:spectral}
	\end{center}
\end{figure}

At the critical value \eqref{eq:critical_g} we also recover the partial amplitude of the first breather $b_1$ and the form factor of the trace of the stress-tensor. They are presented in figures \ref{fig:amplitude} and \ref{fig:ff} respectively. They match precisely the exact analytic expressions \eqref{eq:b1b1-b1b1} and \eqref{eq:theta-b1b1}. The numerical expression for the spectral density is given on figure \ref{fig:spectral}. It is completely saturated by the two particle contribution and  thus it should be regarded as the two particle part of the spectral density.
This is a general feature of our numerical results.

The numerical procedure allows to determine the UV central charges of the sine-Gordon model. For $m_2=\sqrt{3}$ we have
\begin{equation}
\label{eq:numerical_values}
c = c_{b_2} + c_{b_1b_1 + \ldots} = 0.80921+\ldots,\quad
c_{b_2} = 0.72126,\quad
c_{b_1b_1} = 0.08795,
\end{equation}
where $c_{b_2}$ is the single particle contribution of the second breather and $c_{b_1b_1}$ is the two particle contribution of the first breather. The dots stand for other (positive) contributions which are left undetermined by our procedure. The value of $c_{b_2}$ reported in \eqref{eq:numerical_values} is in a perfect agreement with the one obtained from analytic expressions, see figure \ref{fig:c1SG}.

\begin{figure}[t]
	\begin{center}
		\includegraphics[scale=0.8]{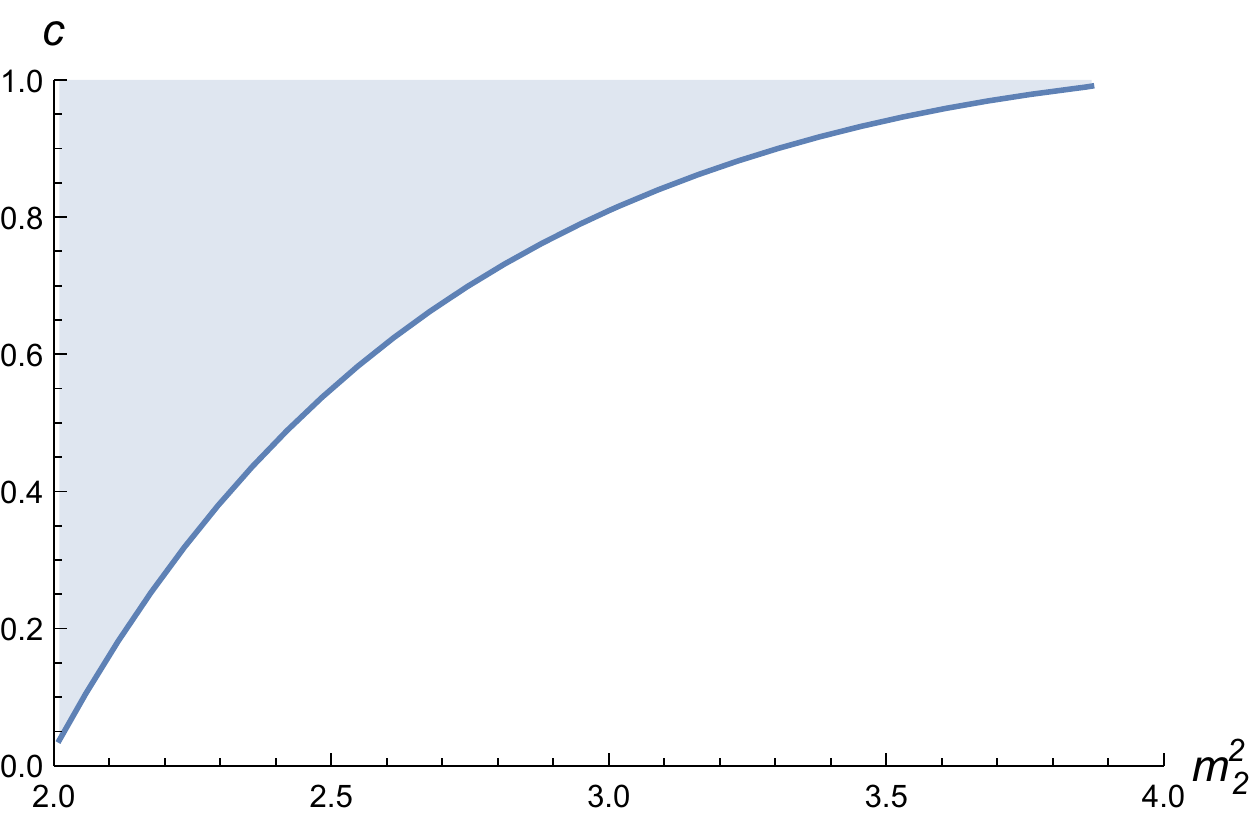}
		\caption{Lower bound on the UV central charge obtained with $N_{max}=30$ as a function of $m_2^2$ with the trilinear coupling $g$ fixed to be the one of the sine-Gordon model. The allowed region is depicted in blue.}
		\label{fig:bound2SG}
	\end{center}
\end{figure}

Finally we vary the mass $m_2$ and fix the trilinear coupling to be precisely the one of the sine-Gordon model \eqref{eq:coupling_SG}. We present the result on figure
\ref{fig:bound2SG}. The values of the central charge and the corresponding partial amplitude and the form factor are precisely the ones of the sine-Gordon model. The bound on the central charge around $m_2^2=4$ approaches $1$, it becomes however very sensitive to $N_{max}$. For the reference we provide here the values of central charges at two extremes of figure \ref{fig:bound2SG}, namely at $m_2^2=2.01$ and $m_2^2=3.87$. We have
\begin{align}
\label{eq:numerical_values_2}
m_2^2=2.01:\quad c &= c_{b_2} + c_{b_1b_1 + \ldots} = 0.03741+\ldots,\quad
c_{b_2} = 0.01456,\quad
c_{b_1b_1} = 0.02285,\\
\label{eq:numerical_values_3}
m_2^2=3.87:\quad c &= c_{b_2} + c_{b_1b_1 + \ldots} = 0.99083+\ldots,\quad
c_{b_2} = 0.55777,\quad
c_{b_1b_1} = 0.43305.
\end{align}
These results are in a full agreement with the discussion of section \ref{sec:SG}, see in particular figure \ref{fig:c1SG}. Notice that at $m_2^2 = 2$ the sine-Gordon contains an infinite number of breathers and thus it is expected that the contributions from the first two breathers account for a very small portion of the central charge. On the contrary,  at $m_2^2 = 4$ the sine-Gordon model becomes a free theory of a   scalar field of mass $m_1=1$ (the coupling $\beta \to 0$ in section \ref{sec:SG}).  
Therefore, the two particle contribution of the first breather accounts for the whole central charge.

\subsubsection{Three poles} 
We would like to study the $E_8$ model also known as the 2d Ising model with magnetic deformation. We assume that the system is described by three asymptotic states with masses
\begin{equation}
\label{eq:spectrum_E8}
m_1=1,\quad
m_2=2\cos(\pi/5),\quad
m_3=2\cos(\pi/30).
\end{equation}
Notice that the $E_8$ model has actually eight asymptotic states but only three of them are   below the two particle threshold $4m_1^2$.
We consider the scattering of particle $m_1$ and allow all three poles due to particles $m_1$, $m_2$ and $m_3$.

We minimize the central charge in this setup not specifying the values of the trilinear couplings first. Unfortunately this turns out not to be enough to single out the $E_8$ model. We further specify the value of the very first trilinear couplings $g_1$ given in \eqref{eq:couplings_E8}. The values of $g_2$ and $g_3$ 
are obtained during the central charge minimization procedure and match precisely the ones in  \eqref{eq:couplings_E8}.

\begin{figure}[t]
	\begin{center}
		\includegraphics[scale=0.6]{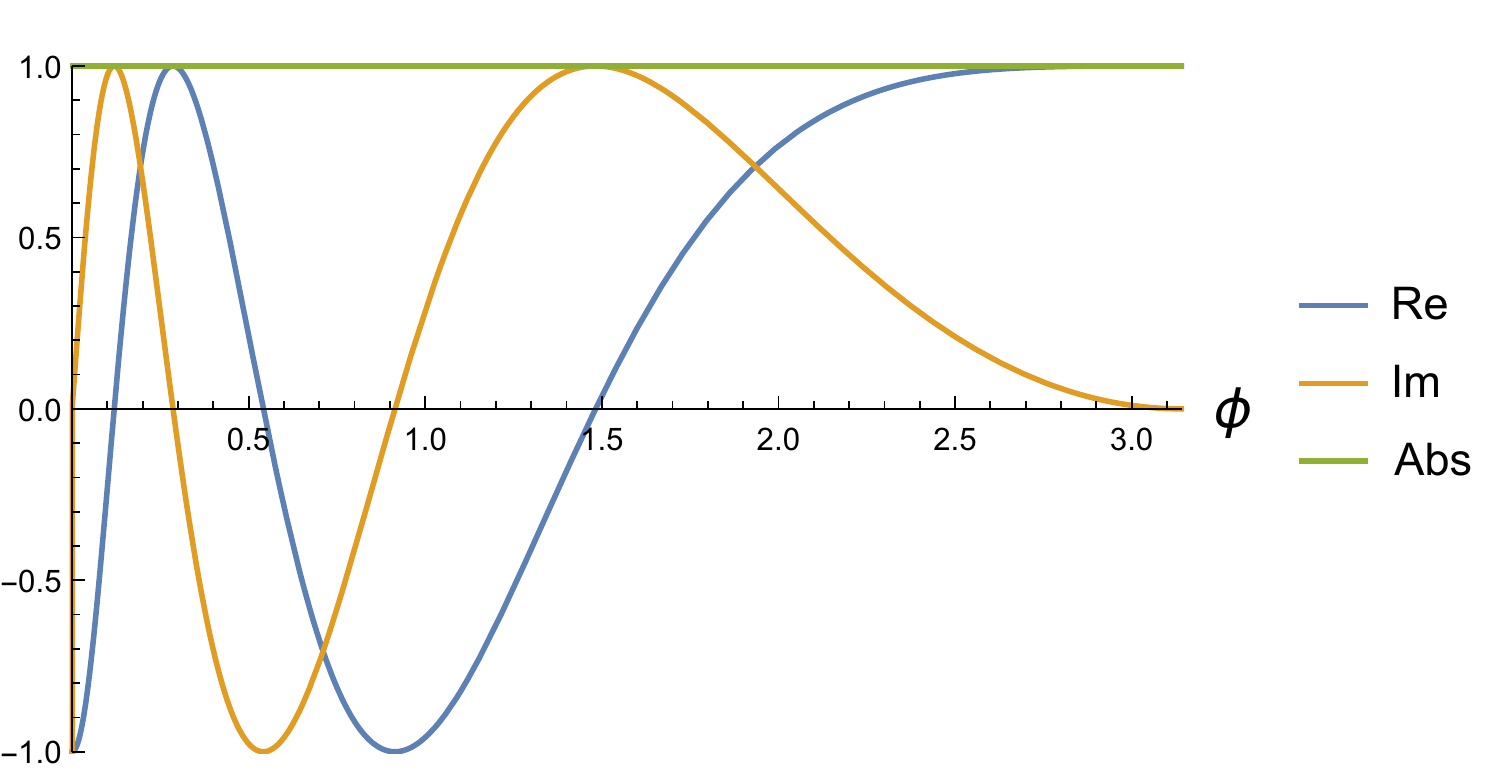}
		\caption{The real part, the imaginary part and the absolute value of the partial amplitude of the scattering of the lightest asymptotic states in the $E_8$ model. The plot is constructed with $N_{max}=50$.}
		\label{fig:ampE8}
	\end{center}
\end{figure}

\begin{figure}[t]
	\begin{center}
		\includegraphics[scale=0.6]{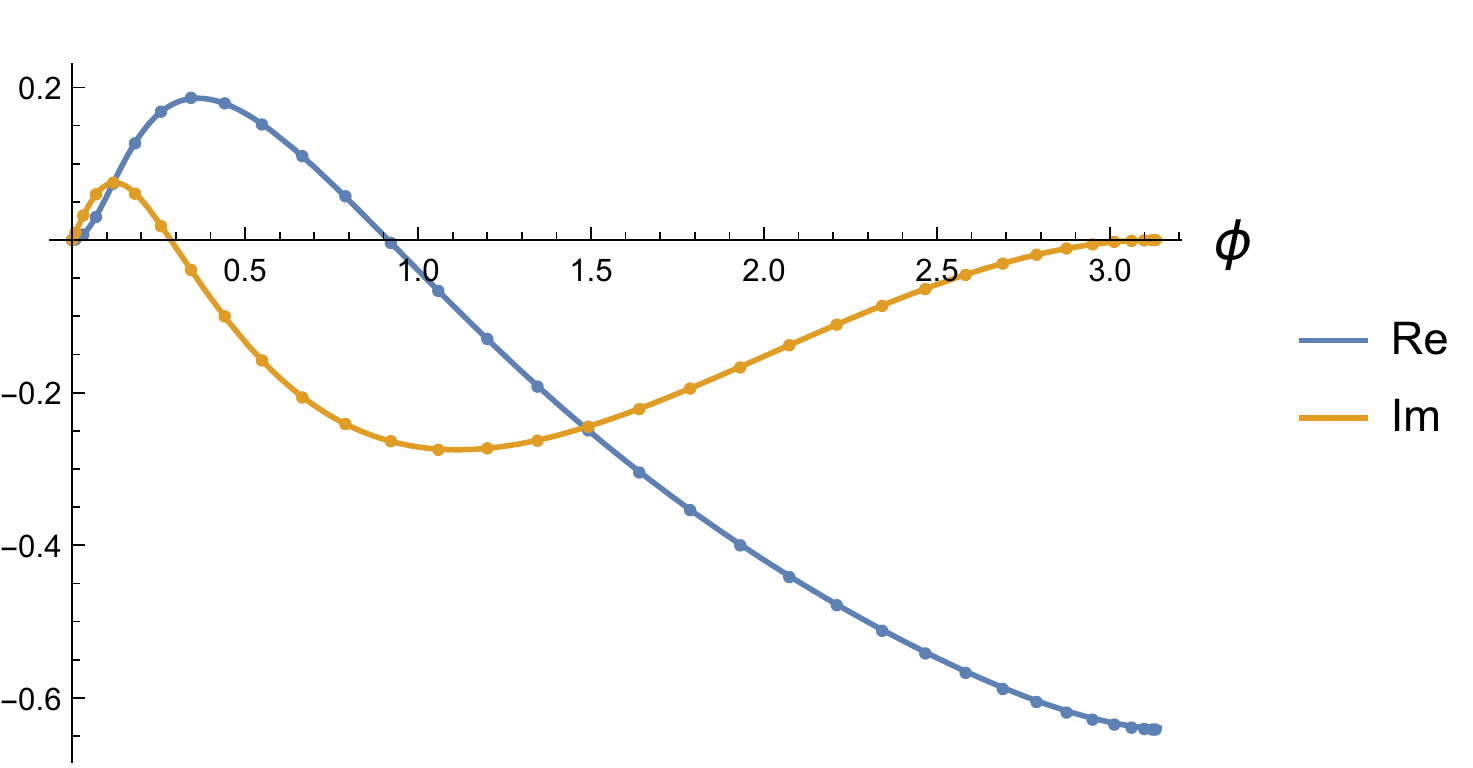}
		\caption{The real and imaginary parts of the form factor of the stress-tensor found numerically in the $E_8$ model. The dots correspond to the trace of the stress-tensor form factor defined in \eqref{eq:fake}. The plot is constructed with $N_{max}=50$.}
		\label{fig:ffE8}
	\end{center}
\end{figure}

As a result of our numerical procedure we also obtain the partial amplitude shown on figure \ref{fig:ampE8} and the two particle form factor of the trace of the stress-tensor shown figure \ref{fig:ffE8}. The partial amplitude perfectly matches the analytic expression \eqref{eq:E8_11-11}. 
This was expected since in \cite{Paulos:2016but}, it was shown that this is the unique amplitude with maximal trilinear coupling $g_1$ given in \eqref{eq:couplings_E8}.
The form factor however does not match the analytic expression \eqref{eq:ff_E8} with the coefficients \eqref{eq:values_AB}. It matches however the following linear combination of the $\sigma$ and $\epsilon$ form factors 
\begin{equation}
\label{eq:fake}
\mathcal{F}_{2,\text{fake}}^\Theta(s) \approx
-0.79\,\mathcal{F}_{11}^\sigma(s)+
4.06\,\mathcal{F}_{11}^\epsilon(s),
\end{equation}   
given by \eqref{eq:ff_E8} with the coefficients \eqref{eq:sigma_epsilon_coefficients} and \eqref{eq:norm_sigma_epsilon}.
We refer to this as the fake trace of the stress-tensor form factor. The appearance of such an object is due to a peculiar situation when the form factors of different scalar operators cannot be easily distinguished (because they have the same large $s$-behaviour). In order to distinguish them one needs a more complicated setup which includes scattering of at least two different asymptotic states. Our results are however consistent with the analytical ones due to \eqref{eq:fake}. It is not a surprise then that the central charge value of the fake form factor $c\approx 0.04945$ does not correspond to the expected values \eqref{eq:c_E8} which follows from the analytic results.

\subsubsection{Zero poles and $O(N)$ global symmetry}
We consider here a single asymptotic state with mass $m=1$ which transforms in the vector representation of the $O(N)$ global symmetry group. We further assume that there are no poles in the scattering amplitude of such states. (For previous works on 2d QFTs with $O(N)$ symmetry see \cite{Paulos:2018fym,He:2018uxa,Cordova:2018uop,Cordova:2019lot}). In section \ref{sec:unitarity_ON} we have shown that this amplitude decomposes into three amplitudes in the trivial, symmetric traceless and antisymmetric representations.
We now write the following ansatz for the partial amplitudes
\begin{align}
\nn
\hat{\mathcal{S}}_{\bullet}(s) &= u^\bullet_{0} + \sum_{n=1}^{N_{max}} \Big(u^\bullet_{n}\, \myRho(s;2)^n+v^\bullet_{n}\, \myRho(4-s;2)^n\Big),\\
\label{eq:ansatz_ON}
\hat{\mathcal{S}}_{\textbf{S}}(s) &= u^{\textbf{S}}_{0} + \sum_{n=1}^{N_{max}} \Big(u^{\textbf{S}}_{n}\, \myRho(s;2)^n+v^{\textbf{S}}_{n}\, \myRho(4-s;2)^n\Big),\\
\hat{\mathcal{S}}_{\textbf{A}}(s) &= u^{\textbf{A}}_{0} + \sum_{n=1}^{N_{max}} \Big(u^{\textbf{A}}_{n}\, \myRho(s;2)^n+v^{\textbf{A}}_{n}\, \myRho(4-s;2)^n\Big),
\nn
\end{align}
where $u_n$ and $v_n$ are some constants. Notice, that contrary to \eqref{eq:ansatz_1} we parametrize here the entire partial amplitude and not only its interacting part. We plug this ansatz into the system of crossing equations \eqref{eq:crossing_ON}. It becomes a system of linear algebraic equation on the linear coefficients $u_n$ and $v_n$.
It can be used for example to express $u^{\textbf{A}}_{0}$ and $v^{\textbf{S}}_{n}$, $u^{\textbf{A}}_{n}$, $v^{\textbf{A}}_{n}$ for $n\geq 1$ in terms of the unknown linear coefficients $u^\bullet_{0}$, $u^{\textbf{S}}_{0}$ and $u^\bullet_{n}$, $v^\bullet_{n}$, $u^{\textbf{S}}_{n}$ for $n\geq 1$.
Plugging this solution back into \eqref{eq:ansatz_ON} we obtain an automatically crossing symmetric ansatz.  We demand then that the ansatz obeys the unitarity constraints \eqref{eq:unitarity_t} and \eqref{eq:unitarity_sa}.
The form factor of the trace of the stress-tensor is defined in \eqref{eq:form_factor_trivial} provided it obeys the following normalization
\begin{equation}
\mathcal{F}_{\bullet2}^\Theta(0) = -2\sqrt{N},
\end{equation}
which follows from \eqref{eq:normalization_trivial} and \eqref{eq:normalization_2FF_2d}.

\begin{figure}[t]
	\begin{center}
		\includegraphics[scale=0.5]{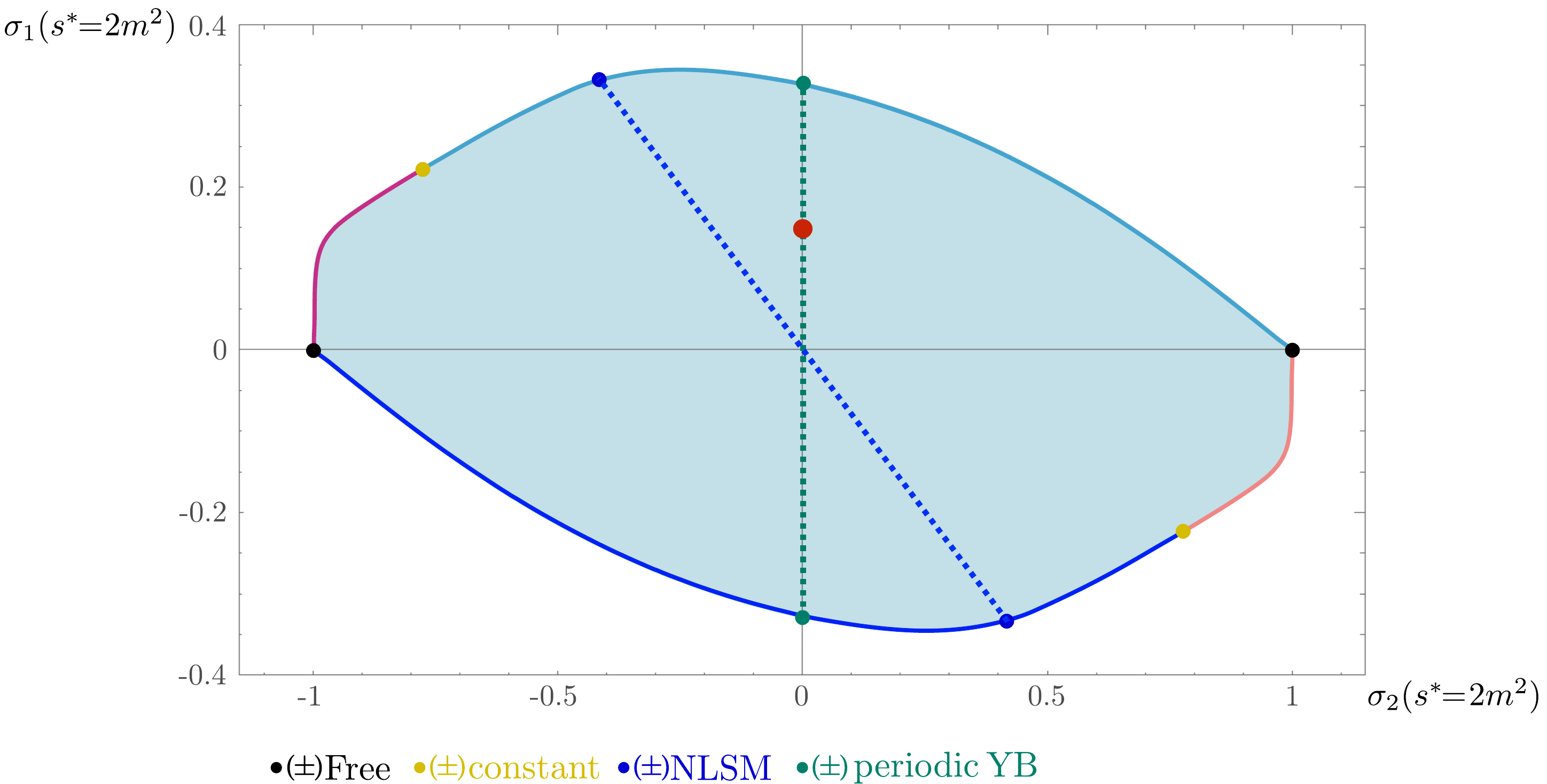}
		\caption{Allowed region in the $(\sigma^*_2,\sigma^*_1)$ plane for $N=7$ (from figure 7 of \cite{Cordova:2019lot}). 
		 The blue dashed line marks the section \eqref{eq:sectoin_1} ending at the integrable $O(N)$ sigma model (NLSM).
		 The green dashed line marks the section \eqref{eq:sectoin_2} ending at the periodic Yang-Baxter solution (pYB).	
		 The red dot marks the value of $(\sigma^*_2,\sigma^*_1)$ that minimizes the UV central charge.	}
		\label{fig:monolith}
	\end{center}
\end{figure}

\begin{figure}[t]
	\begin{center}
		\subfigure[]{\label{fig:neutral33}\includegraphics[scale=0.6]{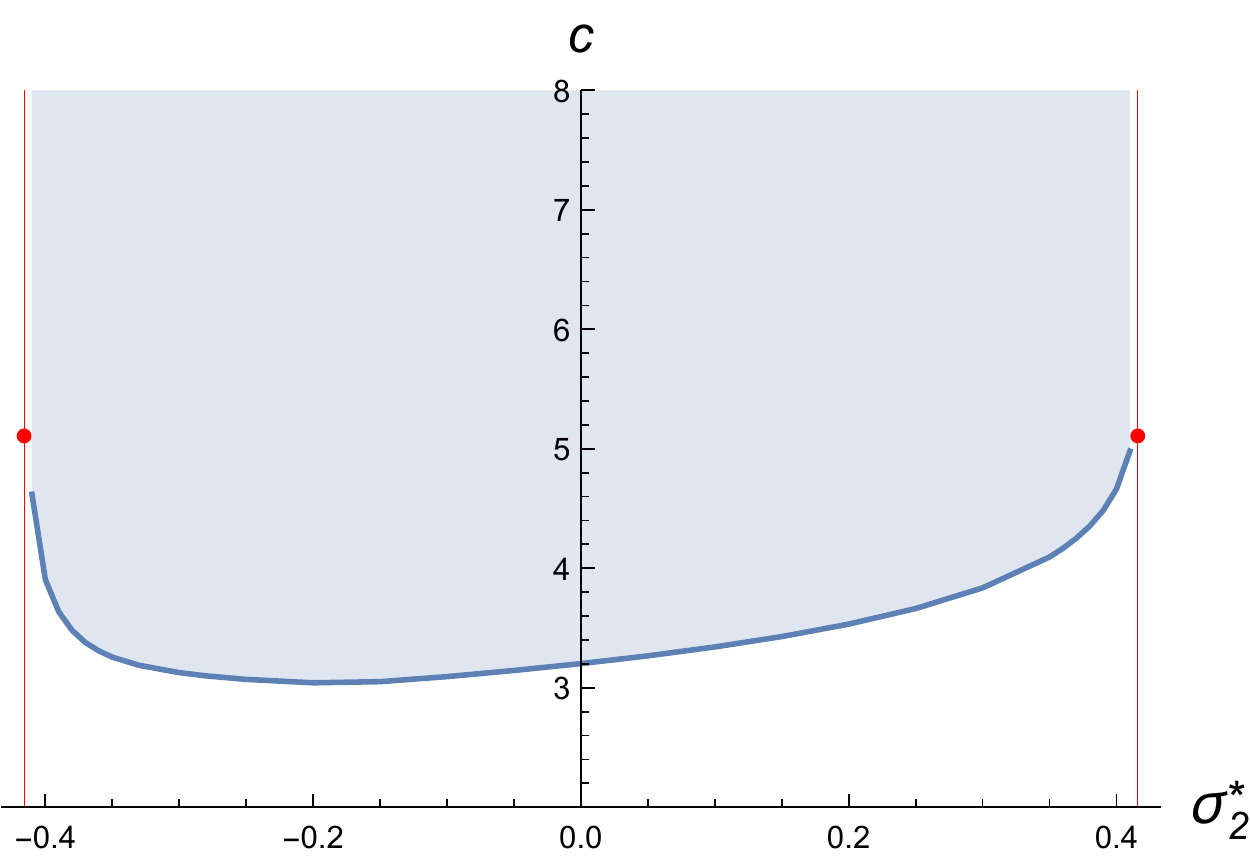}}
		\subfigure[]{\label{fig:neutral44}\includegraphics[scale=0.6]{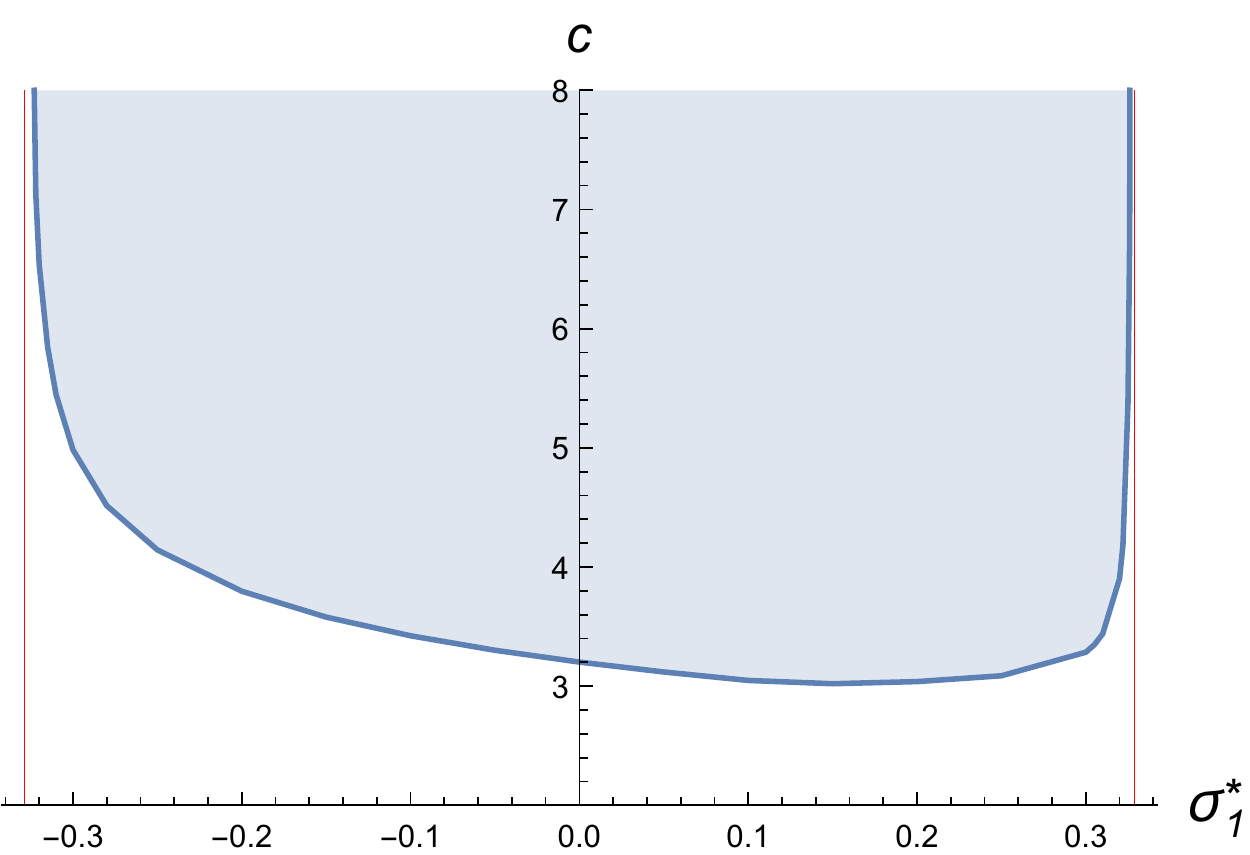}}
		\caption{Bounds on the central charge in $O(N)$ models. The allowed region is depicted in blue. Both plots are constructed with $N=7$ and $N_{max}=30$. \emph{Left}: the bound on the central charge as a function of $\sigma_2^*$ on the section \eqref{eq:sectoin_1}. The vertical lines correspond to $\sigma_2^*\approx \pm0.415927$. The red dots represent the two particle contribution to the central charge $c_2\approx 5.11$ in the NLSM estimated in section \ref{sec:NLSM}. 
		\emph{Right}: the bound on the central charge as a function of $\sigma_1^*$ on the section \eqref{eq:sectoin_2}. The vertical lines correspond to $\sigma_1^*\approx \pm0.329$.}
		\label{fig:ON}
	\end{center}
\end{figure}

Following \cite{He:2018uxa,Cordova:2018uop,Cordova:2019lot} we can minimize the central charge fixing the values of partial amplitudes at the crossing symmetric point. Let us define
\begin{equation}
\label{eq:param_space}
\sigma_1^*\equiv \sigma_1(s=2),\quad
\sigma_2^*\equiv \sigma_2(s=2),\quad
\sigma_3^*\equiv \sigma_3(s=2).
\end{equation}
We remind that the crossing symmetry requires $\sigma_3^*=\sigma_1^*$. Fixing the values \eqref{eq:param_space} is equivalent to fixing the values  $u^\bullet_{0}$, $u^{\textbf{S}}_{0}$ and $u^{\textbf{A}}_{0}$ in \eqref{eq:ansatz_ON} due to \eqref{eq:definitions}. We can now scan for example over $\sigma_1^*$ and $\sigma_2^*$ and minimize the central charge to obtain the 3d plot. 
The allowed values of $\sigma_1^*$ and $\sigma_2^*$ form the bounded domain shown in figure \ref{fig:monolith}. \footnote{This is  figure 7 of \cite{Cordova:2019lot} which we reproduce here for the reader's convenience.}
To decrease the amount of numerical computations we will focus here only on two sections of the plane $\sigma_1^*$ and $\sigma_2^*$, namely
\begin{align}
\label{eq:sectoin_1}
\text{section 1}:\quad
&\sigma^*_1 = -\frac{4}{N-2}\sigma_2^*,\\
\label{eq:sectoin_2}
\text{section 2}:\quad
&\sigma^*_2 = 0.
\end{align}
Our section 1 connects two $(\pm)$NLSM points and our section 2 connects two $(\pm)$pYB points in figure \ref{fig:monolith}.
In the case $N=7$ and $N_{max}=30$ we present the results on figure \ref{fig:ON}. 

Let us discuss the numerical results now. For section 1 the optimization problem is feasible for $\sigma_2^*\in [-0.415,\, 0.415]$. The boundary values from left to right correspond to the ``minus'' NLSM and NLSM respectively. For instance the right boundary value matches the analytic results  \eqref{eq:pa_cros_sym_1} and \eqref{eq:pa_cros_sym_2}. At the right boundary we have  reconstructed the partial amplitudes and the form factor. We have observed that they match very well the analytic results (summarized in section \ref{sec:NLSM}) for $\phi \in[0,\;0.8\pi]$ but differ for $\phi \in[0.8\pi,\;\pi]$. One reason for that is the almost linear growth of the form factor with $s$. According to \eqref{eq:asymptotics_ON}  we have
\begin{equation}
\lim_{s\rightarrow+\infty}\mathcal{F}_2^\Theta(s)\sim s\,(\ln s)^{-\frac{N-1}{N-2}}.
\end{equation}
The ansatz \eqref{eq:ansatz_2} however does not reproduce such behavior well for any finite value of $N_{max}$. As a result the central charge differs from the one expected in the NLSM. Further analysis is required in order to tame the NLSM numerically.

For section 2 the optimization problem is feasible for $\sigma^*_1 \in [-0.328, 0.328]$. The boundary values correspond to the periodic Yang-Baxter (pYB) solutions \cite{Hortacsu:1979pu,Cordova:2018uop}. We observe numerically that the value of the central charge exhibits a divergence like behavior when approaching the boundary. This might be a sign that the pYB solutions are unphysical. One has to however take into consideration poor convergence of the ansatz, thus further analysis is also required to make a definite statement.

\begin{figure}[t]
	\begin{center}
		\subfigure[]{\label{fig:neutral33}\includegraphics[scale=0.6]{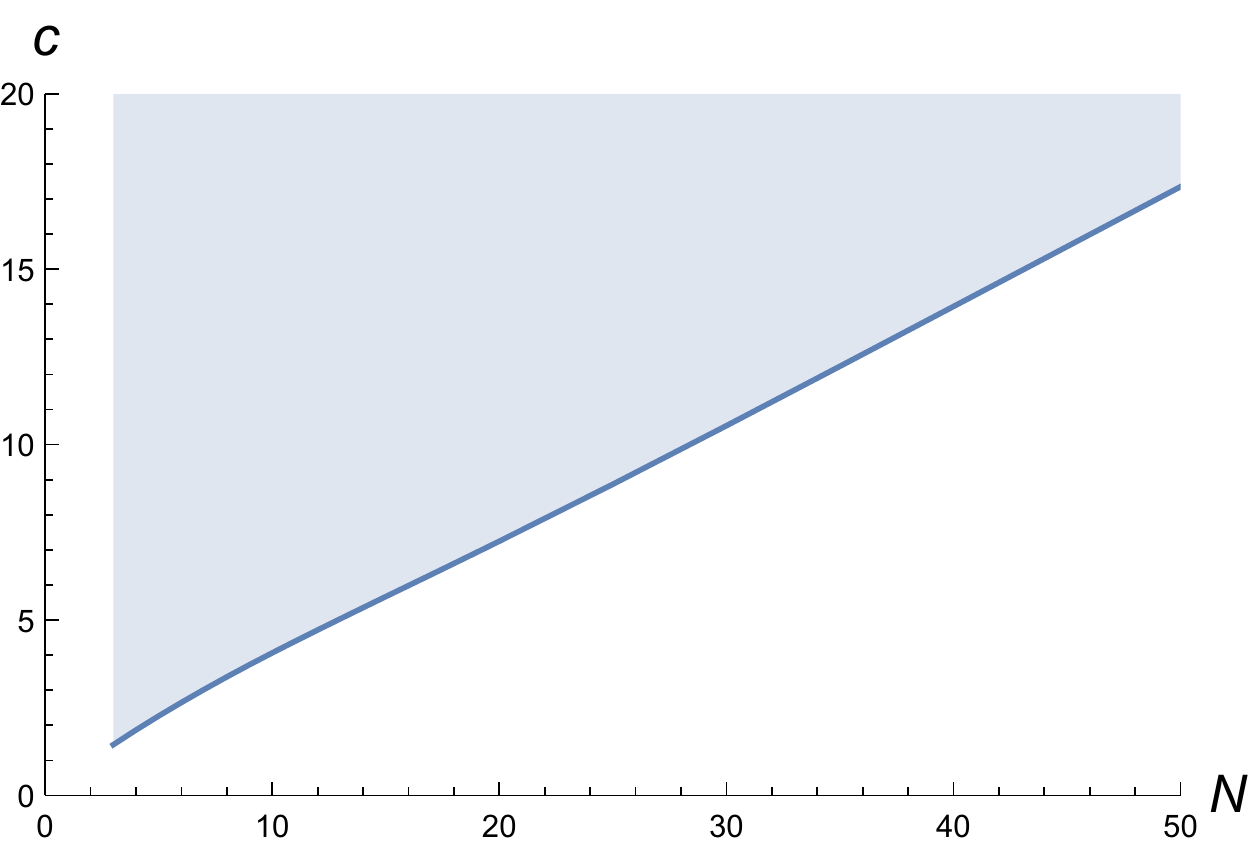}}
		\subfigure[]{\label{fig:neutral44}\includegraphics[scale=0.6]{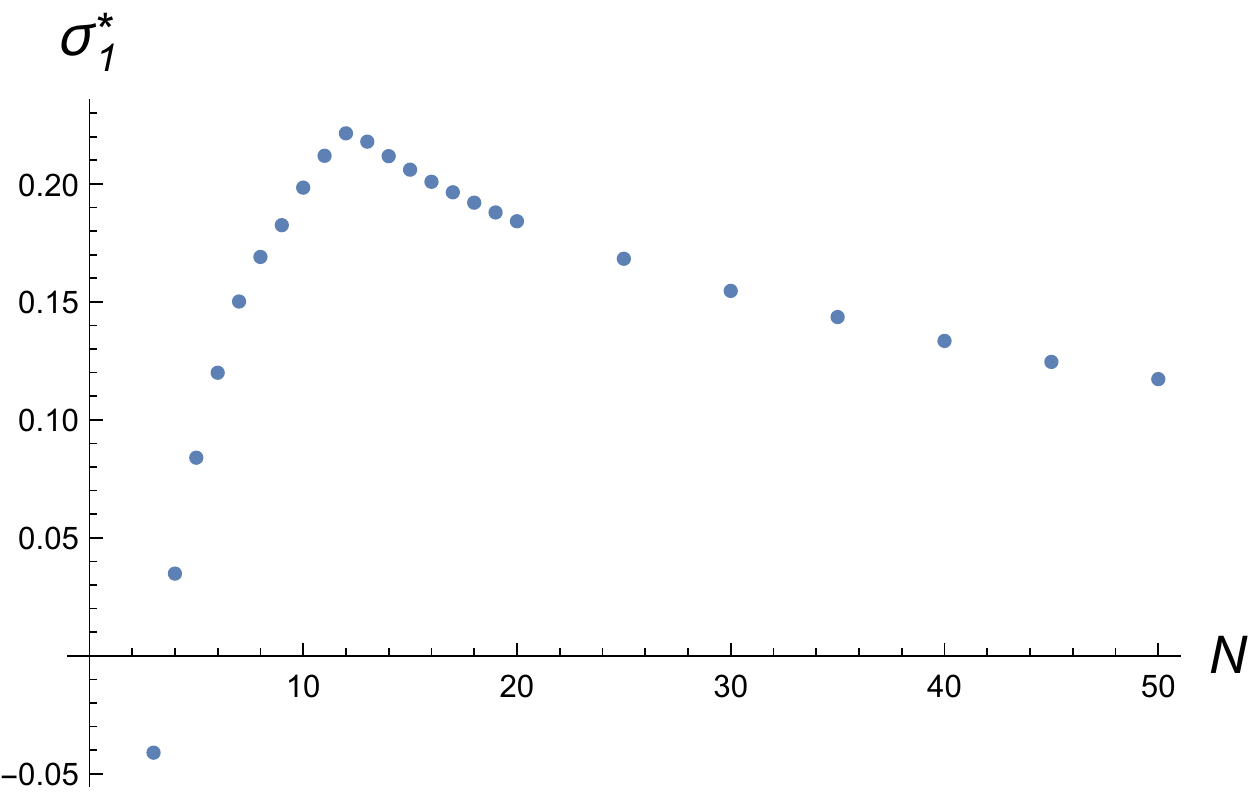}}
		\caption{The plots are constructed with $N_{max}=30$. \emph{Left}: bound on the central charge as a function of $N$. The allowed region is depicted in blue. 
		\emph{Right}: the value of $\sigma_1^*$ of the scattering amplitudes with the minimal central charge.}
		\label{fig:ON_34}
	\end{center}
\end{figure}

To conclude, let us also minimize the central charge for different values of $N$ without fixing $\sigma_1^*$ or $\sigma_2^*$. The result is presented on figure \ref{fig:ON_34}. The bound on the left part of figure \ref{fig:ON_34} is almost linear and can be approximated well by $c\approx0.644+0.334 N$. The values of $\sigma_1^*$ and $\sigma_2^*$ which realize the minimum of the central charge lie on the section 2 \eqref{eq:sectoin_2}. The values of the optimal $\sigma_1^*$ as a function of $N$ are presented on the right part of figure \ref{fig:ON_34}.

\section{Conclusions}
\label{sec:conclusions}
In this paper we have extended the S-matrix bootstrap program to include states created by local operators. This gives rise to a bootstrap setup that mixes scattering amplitudes, form factors and  spectral densities of local operators. 
The latter allows to extract direct information about the UV fixed point which was not  possible so far in the pure S-matrix bootstrap approach.

We have established the groundwork for future explorations and limited ourselves to testing the approach in two dimensional QFTs. Our main result is the derivation of a lower bound for the central charge $c$ of the UV CFT that can flow to a massive phase with a given particle spectrum (and interactions). For example, imposing $O(N)$ global symmetry, in the presence of a single stable particle transforming in the vector representation, we found  the universal lower bound $c \ge c_{min}(N) \approx 0.6+0.3 N$, see figure \ref{fig:ON_34}(a). 

We should however be careful with the meaning of  ``lower bound''. As in all the recent S-matrix bootstrap works \cite{Paulos:2017fhb}, the ``lower bound'' decreases when increasing the number of parameters in the ansatz, \emph{i.e.} $N_{max}$ in equations \eqref{eq:ansatz_1} - \eqref{eq:ansatz_3}. 
Strictly speaking our result is an upper bound for the lower bound.
In all the plots presented in this paper we have taken the value of $N_{max}$ sufficiently large. However, we have not performed a careful convergence analysis. In order to be rigorous, it is important to generalize the functional method of \cite{Cordova:2019lot} to our setup and obtain rigorous lower bounds on the central charge. 

Another direction worth further exploration, is the inclusion of several states created by different local operators. It seems natural to consider the full set of relevant operators of a given CFT. For example, in the 2d Ising model, it would be interesting to consider both $\sigma$ and $\epsilon$. This setup would include form factors for both operators and a $2\times2$ matrix of Wightman two-point functions.

In two spacetime dimensions, it would  be interesting to further explore the connection with integrable models. In our setup, we have observed that the numerical optimization problem  tends to saturate the conditions \eqref{eq:condition_1} - \eqref{eq:condition_3}, which include Watson's equation and absence of particle production. Therefore, similarly to the pure S-matrix bootstrap, we found that the optimal solutions often correspond to integrable theories.
In this work, we encountered amplitudes and form factors of
 the sine-Gordon, $E_8$ and $O(N)$ models.

In the presence of continuous global symmetries, it is natural to study states created by the conserved currents. Notice that form factors of conserved currents also have a natural normalization (at $s=0$) following from the conserved charges.
This should have interesting applications both in $d=2$ and in higher dimensions. 
In $d=2$, it seems clear that a detailed study of the $O(N)$ model with our approach will benefit from the inclusion of states created by the non-abelian currents. Moreover, it would be useful to obtain the 3D plot of the central charge lower bound above the allowed region in figure \ref{fig:monolith}.

Let us conclude by discussing our new bootstrap method in higher dimensions. In $d\geq 3$ one has to consider the Wightman two-point function of the full stress-tensor $T^{\mu\nu}$ and not only its trace $\Theta =T_\mu^\mu$. Such two-point function can be decomposed into two spectral densities
\begin{equation}
\label{eq:sp_dens}
\rho_{\Theta}(s),\quad
\rho_2(s),
\end{equation}
where the first spectral density represents the trace of the stress-tensor exactly as in $d=2$ and the second spectral density is the new object special to $d\geq 3$. In all dimensions we have the following asymptotic behavior at large energies\footnote{
This follows from $\Theta(x) \propto g\, \mathcal{O}(x)$ and \eqref{eq:spectral_density_CFT}.
 }
\begin{equation}
\label{eq:prop_1}
\lim_{s\rightarrow+\infty} \rho_{\Theta}(s) \propto  g^{2} s^{\Delta_{r}-\frac{d}{2}}\,,
\end{equation}
corresponding to the relevant deformation $g \int d^dx\, \mathcal{O}(x)$ of the UV CFT by an operator of dimension $\Delta_r <d$ and $g$ is a dimensionful coupling constant with the mass dimension $[g]=d-\Delta_r$.
The value of the central charge in $d=2$ is hidden inside $\rho_{\Theta}$ and can be extracted using \eqref{eq:c_final}. We reproduce it here for convenience
\begin{equation}
\label{eq:prop_2}
c_{UV} = (2\pi)^2\times\frac{3}{\pi}
\int_0^\infty ds\, \frac{\rho_\Theta(s)}{s^2}.
\end{equation}
In $d\geq 3$ there is no known analogous integral expression and the value of the central charge is hidden instead in the asymptotics of the second spectral density in \eqref{eq:sp_dens}. More precisely
\begin{equation}
\label{eq:prop_3}
\lim_{s\rightarrow +\infty} \rho_{2}(s) = const\times C_T^{UV} s^{d/2}.
\end{equation}
Here $const$ is a numerical factor which depends on the precise definition of $\rho_2$ and is irrelevant for the present discussion. Reiterating, even though the spectral densities \eqref{eq:sp_dens} allow to access the values of the central charge in any number of dimensions, the information about it is encoded differently in  them in $d=2$ and in $d\geq 3$. Interestingly enough, the integral formula \eqref{eq:c_2d} captures both \eqref{eq:prop_2} and \eqref{eq:prop_3} in a uniform way. For more details see \cite{Karateev:2020axc}.

Our numerical bootstrap approach can easily bound integrals like \eqref{eq:prop_2} but it cannot bound coefficients in the asymptotic behaviour of spectral densities like in  \eqref{eq:prop_3}. For this reason, we cannot use our bootstrap method to put non-trivial bounds on the central charge $C_T^{UV}$ in $d\geq 3$.
What we can do instead in higher dimensions  is to put lower bounds on the following dimensionless quantity
\begin{equation}
\label{eq:higher_d}
\int_0^{\infty} \frac{ds}{s^{d/2+1}}\,\rho_{\Theta}(s),
\end{equation}
which is a simple generalization of \eqref{eq:prop_2} to higher dimension. We stress however that contrary to  \eqref{eq:prop_2} this quantity 
is not directly related to a property of the UV CFT and is thus less interesting than \eqref{eq:prop_2}, see \cite{Cappelli:1990yc} for further discussion. Focusing to $d=4$ there is a more interesting quantity we can bound which is the $a$-anomaly. This requires however some modifications in the present formalism. We plan to address this question in the near future.


\section*{Acknowledgements}

We are grateful to all our colleagues at EPFL for  numerous helpful discussions. In particular we thank Marc Gillioz, Aditya Hebbar, Andrei Khmelnitsky, Alexander Monin, Riccardo Rattazzi and Matt Walters. The authors are very grateful to Slava Rychkov for discussions and several crucial suggestions. DK and JP thank the organizers and participants of the Bootstrap 2019 Workshop at Perimeter Institute, of the Workshop on Non-Perturbative Methods in Quantum Field Theory at ICTP in Trieste and of the Workshop on S-matrix Bootstrap at ICTP-SAIFR in S\~ao Paulo. In particular we thank Joan Elias Miro, Victor Gorbenko, Andrea Guerrieri, Petr Kravchuk, Balt van Rees, Pedro Vieira and Bernardo Zan. DK also thanks the organizers of the Workshop on Challenges in Theoretical High-Energy Physics at Nordita in Stockholm for hospitality and Gabriele Ferretti for useful discussions.

DK and JP are supported by the Simons Foundation grant 488649 (Simons Collaboration on the Nonperturbative Bootstrap) 
and by the Swiss National Science Foundation through the project
200021-169132 and through the National Centre of Competence in Research SwissMAP.

\appendix

\section{Definitions and auxiliary results}
\label{app:auxiliary}
Here we summarize basic definitions and various auxiliary results used throughout the paper 

\paragraph{Fourier transformation}
The Fourier transform $\hat f(p)$ of a function $f(x)$ is given by
\begin{equation}
\hat f(p) = \int d^dx\, e^{-ix\cdot p}f(x),\quad
f(x) = \int \frac{d^dp}{(2\pi)^d}\, e^{ix\cdot p} \hat f(p).
\end{equation}
The Dirac $\delta$-function is
\begin{equation}
(2\pi)^d \delta^{(d)}(p)=\int d^d x\,e^{ip\cdot x}.
\end{equation}

\paragraph{Spherical coordinates}
We will need to evaluate $n$-dimensional integrals in Euclidean signature. It is best done in spherical coordinates which we introduce here. The $n$-dimensional spherical coordinates consist of the radius $r$ and a set of $n-1$ angles with the following ranges
\begin{equation}
\theta_1,\ldots,\theta_{n-2}\in[0,\pi],\quad
\theta_{n-1}\in[0,2\pi].
\end{equation}
They are related to the Cartesian coordinates as
\begin{equation}
\label{eq:axis_1}
\begin{aligned}
x^1 &=r \cos\theta_1,\\
x^2 &=r \cos\theta_2\sin\theta_1,\\
x^3 &=r \cos\theta_3\sin\theta_2\sin\theta_1,\\
&\ldots\\
x^{n-1} &=r \cos\theta_{n-1}\sin\theta_{n-2}\ldots\sin\theta_1,\\
x^{n} &=r \sin\theta_{n-1}\sin\theta_{n-2}\ldots\sin\theta_1.
\end{aligned}
\end{equation}
The Jacobian $J$ of the variable change from Cartesian to spherical coordinates reads as
\begin{equation}
\label{eq:jacobian}
J (r;\theta_1,\ldots,\theta_{n-2}) = r^{n-1}\sin^{n-2}\theta_1\sin^{n-1}\theta_2\ldots\sin^{2}\theta_{n-3}\sin\theta_{n-2}.
\end{equation}
The infinitesimal spherical angle in $n$-dimensional space is then
\begin{equation}
d\Omega_n = J(1;\theta_1,\ldots,\theta_{n-2}) \times d\theta_1\ldots d\theta_{n-1}.
\end{equation}
It is then straightforward to evaluate the spherical angle $\Omega_n$\footnote{Notice that often the spherical angle is denote by $\Omega_{n-1}$ in $n$-dimensions.}
\begin{equation}
\label{eq:spherical_angle}
\Omega_n=\frac{n\,\pi^{n/2}}{\Gamma(n/2+1)}.
\end{equation}
The $n=1$ case is special. We do not have angles, however we have two (equivalent) directions with $x\geq 0$ and $x<0$. This fact is already contained in \eqref{eq:spherical_angle} since $\Omega_1=2$.
Finally, the spherical $\delta$-function is given by
\begin{equation}
\label{eq:delta_spherical}
\delta^{(n-1)}(\Omega) = \frac{\delta(\theta_1)\ldots \delta(\theta_{n-2})}{J(1;\theta_1,\ldots,\theta_{n-2})} \times\delta(\theta_{n-1}).
\end{equation}

\paragraph{Spherical harmonics}
Any normalizable function on the sphere $S^{n-1} \subset \mathbb{R}^n$ can be decomposed in the basis of spherical harmonics.
If the function is invariant under $SO(n-1)$ rotations (therefore only depends on $\theta_1$) then it can be expanded in  Gegenbauer polynomials
\begin{equation}
\label{eq:spherical_harmonics}
C_j^{(n-2)/2}(x),\quad x\equiv\cos\theta_1,
\end{equation}
where $j=0,1,2,\ldots$ is a non-negative integer which is often referred to as spin.
In the $n=3$ case, the Gegenbauer polynomials coincide with the Legendre polynomials
\begin{equation}
P_j(x) = C_j^{1/2}(x).
\end{equation}
The Gegenbauer polynomials satisfy the following   orthogonality property
\begin{equation}
\label{eq:orthogonality_1}
\int_{-1}^{+1}C_{j'}^{k}(x) C_j^{k}(x)(1-x^2)^{k-1/2} dx = \nu^k_j\times\delta_{j'j},\quad
\nu^k_j\equiv\frac{2^{1-2k}\pi\Gamma(2k+j)}{j!(k+j)\Gamma(k)^2},
\end{equation}
and completeness relation
\begin{equation}
\label{eq:orthogonality_2}
\sum_{j}
\frac{1}{\nu^k_j}\times
C_{j}^{k}(x) C_{j}^{k}(y) = (1-x^2)^{1/2-k}\times\delta(x-y).
\end{equation}

\paragraph{Change of variables}
Consider a two particle state with the $(d-1)$-momenta $\vec p_1$ and $\vec p_2$. Let us perform the following change of variables\footnote{In writing this we have used the center of mass frame $\vec p_1 = - \vec p_2$, where $\vec p_1$ is aligned with $x^1$ axis.}
\begin{equation}
\label{eq:change_variables}
d^{d-1}\vec p_1\times d^{d-1}\vec p_2 =
d^{d-1}(\vec p_1 + \vec p_2)\times d^{d-1}\vec p_1=
d^{d-1}(\vec p_1 + \vec p_2)\times |\vec p_1|^{d-2}d|\vec p_1|\;d\Omega_{d-1},
\end{equation}	
where $p_1^0$ and $p_2^0$ are given by the mass shell condition \eqref{eq:on-shell_condition} and thus
\begin{equation}
d|\vec p_1| = \frac{p_1^0p_2^0}{|\vec p_1|\,(p_1^0+p_2^0)}\;d(p_1^0+p_2^0).
\end{equation}
Defining the total $d$-momentum
\begin{equation}
p^\mu \equiv p_1^\mu + p_2^\mu,
\end{equation}
we can write
\begin{equation}
d^{d-1}\vec p_1\times d^{d-1}\vec p_2 =
d^{d}p\times |\vec p_1|^{d-3}
\frac{p_1^0p_2^0}{p^0}\times d\Omega_{d-1}.
\end{equation}	
Equivalently we have
\begin{equation}
\label{eq:change_of variables}
\frac{d^{d-1}p_1}{(2\pi)^{d-1}}\frac{1}{2p^0_1}
\times
\frac{d^{d-1}p_2}{(2\pi)^{d-1}}\frac{1}{2p^0_2}
=
\frac{1}{\mathcal{N}_d}\,
\frac{d^d p}{(2\pi)^d}\,
\frac{d\Omega_{d-1}}{(2\pi)^{d-2}},
\end{equation}
where we have defined
\begin{equation}
\label{eq:factor_N_full}
\mathcal{N}_d \equiv 4p^0|\vec p_1|^{3-d} =
2^{d-1} \sqrt{s} \,\Big(s-2\,(m_1^2+m_2^2)+s^{-1}(m_1^2-m_2^2)^2\Big)^{(3-d)/2},\quad
p^0=\sqrt{s}.
\end{equation}

\paragraph{Phase space}
Let us write explicitly the phase space for scalar identical particles
\begin{align}
\nn
d\Phi_n &= \frac{1}{n!}
\frac{d^dp_1}{(2\pi)^d}\ldots\frac{d^dp_\n}{(2\pi)^d}\times
2\pi\theta(p_1^0)\delta(p^2_1+m^2)\ldots 2\pi\theta(p_\n^0)\delta(p^2_\n+m^2)\\
&= \frac{1}{n!}
\frac{d^{d-1}p_1}{(2\pi)^{d-1}}\frac{1}{2p^0_1}\ldots
\frac{d^{d-1}p_\n}{(2\pi)^{d-1}}\frac{1}{2p^0_\n}.
\label{eq:phase_space}
\end{align}
Here all the energies $p^0_i$ satisfy the mass-shell condition \eqref{eq:on-shell_condition}. Notice the presence of the $1/n!$ factor which removes overcounting of indistinguishable (identical) particles.

\paragraph{Wightman functions} In Euclidean signature only the time-ordered correlation functions make sense. In Lorentzian signature we also have non time-ordered correlators known as the Wightman functions.\footnote{See for example appendix B in \cite{Simmons-Duffin:2016gjk}.}  They are defined as follows
\begin{equation}
\label{eq:wightman}
\<0|\phi(\hat t_1,\vec x_1)\ldots\phi(\hat t_n,\vec x_n)|0\>,\quad \hat t_j\equiv t_j-i\epsilon_j,\quad
\epsilon_1>\ldots\epsilon_n>0.
\end{equation}
Roughly speaking, the presence of $\epsilon$'s is required in order to introduce a damping factor. For instance consider the 2-point Wightman function. Using \eqref{eq:operator_transformation} we can write
 \begin{equation}
\<0|\phi(\hat t_1,\vec x_1)\phi(\hat t_2,\vec x_2)|0\>=
\<0|\phi(0,\vec x_1)e^{iH(t_2-t_1)}e^{-H(\epsilon_1-\epsilon_2)}\phi(0,\vec x_2)|0\>.
 \end{equation}

\section{K\"{a}ll\'en-Lehmann representation in Euclidean signature}
\label{app:KL_euclidean}
In Euclidean signature there is no notion of Wightman correlation functions, only time-ordered correlators are well defined. Here we obtain the K\"{a}ll\'en-Lehmann representation of a Euclidean two-point function by applying the Wick rotation to \eqref{eq:spectral_representation_3}.

Consider the Feynman propagator \eqref{eq:feynman_propagator}. The integrand has poles at
\begin{equation}
\label{eq:poles}
q^0=\pm\sqrt{\vec q\,^2+\mu^2}\mp i\epsilon.
\end{equation}
The integration goes along the real $q^0$ values and thus the $i\epsilon$'s are needed to avoid the presence of poles on the line of integration. We can now rotate the line of integration by $+\pi/2$. This is done in such a way that we do not cross the poles. It is represented by the change of variables
\begin{equation}
\label{eq:wick_rotation}
x^0_E=ix^0,\qquad
q^0_E=iq^0,
\end{equation}
which is known as the Wick rotation.\footnote{More precisely the rotation of the energy $p^0$ to purely imaginary values is done by $p^0\rightarrow p^0 e^{i\phi}\rightarrow ip^0\equiv p^0_E$, where the angle $\phi$ changes from $0$ to $+ \pi/2$. The rotation of the time is determined by the condition to keep the scalar product $(p\cdot x)=p^0 x_0+\vec p\cdot \vec x$ real, which is important to require in order not to introduce any divergences in the Fourier transform. This leads to the following $x_0\rightarrow x_0 e^{-i\phi}\rightarrow -ix_0\equiv x_{E0}=x_E^0$.} The subscript $E$ stands for Euclidean. Applying the change of coordinates \eqref{eq:wick_rotation} to \eqref{eq:feynman_propagator} we obtain the Euclidean propagator
\begin{equation}
\label{eq:euclidean_propagator}
\Delta_F(x;\mu^2) = i\Delta_E(x_E;\mu^2),\quad
\Delta_E(x_E;\mu^2)\equiv\int \frac{d^d q_E}{(2\pi)^d}\;e^{iq_E\cdot x_E}\; \frac{1}{q_E^2+\mu^2},
\end{equation}
where we have
\begin{equation}
q_E\cdot x_E = q_E^0 x_E^0+\vec q_E\cdot \vec x_E,\quad
q_E^2 = (q_E^{0})^2+\vec q_E^{\;2}.
\end{equation}
Notice the absence of $i\epsilon$'s. They can now be set to zero since there are no poles on the line of integration.
Plugging \eqref{eq:euclidean_propagator} into \eqref{eq:spectral_representation_3} we obtain the Euclidean K\"{a}ll\'en-Lehmann representation
\begin{equation}
\label{eq:spectral_representation_3_euclidean}
\<0| \cO^\dagger(x_E) \cO(0)|0\>_T = \int_0^\infty d\mu^2 \rho(\mu^2) \Delta_E(x_E;\mu^2).
\end{equation}

\section{Free scalar theory}
\label{sec:free_boson}
Let us consider the free field theory with a single real scalar field $\phi(x)$. It is defined via the Lagrangian density
\begin{equation}
\mathcal{L}_{\text{free}}(x) = -\frac{1}{2} \left(\partial \phi(x)\right)^2-\frac{1}{2} m^2\phi(x)^2.
\end{equation}
From this Lagrangian density the Klein-Gordon equation of motion follows
\begin{equation}
\left(-\partial^2+m^2\right)\phi(x) = 0.
\end{equation}
It has the following general solution
\begin{equation}
\label{eq:free_field}
\phi(x) = \int \frac{d^d p}{(2\pi)^d} \theta(p^0)(2\pi)\delta(p^2+m^2)
\left(a(p)\,e^{ip\cdot x}+b^\dagger(p)\,e^{-ip\cdot x}\right),
\end{equation}
where $a(p)$ and $b^\dagger(p)$ are some operator valued functions of the $d$-momenta $p^\mu$ called the annihilation and creation operators respectively. The reality condition $\phi^\dagger(x)=\phi(x)$ implies $a(p)=b(p)$. The operators $a$ and $a^\dagger$ are required to satisfy the standard commutation relations
\begin{equation}
\label{eq:commutator}
[a(p),a(p')]=0,\quad
[a(p),a^\dagger(p')]=2p^0\times(2\pi)^{d-1}\delta^{(d-1)}(\vec p\,{}'-\vec p).
\end{equation}
Acting on the vacuum they create  $n$-particles states
\begin{equation}
\label{eq:n_particle_free}
|\n\> = a^\dagger(k_1)\ldots a^\dagger(k_n) |0\>,\quad
a(k) |0\> =0.
\end{equation}
Due to \eqref{eq:commutator}, the $n$-particles states \eqref{eq:n_particle_free} are normalized exactly as required by \eqref{eq:normalization_1PS} and \eqref{eq:normalization_2PS} for $n=1$ and $n=2$ respectively. Free theory provides an explicit construction of the Hilbert space discussed in the beginning of section \ref{eq:states}.

The stress-tensor for the free real scalar field reads as
\begin{equation}
\label{eq:ST_free}
T^{\mu\nu} (x) =\; :(\partial^\mu\phi)(\partial^\nu\phi)-\frac{1}{2}\,\eta^{\mu\nu}
\left( (\partial\phi)^2+m^2\phi^2\right):.
\end{equation}
The operators are enclosed between two symbols ``:'' denoting normal-ordering. The expression for the trace of the stress-tensor follows from \eqref{eq:ST_free} and reads as
\begin{equation}
\Theta(x)\equiv \eta_{\mu\nu}T^{\mu\nu} (x)=\;:\left(1-\frac{d}{2}\right)(\partial\phi)^2-\frac{d}{2}\,m^2\phi^2:.
\end{equation}

In the remainder of this appendix let us focus on the the case of $d=2$. The trace of the stress-tensor then reads  
\begin{equation}
\label{eq:trace}
\Theta(x) = -m^2:\phi(x)^2:.
\end{equation}
Let us compute the one and two particle form factors for the operator \eqref{eq:trace}. In the former case we have
\begin{equation}
\label{eq:1ff_free}
\mathcal{F}_1^\Theta = -m^2\,\<0|a_{p_1}:\phi(0)^2:|0\>=0.
\end{equation}
In the latter case we have
\begin{align}
\nn
\mathcal{F}_2^\Theta(p_1,p_2)
&= -m^2\,\<0|a_{p_1}a_{p_2}:\phi(0)^2:|0\>\\
\nn
&=-m^2
\int \frac{d \vec k_1}{2\pi}\frac{1}{2k_1^0}
\int\frac{d \vec k_2}{2\pi}\frac{1}{2k_2^0}
\<0|a_{p_1}a_{p_2}a^\dagger_{k_1}a^\dagger_{k_2}|0\>\\
\nn
&=-m^2
\int \frac{d \vec k_1}{2\pi}\frac{1}{2k_1^0}
\int\frac{d \vec k_2}{2\pi}\frac{1}{2k_2^0}
\<0|[a_{p_1},a^\dagger_{k_1}][a_{p_2},a^\dagger_{k_2}]+[a_{p_1},a^\dagger_{k_2}][a_{p_2},a^\dagger_{k_1}]|0\>\\
&=-2m^2.
\label{eq:2ff_free}
\end{align}
Here we have plugged \eqref{eq:free_field} and used the properties of the creation and annihilation operators \eqref{eq:commutator} and \eqref{eq:n_particle_free}. The expression above is obviously consistent with the normalization \eqref{eq:normalization_2FF_2d}. It is clear that all the other $n\geq 3$ particle form factors of the trace of the stress-tensor vanish in free theory.

We can use \eqref{eq:2ff_free} to compute the spectral density \eqref{eq:spectral_2part},  we get
\begin{equation}
\label{eq;spectral_free_1}
2\pi\rho(s) = \frac{4m^4}{\mathcal{N}_2}.
\end{equation}
Plugging it into \eqref{eq:c_final} we obtain the standard central charge value of the free boson in $d=2$
\begin{equation}
\label{eq:c_free_boson}
c = 1.
\end{equation}
Notice that the two particle states reproduce the entire value of the central charge.

\bibliographystyle{JHEP}
\bibliography{refs}

\end{document}